\documentclass[aps,prx,twocolumn,superscriptaddress,amsmath,amssymb,noeprint,tbtags]{revtex4-2}
\usepackage[english]{babel}

\usepackage{graphicx}
\usepackage{mathtools}
\usepackage{feynmp-auto}
\usepackage{bm}
\usepackage{xr}
\allowdisplaybreaks[1]

\usepackage[colorlinks=True,urlcolor=black]{hyperref}
\usepackage{cleveref}

\makeatletter
\newcommand*{\addFileDependency}[1]{\typeout{(#1)}\@addtofilelist{#1}
\IfFileExists{#1}{}{\typeout{No file #1.}}
}\makeatother

\newcommand{\nzero}[1]{\overline{n}_{0#1}}
\newcommand{\cprime}[2]{#2}
\renewcommand{\bar}[1]{\bm{#1}}

\newcommand{\Sfields}{S( { \tilde \phi} , \phi  )}
\newcommand{\Sfieldsbaseline}{S_0( { \tilde \phi} , \phi  )}

\newcommand{\leglabel}[1]{$#1$}

\begin{document}
	\title{Dynamical Variational Theory of Stochastic Reaction Networks
	}

\author{Moshir Harsh}
\address{Institut f{\"u}r Theoretische Physik, Georg-August-Universit{\"a}t G{\"o}ttingen, Göttingen 37073, Germany}
\thanks{Present affiliation: Harvard Medical School, Boston 02115, USA}

\author{Peter Sollich}
\email{peter.sollich@uni-goettingen.de}
\address{Institut f{\"u}r Theoretische Physik, Georg-August-Universit{\"a}t G{\"o}ttingen, Göttingen 37073, Germany}
\address{Department of Mathematics, King's College London, London WC2R 2LS, UK}

\date{\today}

\begin{abstract}

Discrete stochastic population dynamics describe systems ranging from biochemical reaction networks and gene regulation to epidemic spreading, ecology, and population genetics, where finite populations generate strong intrinsic fluctuations. Predicting their dynamics, including mean populations, fluctuations, time-lagged correlations, response functions, and steady states, remains challenging: the chemical master equation admits no general analytical solution, numerical solutions scale poorly with system size, and standard approximations fail at low copy numbers.
Using a field-theoretic representation of the reaction dynamics, we develop a dynamical variational approximation for stochastic reaction networks based on the Plefka expansion from spin-glass physics. For reaction systems with general stoichiometry and time-dependent rates, the expansion yields
tractable non-Markovian effective dynamics with self-consistent memory. We further introduce a generalized self-consistent bubble resummation that provides closed predictions for means, correlations, and response functions.
The formulation applies broadly to discrete stochastic population dynamics on networks, and we provide a numerical implementation with adaptive time stepping for user-defined systems.
We benchmark the method on stochastic gene regulation, enzyme kinetics, and infection dynamics, where it accurately captures both time-dependent behaviour and steady states, including fluctuation-driven effects missed by conventional approximations. Together, these results establish a general analytical framework for stochastic reaction dynamics across transient and stationary regimes in strongly fluctuating low-copy-number systems.

\end{abstract}

\maketitle

\begin{fmffile}{diagrams_plefka/fields}

	\fmfcmd{
		style_def end_arrow expr p =
		cdraw p;
		cfill (harrow (p, 0.5));
		enddef;}

	\fmfcmd{
		style_def end_arrow_two expr p =
		cdraw p;
		cfill (harrow (p, 0.6));
		enddef;}

	\fmfcmd{
		style_def end_arrow_dash expr p =
		draw_dashes p;
		cfill (harrow (p, 0.4));
		enddef;}

	\fmfcmd{
		style_def end_arrow_dash_two expr p =
		draw_dashes p;
		cfill (harrow (p, 0.5));
		enddef;}

	\fmfcmd{
		style_def double_arrow expr p =
		draw_double p;
		cfill (arrow p);
		enddef;
	}

	\fmfcmd{
		style_def mixed_response_GB expr p =
		save oldpen;
		pen oldpen;
		oldpen := currentpen;
		pickup oldpen scaled 3;
		draw (subpath( (0 ,0.5)*length(p)) of p) withcolor green;
		draw (subpath( (0.5,1)*length(p)) of p) withcolor blue;
		pickup oldpen;
		cullit ; undraw p; cullit;
		cfill (marrow (p, 0.5));
		enddef;}

	\fmfcmd{
		style_def mixed_response_BG expr p =
		save oldpen;
		pen oldpen;
		oldpen := currentpen;
		pickup oldpen scaled 3;
		draw (subpath( (0 ,0.5)*length(p)) of p) withcolor blue;
		draw (subpath( (0.5,1)*length(p)) of p) withcolor green;
		pickup oldpen;
		cullit ; undraw p; cullit;
		cfill (marrow (p, 0.5));
		enddef;}

\section{Introduction}

Modelling classical stochastic population dynamics systems in a way that retains the intrinsic noise due to the discrete nature of its constituents has gained prevalence in recent years, mainly on account of improvements in measurement technologies that can resolve these systems down to each individual constituent. The interesting regime where new phenomenology due to such discreteness can be observed, via significant deviations from mass-action-like descriptions in terms of concentrations, is the one of small populations. These engender large stochastic effects, i.e.\ large fluctuations and small signal-to-noise ratios in observed trajectories.

An important example of such situations arises in chemical reactions~\cite{cardelli_stochastic_2016,gillespie_perspective_2013,lecca_stochastic_2013} inside cells, with atto- to femtolitre sized compartments for which, at physiological concentrations, some chemical species occur in low molecular copy numbers. The resulting intrinsic stochasticity has been experimentally shown to be a key feature in biochemical reaction networks such as gene regulation networks~\cite{ozbudak_regulation_2002,bar-even_noise_2006, elowitz_stochastic_2002,golding_real-time_2005,blake_noise_2003}. Experiments have also demonstrated vastly different outcomes in otherwise identical cells~\cite{elowitz_stochastic_2002, ozbudak_regulation_2002}, which have been attributed to random fluctuations in gene expression. Cellular functions are designed to operate with this stochasticity, as can be seen for example by the fact that in many cases, a few copies of a signalling molecule or transcription factor suffice to trigger signalling cascades in a cell~\cite{ladbury_noise_2012}.

\begin{figure*}[t]
	\includegraphics[width=\linewidth]{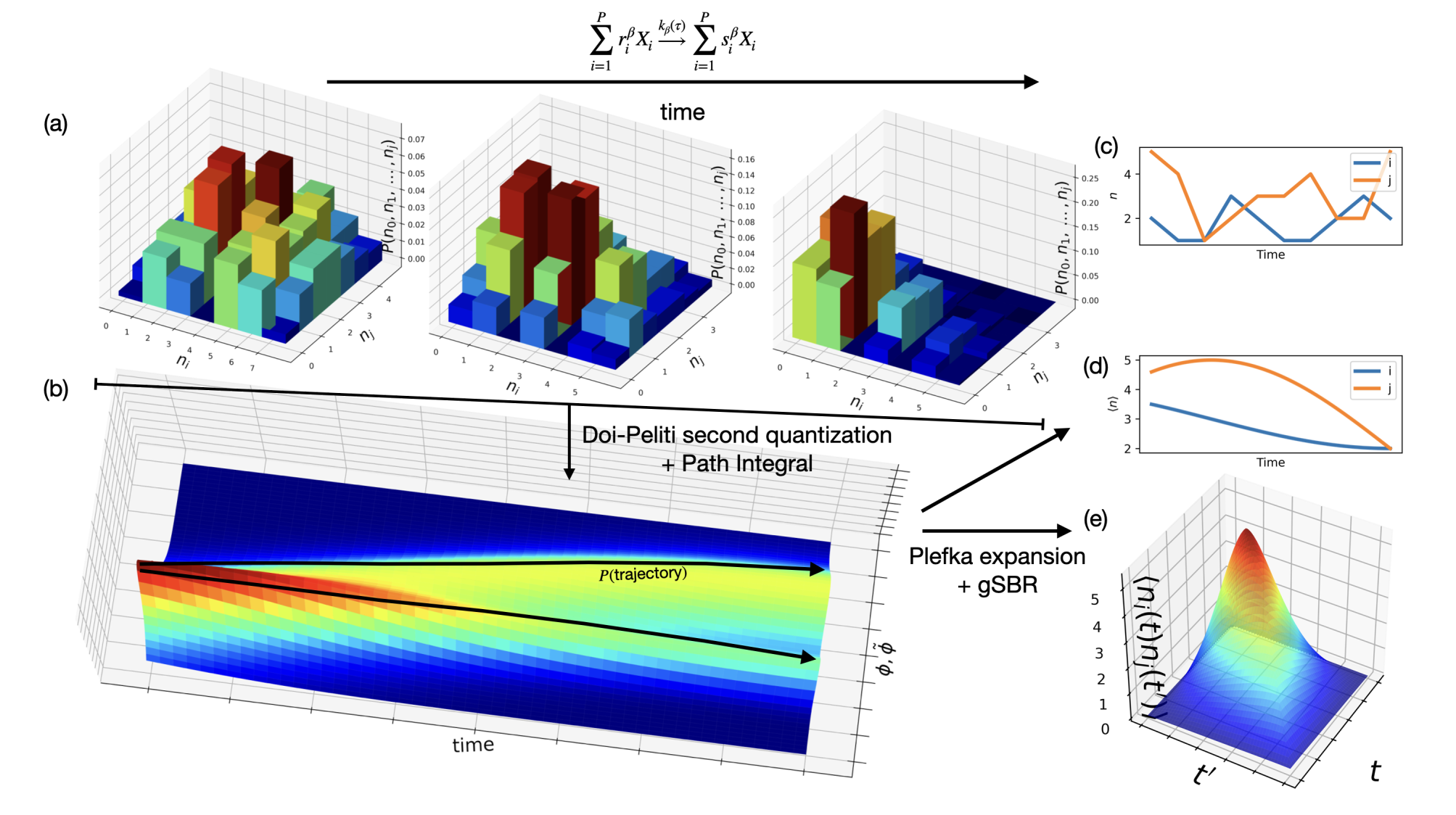}
	\caption[Method schematic]{Method schematic: (a) The $P$-species discrete distribution evolves under the reaction network defined by~\cref{general_reaction_form} starting from some initial conditions. (b) Using Doi-Peliti second quantization this is mapped to a continuous field theory with a coherent state path integral defining the probability of observing any given trajectory $(n_0(t),n_1(t),\dots,n_P(t))$ from $t=0$ to some final time. (c) In the regime of small copy numbers, the time traces of the system show large intrinsic fluctuations where most approaches like mean field approximations fail. (d) The Plefka expansion with the gSBR Ansatz can be used to accurately approximate statistical quantities like time evolution of mean copy numbers and non-equilibrium temporal correlation functions.}
	\label{fig_schematic}
\end{figure*}

Other examples of stochastic population dynamics range from ecological or agent based models with small populations~\cite{gardiner_handbook_1985} in ecological niches, to fluctuations in species close to extinction levels where the discrete numbers of individuals is of obvious importance~\cite{assaf_extinction_2010, assaf_wkb_2017}. The associated rare events also affect, for example, the outbreak size distributions~\cite{hindes_outbreak_2022} in epidemics. New phenomena such as noise-induced Turing patterns~\cite{karig_stochastic_2018} can also be observed; these arise from the intrinsic stochasticity in small populations and so are absent in the deterministic limit of large populations.

Although the stochastic process models described above are relatively straightforward to formulate, in terms of a master equation on the joint probability distribution of copy numbers of all species, no general methods exist for a time-dependent solution; and only very few approximation schemes~\cite{schnoerr_approximation_2017} are useful in our regime of interest, for calculating e.g.\ the time-dependent moments of the copy number distribution. This problem is particularly challenging because, in a Langevin description, the intrinsic noise is in general imaginary and state-dependent
as discussed below. Numerical methods such as stochastic simulation algorithms (SSA),
of which the Gillespie algorithm is the most popular, can exactly sample the system trajectories but the computational effort for systems containing many distinct species limits their application. Sophisticated methods such as tensor networks~\cite{nicholson_quantifying_2023} can sample rare fluctuations, but remain expensive for reliable calculations of copy number statistics. In addition, likelihood-based inference is impossible from these numerical methods. The commonly employed analytical approximation of mass action kinetics (MAK), which uses deterministic mean-field rate laws, assumes continuous concentration variables and is known to produce severe deviations~\cite{mcquarrie_stochastic_1967, renyi_treating_1953} from the true stochastic dynamics in populations with small copy numbers. Other commonly employed approaches like the Linear Noise Approximation (LNA)~\cite{cardelli_stochastic_2016}, Moment Closure~\cite{sukys_momentclosurejl_2021,schnoerr_approximation_2017}, Fokker-Plank equations~\cite{grima_how_2011}, Kramers-Moyal~\cite{kramers_brownian_1940,moyal_stochastic_1949} expansion and Effective Mesoscopic Rate Equations (EMRE)~\cite{grima_effective_2010} typically also fail~\cite{schnoerr_approximation_2017} in this challenging regime of small copy numbers with large fluctuations.

The problem of approximating the solutions of master equations for stochastic discrete population dynamics is thus an outstanding open challenge with significant ramifications for the study of (bio-)chemical reaction networks, ecological models etc. Our work here presents a leap forward in its analysis by applying a dynamical approach based on the Plefka expansion~\cite{plefka_convergence_1982,bravi_extended_2016}, originally founded in spin glass physics to derive the celebrated Thouless-Anderson-Palmer (TAP)~\cite{thouless_solution_1977} equations, to a field theory~\cite{tauber_applications_2005,doi_second_1976,peliti_path_1985} for stochastic population dynamics or chemical reactions.
The Plefka free energy is a constrained free-energy functional, whose stationarity conditions provide a variational principle for deriving the equations governing the dynamical observables, i.e.\ (generally non-stationary, out-of-equilibrium) averages and correlations in the system. A key feature of the resulting effective dynamics is the emergence of memory: although the underlying stochastic reaction process is Markovian, the effective corrections are non-Markovian and appear as self-consistent memory terms that encode the feedback of unresolved higher-order moments onto the retained means, two-time correlations, and response functions.

We present our analysis in the context of chemical reaction networks with dynamics described by the chemical master equation~\cite{gillespie_rigorous_1992}, but the formalism extends straightforwardly to other Markovian jump processes or interacting population dynamics systems with polynomial dependence of transition rates on the population size.
In the most important case of reaction networks with at most binary interactions, we provide the expressions for the variational dynamical equations in a convenient closed form, which only depends on the system's stoichiometry and so can be evaluated for arbitrary (binary) networks. Interactions of higher order can then be captured indirectly as sequences of binary reactions proceeding through short-lived intermediate states~\cite{Atkins2018, Laidler1987, Steinfeld1999}.

To construct the approximation, we first demonstrate how the Mode Coupling Approximation (MCA) emerges systematically at the second order in the extended Plefka expansion. Going beyond the MCA, we propose a well-motivated regularizing ansatz for the fields, the generalized Self-consistent Bubble Resummation (gSBR), which is inspired by diagrammatic perturbation theory~\cite{harsh_accurate_2023, harsh_dynamics_2024}. This extends the applicability of our method to a very large range of reaction network structures and reaction rates.
This approach not only outperforms other approximation methods in predicting the marginal time-dependent distributions and two-time correlations; it also gives access to response functions and, ultimately, the probability of observing any dynamical system trajectory, making it a potential basis for likelihood-based inference.

To accompany the theoretical considerations in this paper, we have developed a plug-and-play computational implementation, namely \textit{Adaptive Chemical Dynamics from Memory} or \href{https://github.com/btemoshir/Achedmy.jl}{\texttt{Achedmy.jl}} (\href{https://github.com/btemoshir/Achedmy.jl}{https://github.com/btemoshir/Achedmy.jl}). This allows the user to define the system of interest using the popular framework \texttt{Catalyst.jl}~\cite{CatalystPLOSCompBio2023} in the \texttt{Julia} scientific programming language~\cite{julia_prog_lang} and then to obtain the solution within our approximation(s). Under the hood, the code numerically solves the two-time equations appearing in our approximation method, using an adaptive time-step scheme~\cite{meirinhos_adaptive_2022}. The resulting computational efficiency makes this particularly useful for capturing both initial fast relaxations and a potentially slow approach to equilibrium, and when dealing with large variations among reaction rates in systems with multiple reactions. The three paradigmatic examples illustrating our method also use this computational tool; we simply define the relevant systems and then run the analysis using the general purpose code. ~\texttt{Jupyter} notebooks for these examples are provided with the code.

In concrete terms, we first apply our technique to a model of a gene regulation network, where stochasticity is crucial due to the handful of copies of genes corresponding to each protein in the cell resulting in the well-known bursting behaviour. We show how our approach leads to exponential corrections to mass action kinetics and closely reproduces the true, non-stationary time-lagged correlations.
We also apply the method to stochastic enzyme kinetics with only a single enzyme molecule present on average, again resulting in large fluctuations that are captured by our approach.
We finally consider a model of infection dynamics in a small population where we demonstrate how our method corrects the mean field dynamics, which severely overestimates the size of the infection outbreak. Infection models can be thought of as chemical reactions, but they are typically studied in the fields of population dynamics or ecology or epidemiology. Thus, this last example is included to demonstrate the versatility of our method, including applications to models that are not typically interpreted as chemical reactions.

The manuscript is organised as follows: in~\cref{sec:stochastic dynamics} we introduce the description of the discrete stochastic dynamics, namely the Master equation, construct its path integral representation and tailor it to chemical reaction networks. In~\cref{sec:effective_action_section} we introduce the Plefka expansion and in~\cref{sec:extended_Plefka} its extended version. Both are applied in turn to reaction networks, obtaining dynamical equations for the order parameters of the system at different levels of approximation. In~\cref{sec:numerical_solution} we provide details of our numerical solution method. Finally, in~\cref{sec:applications}, we apply our method to stochastic models of gene regulation, enzyme kinetics and infection dynamics. We end with conclusions and discussions in~\cref{sec:conclusions}.

Readers familiar with Doi-Peliti field theory and the associated path integral formulation can safely skip directly to~\cref{sec:effective_action_section} for the Plefka expansion and the subsequent effective field calculations. On the other hand,
readers more interested in the results and the application of the method can bypass the derivations and jump to~\cref{sec:numerical_solution}, referring to the example \texttt{Jupyter} notebooks provided alongside the code for details.

\section{Stochastic chemical reaction dynamics}{\label{sec:stochastic dynamics}}

\subsection{Chemical reaction network (CRN)}

\label{sec:CRN}

We consider a system of $P$ molecular species $X_i$ such that the state of the system is defined by specifying the number of molecules of all species, $ \bm{n} = (n_1,n_2,\dots n_P)$ with $n_i$ the number of molecules of species $X_i$. Following, and expanding on, the presentation in ~\cite{harsh_accurate_2023},
we allow a general system of reactions where reaction $\beta$ converts $r_1^\beta$ copies of species $X_1$ together with $r_2^\beta $ copies of $X_2$ etc.\ into $s_1^\beta$ copies of  $X_1$, $s_2^\beta$ copies of $X_2$ etc., or in shorthand
\begin{equation}{\label{general_reaction_form}}
	\sum_{i=1}^{P} r_i^\beta X_i \xrightarrow{k_\beta(\tau)} \sum_{i=1}^{P} s_i^\beta X_i \,.
\end{equation}
Here $k_\beta(\tau)$ is the rate constant for the reaction at time $\tau$, with units of inverse time; we allow it to have an essentially arbitrary time dependence. The order of reaction $\beta$ is then defined to be~\cite{laidler_glossary_1996} $\sum_i r_i^\beta$.
The entries $S_{i\beta}=s_i^\beta-r_i^\beta$ of the stoichiometric matrix $\bm{S}$ give the net change in molecule numbers for each reaction. The stoichiometric matrix together with the reaction rates
defines a chemical reaction network (CRN).

The probability that reaction $\beta$ will take place in a small time interval $(\tau,\tau+\Delta \tau)$ is given by $f_\beta(\bm{n},\tau)\Delta\tau$ in terms of the microscopic \textit{propensity function}~\footnote{The ratio of factorials takes care of the appropriate combinatorics, whereby e.g.\ for the reaction $2X_1\to X_2$ the reaction probability is proportional to the number $n_1(n_1-1)/2=n_1!/(n_1-2)!/2$ of pairs of $X_1$ molecules that can react. The factor 1/2 that appears here compared to \cref{propensity_1}, which in the general case would be $1/\prod_i r_i^\beta!$, has been included in the definition of $k_\beta$ to make the following expressions shorter.}
\begin{equation}{\label{propensity_1}}
	f_\beta(\bm{n},\tau) = k_\beta(\tau) \prod_i \frac{n_i !}{(n_i - r_i^\beta)! } \,.
\end{equation}
To illustrate the setup, we consider a single chemical reaction $A + B \xrightarrow{k} C$. This has
$r_A = 1,\ r_B = 1,\ s_C =1$ and $r_C = s_A = s_B = 0$, with rate constant $k_\beta (\tau) = k$ and propensity function $f_\beta(n_A,n_B,n_C) = k \, n_A n_B$. The propensity function in \cref{propensity_1} contains the rate constant $k_\beta$ with physical dimension of inverse time, $1/t$. This is because we are working with absolute molecule numbers; see~\cref{supp-SM:sec:propensity_function} for further discussion and how to convert such rate constants to their mass action equivalents.

\subsection{Dynamics}

The common approach to dynamics of CRNs consists of considering large reaction volumes $V$, working with approximately continuous concentration variables i.e.\ number densities $\bm{x}=(x_1,\ldots,x_P)$ with $x_i = n_i/V$ and neglecting the noise arising from the discrete numbers of molecules. The resulting mass-action kinetics (MAK)~\cite{schnoerr_approximation_2017,waage_studier_1864} is given by
\begin{equation}
	\partial_\tau \bm x = \bm S \bm f \,,
	\label{eq:compact_time_evolution}
\end{equation}
where the vector $\bm{f}$ of reaction fluxes for the different reaction contains the MAK propensity function defined as~\cite{schnoerr_approximation_2017, waage_studier_1864}
\begin{equation}
	f_\beta^{\text{MAK}} (\bm x) = j_\beta \prod_{i=1}^{P} x_i^{r_{i\beta}},\
	\label{eq:MAK_propersity}
\end{equation}
where the $j_\beta$ are proportional to the $k_\beta$ and differ only by factors of system volume, see \cref{supp-SM:sec:propensity_function}.

Retaining the discrete nature of particle number and the full stochastic problem, we start by considering the full time evolution of the probability that the (well-mixed) system is in state $\bm n$ at time $\tau$,
$P(\bm{n},\tau)$. This distribution evolves according to the \textit{Chemical Master Equation} (CME)~\cite{gillespie_rigorous_1992}
\begin{eqnarray}
\partial_\tau P(\bm{n},\tau) & = & \sum_\beta f_\beta(\bm{n} - \bm{s}^\beta + \bm{r}^\beta,\tau) P(\bm{n}- \bm{s}^\beta + \bm{r}^\beta,\tau) \nonumber \\
		& & {}- \sum_\beta f_\beta(\bm{n},\tau)P(\bm{n},\tau) \,.
		 \label{Master_equation}
\end{eqnarray}
Such CMEs are not generally solvable for arbitrary stoichiometry and non-linearity, and it is only for special cases that time-dependent~\cite{renyi_treating_1953,mcquarrie_kinetics_1964,mcquarrie_stochastic_1967,aranyi_full_1977} or steady state solutions~\cite{gardiner_handbook_1985,schnoerr_approximation_2017} are known.

A Fokker-Plank equation can be derived from the CME by truncating the Kramers-Moyal expansion~\cite{moyal_stochastic_1949,kramers_brownian_1940} after the second order, thus approximating the discrete molecular number Markov jump process by diffusive dynamics of continuous concentrations. However, this approach fails as expected in the regime of small molecular numbers~\cite{grima_how_2011}. Moreover, the associated Chemical Langevin Equation~\cite{schnoerr_complex_2014} contains noise terms that are in general state-dependent
and in fact imaginary. The Chemical Langevin Equation also lacks natural boundary condition at zero concentration, making its analysis far from straightforward.

In this paper, we use a path integral approach to the dynamics of CRNs that retains the discrete nature of constituents and provides the probability of observing any trajectory of the system, while being amenable to tools from field theory. \linebreak

\subsection{The path integral}

Following the seminal work of Doi and Peliti~\cite{doi_second_1976,peliti_path_1985}, one can cast the master equation in a quantum mechanical ``second quantized'' form, see~\cref{supp-SM:sec:second_quant}, and write a corresponding path integral in terms of two fields per species, $\phi_i$ and $\tilde \phi_i$. This procedure is standard~\cite{tauber_applications_2005} but somewhat involved, so we defer the details to~\cref{supp-SM:sec:path_integral}. Here, we directly state the results and show some properties that will be used later. The Doi-Peliti field theory has found successful applications in many areas of physics, especially for diffusing interacting particle systems~\cite{andreanov_field_2006,lefevre_dynamics_2007} and even in active matter~\cite{garcia-millan_run-and-tumble_2021,bothe_field_2021} as well as of course to chemical reaction dynamics, with Refs.~\cite{vastola_solving_2021,tauber_applications_2005,falkovich_john_2008,
greenman_reaction_2023,del_razo_field_2026} being some examples amongst numerous others. This field theory retains the discrete nature~\cite{bothe_particle_2023} of its constituents, like chemical molecules in our case.

Defining a state vector $| \bm{n} \rangle = |n_1,n_2,\dots, n_i,\dots,n_P\rangle$ (in an in principle infinite-dimensional state space) and the associated probability vector $|P(\tau)\rangle = \sum_{\bm{n}} P(\bm{n},\tau)|\bm{n}\rangle$, the time evolution of the system can be expressed as
\begin{equation}
	\partial_\tau | P(\tau) \rangle = \hat{H} | P(\tau) \rangle \,,
\end{equation}
with Hamiltonian (see~\cref{supp-SM:sec:second_quant})
	\begin{equation}{\label{Hamiltonian_operator}}
		\hat{H} = \sum_\beta k_\beta (\tau) \left[ \prod_i (\hat{a}_i^\dagger)^{s_i^\beta} (\hat{a}_i)^{r_i^\beta} - \prod_i (\hat{a}_i^\dagger)^{r_i^\beta} (\hat{a}_i)^{r_i^\beta} \right] \,,
	\end{equation}
where $\hat{a}_i^\dagger$ and $a_i$ are creation and annihilation operators for particles of species $i$. Here and below we do not explicitly write the time-dependence of $\hat{H}$, which comes from the $k_\beta(\tau)$.
The Hamiltonian $\hat{H}$ conserves probability but is in general non-Hermitian; only in the special case where the reaction dynamics obeys detailed balance can it be made Hermitian by a similarity transform.

Discretizing time into steps $\Delta t$ and inserting identity operators constructed as integrals over projectors onto coherent states, one finds an expression for the generating function of the process as a path integral:
	\begin{equation}{\label{partition_function}}
	\mathcal{Z}( \bm{\tilde{\theta}}  , \bm \theta ) = \lim_{\Delta t \rightarrow 0} \mathcal{N}^{-1} \int \prod_\tau d \tilde{\bm{\phi}}_\tau d\bm{\phi}_\tau \,e^{\Sfields} \,.
\end{equation}
The action appearing here is given in~\cref{supp-Action_discrete} of~\cref{supp-SM:sec:path_integral}. In the limit $\Delta t \rightarrow 0$ it becomes,
\begin{widetext}
\begin{equation}{\label{action_continuous}}
	\begin{split}
		\Sfields &=  \int_0^{t} d\tau\, H(\bm{\tilde{\phi}}(\tau), \bm{\phi}(\tau_-))  + \sum_i \Big( \nzero{i} \tilde{\phi}_i(0) - \phi_i(0) \tilde{\phi}_i(0) + \int_0^{t} d\tau \left[ -\tilde{\phi}_i(\tau) \partial_\tau \phi_i(\tau) + \tilde{\theta}_i(\tau) \tilde{\phi}_i(\tau)  + \theta_i(\tau) \phi_i(\tau)  \right] \Big) \,,
	\end{split}
\end{equation}
\end{widetext}
where $t$ is the total time of the trajectory. The Hamiltonian $H$ in the path integral is (see~\cref{supp-SM:sec:second_quant})
\begin{multline}{\label{Hamiltonian_operator_fields}}
H = \sum_\beta k_\beta(\tau_-) \left[ \prod_i (\tilde{\phi}_i(\tau) + 1)^{s_i^\beta} \right. \\
		\left. - \prod_i (\tilde{\phi}_i(\tau)+1)^{r_i^\beta} \right]  \prod_i (\phi_i(\tau_-))^{r_i^\beta}
\end{multline}
and corresponds to~\cref{Hamiltonian_operator}, with annihilation and creation operators replaced by the fields $\phi$ and $\tilde{\phi}+1$, respectively. The $+1$ here is the so-called Doi shift and simplifies later manipulations. As indicated by the notation $\tau_-=\tau-\Delta t$, the $\tilde{\phi}$ fields are a time step $\Delta t$ ahead of the $\phi$ fields in $H$, which imposes a useful causality property on the dynamics as discussed below.
In~\cref{action_continuous} we have also
inserted generating fields $\bm \theta$, $ \bm{\tilde \theta}$ that will allow us to easily take averages by differentiating the generating function. Finally, in writing~\cref{action_continuous} we have taken the initial state of the system to be a Poisson mixture, such that the initial copy number of each molecular species $i$ has independent Poisson fluctuations around a mean $\nzero{i}$.

\subsection{Observables in the path integral}{\label{sec:observables}}
Once the path integral is defined over the fields $\bm{\phi}$ and $\bm{\tilde{\phi}}$, we can relate averages of observables to the statistics of these fields. We start by taking derivatives w.r.t.\ $\tilde\theta_i(\tau_+)$ (with $\tau_+ = \tau+\Delta t$) and $\theta_i(\tau)$ in \cref{supp-path_integral1} of section~\cref{supp-SM:sec:path_integral}
such that the corresponding creation and annihilation operators are normal ordered, giving
\begin{equation}
	\begin{split}
		\frac{1}{(\Delta t)^2} \left( \frac{\partial }{\partial \tilde\theta_i(\tau_+)} + 1 \right) \frac{\partial }{\partial \theta_i(\tau)} \mathcal{Z} \Bigg|_{\theta,\tilde{\theta}=0} = \langle n_i(\tau) \rangle \,,
	\end{split}
\end{equation}
where $n_i(\tau)$ is the copy numbers of species $i$ at time $\tau$.
In the path integral one can take the same derivatives, which gives	$\langle [\tilde\phi_i(\tau_+)+1]\phi_i(\tau)\rangle$.
This average simplifies to  $\langle \phi_i(\tau)\rangle$ because of a {\em causality property} of the path integral: any average of a product of $\phi$ or $\tilde\phi$ fields vanishes when the last factor -- the one associated with the latest of all times that occur -- is a $\tilde\phi$; see~\cref{supp-SM:sec:causality_prop}.

Introducing the symbol $\mu$ for the means of the fields, we have
\begin{eqnarray}
		\mu_i(\tau) &=& \langle \phi_i(\tau)\rangle = \langle n_i(\tau) \rangle \,, \quad \mbox{and} \\
		\tilde{\mu}_i(\tau) &=& \langle \tilde{\phi}_i(\tau)\rangle = 0 \,.
\end{eqnarray}
In fact, applying the causality property inductively, one sees that the average of any product of $\tilde\phi$-factors vanishes, e.g.\ $\langle \tilde{\phi}(\tau) \tilde{\phi}(\tau') \rangle = 0 \text{ } \forall \text{ } \tau,\tau'$.

Similarly calculating the fourth order derivatives in terms of the number operator and from the path integral we obtain for $\tau' < \tau$,
\begin{equation}{\label{fourth_order_derivatives}}
	\begin{split}
	\langle n_i(\tau) n_j(\tau') \rangle = \langle [\tilde{\phi}_i(\tau_+)+1]\phi_i(\tau) [\tilde{\phi}_j(\tau'_+)+1] \phi_j(\tau') \rangle \,.
	\end{split}
\end{equation}
For $j \neq i$ the equal-time value ($\tau'=\tau$) has the same form
\begin{equation}{\label{fourth_order_derivatives_equal_time_neq}}
	\begin{split}
		\langle n_i(\tau) n_j(\tau) \rangle = \langle [\tilde{\phi}_i(\tau_+)+1]\phi_i(\tau) [\tilde{\phi}_j(\tau_+)+1] \phi_j(\tau) \rangle \,.
	\end{split}
\end{equation}
while for $j=i$ one has to substract the mean on the l.h.s.:
\begin{equation}{\label{fourth_order_derivatives_equal_time}}
	\begin{split}
	 \langle n_i^2(\tau) \rangle - \langle n_i(\tau) \rangle = \langle [\tilde{\phi}_i(\tau_+)+1]^2 \phi^2_i(\tau) \rangle \,.
	\end{split}
\end{equation}
These can now be linked to the connected response and correlation functions for all pairs of species $i,j$,
\begin{eqnarray}{\label{mean_resp_corr}}
R_{ij}(\tau,\tau') &=& \langle \delta \phi_i(\tau) \delta \tilde{\phi}_j(\tau') \rangle \,, \; \mbox{and} \\
	C_{ij}(\tau,\tau') &=& \langle \delta \phi_i(\tau) \delta \phi_j(\tau') \rangle \,,
\end{eqnarray}
where $\delta\phi_i = \phi_i-\mu_i$ and similarly $\delta\tilde{\phi}_i = \tilde{\phi}_i-\tilde{\mu}_i=\tilde{\phi}_i$.
The corresponding disconnected functions do not contain the subtraction of the product of the averages; note though that since $\langle \tilde{\phi}\rangle = 0$, the connected and the disconnected {\em response} functions are identical.  In the following we will abbreviate diagonal contributions with a single species subscript,  e.g.\ $R_{i}(\tau,\tau') \equiv R_{ii}(\tau,\tau')$.

On the r.h.s.\ of \cref{fourth_order_derivatives_equal_time} we have, by causality again, the disconnected correlator $\langle \phi_i^2(\tau)\rangle = C_i(\tau,\tau)+\mu_i^2(\tau)$. Subtracting the squared mean shows
\begin{equation}
	\begin{split}{\label{num_correlator_relation_1}}
		\text{Var}(n_i(\tau)) - \langle n_i(\tau)\rangle &= C_i(\tau,\tau) \,.
	\end{split}
\end{equation}
The l.h.s.\ vanishes for Poisson copy number fluctuations, so a non-zero equal-time correlator of the $\phi$-field indicates deviations from Poisson statistics.

For the cases in \cref{fourth_order_derivatives,fourth_order_derivatives_equal_time_neq}, causality implies
\begin{equation}
	\begin{split}
		\langle \delta n_i(\tau) \delta n_j(\tau') \rangle =
		C_{ij}(\tau,\tau') +
		\langle \phi_i(\tau) \tilde{\phi}_j(\tau'_+) \phi_j(\tau') \rangle \,,
	\end{split}
\end{equation}
where the copy number fluctuations are defined as $\delta n(\tau) = n(\tau) - \mu(\tau)$.
The final three-point correlator cannot be simplified in general, but for Gaussian field statistics one can use Wick's theorem and causality to express it as
$\langle \phi_i(\tau) \tilde{\phi}_j(\tau'_+) \phi_j(\tau') \rangle = \langle \phi_i(\tau) \tilde{\phi}_j(\tau'_+)\rangle \langle \phi_j(\tau') \rangle $ so that
\begin{equation}
	\begin{split}{\label{num_correlator_relation_2}}
		N_{ij}(\tau,\tau') &\equiv \langle \delta n_i(\tau) \delta n_j(\tau') \rangle \\
		&= R_{ij}(\tau,\tau')\mu_j(\tau') + C_{ij}(\tau,\tau') \; \text{for } \tau'<\tau \,,
	\end{split}
\end{equation}
and
\begin{equation}
	\begin{split}{\label{num_correlator_relation_3}}
		N_{ij}(\tau,\tau) \equiv \langle \delta n_i(\tau) \delta n_j(\tau) \rangle
		= C_{ij}(\tau,\tau) \; \text{ for } j \neq i,
	\end{split}
\end{equation}
while
\begin{equation}
	\begin{split}{\label{num_correlator_relation_4}}
		N_{ii}(\tau,\tau) \equiv \text{Var}(n_i(\tau))
		\equiv \langle \delta n^2_i(\tau) \rangle
		= \mu_i(\tau) + C_{ii}(\tau,\tau) \,.
	\end{split}
\end{equation}
All moments of the marginal copy number distribution for each species $i$ can in fact also be calculated in terms of moments of $\delta\phi_i(\tau)$, see~\cref{supp-SM:sec:marginal_copy_dist}. This again simplifies to a Poisson distribution if the path integral is Gaussian and $\phi\phi$-correlation functions are zero, which is a useful special case that we will encounter shortly in~\cref{baseline_action_section}: the free (Gaussian) theory gives Poisson copy number distribution for all species with time-dependent means.

\subsection{General reaction network Hamiltonian}
We use a convenient form for the Hamiltonian of \cref{Hamiltonian_operator_fields} for a general reaction network with given stoichiometry~\cite{MovillaMiangolarra2017}, as a first step towards a general expression for the effective fields that we will find below as correction terms.
We express the fundamental fields of the theory in terms of their deviations from the means $\phi = \mu + \delta\phi$ and $\tilde \mu = \tilde \mu + \delta \tilde \phi$ as in the correlation and response functions above, and then apply a binomial expansion to each term in~\cref{Hamiltonian_operator_fields} to get
\begin{equation}{\label{H_expression}}
	H(\bm{\tilde{\phi}}(\tau), \bm{\phi}(\tau_-))  = \sum_{\bar{m},\bar{n}} \cprime{c_{\bar{m},\bar{n}}(\tau) }{\frac{c_{\bar{m},\bar{n}}(\tau)}{\bar{m}!\bar{n}!}}\prod_i \delta\tilde{\phi}_i^{m_i}(\tau) \delta\phi_i^{n_i}(\tau_-) \,,
\end{equation}
\cprime{}{where $\bar{m}!=\prod_i m_i!$ and similarly $\bar{n}!=\prod_i n_i!$.}
The coefficients appearing above are defined as
\begin{multline}{\label{c_mn_defn}}
		c_{\bar{m},\bar{n}}(\tau) = \sum_\beta k_\beta(\tau_-) \left[ \prod_i \cprime{}{m_i!}\binom{s_i^\beta}{m_i} \left( \tilde{\mu}_i(\tau) + 1 \right)^{s_i^\beta - m_i} - \right. \\
		\left. \prod_i \cprime{}{m_i!}\binom{r_i^\beta}{m_i} \left( \tilde{\mu}_i(\tau) + 1  \right)^{r_i^\beta-m_i} \right]
		\prod_i  \cprime{}{n_i!}\binom{r_i^\beta}{n_i} \mu_i^{r_i^\beta-n_i}(\tau_-) \,,
\end{multline}
with the convention that $\binom{n}{m} = 0$ for $m>n$. The sum over $\bm{m}$ runs over all vectors with integer components $m_i=0, 1,\ldots$
such that, for at least one reaction $\beta$, $m_i \leq \max(r_i^\beta,s_i^\beta)\,\forall \, i$. We will denote the set of such index vectors by $\mathcal{S}$. The sum over $\bm{n}$ also runs over $\mathcal{S}$, and the same will be true for all sums over index vectors below unless otherwise specified. Note that the index vector $\bm{n}$ here is distinct from the vector of particle copy numbers introduce above; this distinction will always be clear from the context in what follows.

We will mostly be concerned with networks of at most {\em binary} reactions like $X_1+X_2\to X_3$ or $X_1+X_2\to X_3+X_4$, which are defined by
$\sum_i r_i^\beta \leq 2$ and $\sum_i s_i^\beta\leq 2$.
For networks of such binary reactions,
the $c_{\bar{m},\bar{n}}(\tau)$ vanish unless $\sum_i m_i\leq 2$ and $\sum_i n_i\leq 2$. Given this constraint, the set $\mathcal{S}$ can contain only index vectors of the form $\bm{m}=\bm{0}$, $\bm{e}_i$ or $\bm{e}_i+\bm{e}_j$ with $1\leq i,j\leq P$, where $\bar{e}_i$ is the $i$-th Cartesian unit vector, defined as having entry $1$ at position $i$ and zeros elsewhere. There are $(P+1)(P+2)/2$ different index vectors of the form above; in practice the set $\mathcal{S}$ will be even smaller depending on the specific structure of the reaction network.

We will occasionally use a diagrammatic representation in the following, where each term
proportional to $c_{\bar{m},\bar{n}}(\tau)
\prod_i \delta\tilde{\phi}_i^{m_i}(\tau) \delta\phi_i^{n_i}(\tau_-)$  in~\cref{H_expression} is represented as a diagram so that the Hamiltonian becomes
\begin{equation}
	H(\bm{\tilde{\phi}}(\tau), \bm{\phi}(\tau_-))  =  \quad \quad
	\parbox{30mm}{
		\begin{fmfgraph*}(60,55)

			\fmfleft{e1,e2,e3,e4,e5}
			\fmfright{e6,e7,e8,e9,e10}

			\fmf{dashes_arrow,tension=1}{v1,e1}
			\fmf{dashes_arrow,tension=1}{v1,e2}
			\fmf{phantom,tension=1}{v1,e3}
			\fmf{phantom,tension=1}{v1,e4}
			\fmf{dashes_arrow,tension=1}{v1,e5}

			\fmfv{label=$m_1$,label.dist=1}{e1}
			\fmfv{label=$m_2$,label.dist=1}{e2}
			\fmfv{decor.shape=circle, decor.filled=full, decor.size= 1thick}{e3}
			\fmfv{decor.shape=circle, decor.filled=full, decor.size= 1thick}{e4}
			\fmfv{label=$m_P$,label.dist=1}{e5}

			\fmf{dashes,tension=1}{v1,e6}
			\fmf{dashes,tension=1}{v1,e7}
			\fmf{phantom,tension=1}{v1,e8}
			\fmf{phantom,tension=1}{v1,e9}
			\fmf{dashes,tension=1}{v1,e10}

			\fmfv{label=$n_1$,label.dist=1}{e6}
			\fmfv{label=$n_2$,label.dist=1}{e7}
			\fmfv{decor.shape=circle, decor.filled=full, decor.size= 1thick}{e8}
			\fmfv{decor.shape=circle, decor.filled=full, decor.size= 1thick}{e9}
			\fmfv{label=$n_P$,label.dist=1}{e10}

\fmfv{label=$\tau$,label.dist=8,decor.shape=circle,decor.filled=full, decor.size=4thick,label.angle=-90}{v1 }
	\end{fmfgraph*}} \,.
\end{equation}
In each diagram we have $m_i$ legs with arrows corresponding to $\delta\tilde{\phi}_i(\tau_-)$ legs, $n_i$ legs without arrows corresponding to $\delta{\phi}_i(\tau)$ legs, and the centre blob indicating $c_{\bar{m},\bar{n}}(\tau)$, which includes the sum over different reactions $\beta$ and the factors of $\mu_i^{r_i^\beta-n_i}(\tau_-)$ corresponding to each species $i$.

Note that $\tilde{\mu}_i(\tau) = 0 \ \forall i,\,\tau$ in the final solution of the equations of motion, which then also implies that ${c_{\bar{0},\bar{n}}(\tau) = 0 \ \forall \, \bar{n},\tau}$.
We keep the factors of $\tilde{\mu}$ in $c_{\bar{m},\bar{n}}(\tau)$ and elsewhere, however, because we will have to take derivatives w.r.t.\ them later.

	\subsection{Baseline action and Gaussian path integral}{\label{baseline_action_section}}

	Equipped with the path integral for describing chemical reactions, we now look at a concrete set of chemical reactions that we call the \textit{baseline}, because the Hamiltonian associated with these reactions by \cref{Hamiltonian_operator_fields} will be quadratic in the fields. We will treat other reactions as perturbations to this quadratic baseline Hamiltonian. If we consider the set of only creation and (unary) destruction reactions for each of the species, which in ecological models would be interpreted as immigration and exodus of species, one obtains the following system with rate constants as indicated:
	\begin{eqnarray}
		\emptyset \xrightarrow{k_{1i}} X_i ,&\quad  X_i \xrightarrow{k_{2i}} \emptyset \,.
	\end{eqnarray}
From~\cref{Hamiltonian_operator_fields} the Doi-shifted Hamiltonian in the action is  then
	\begin{equation}{\label{eq:H0}}
		H_0(\bm{\tilde{\phi}}, \bm{\phi} ) = \sum_i
		\left( k_{1i}\tilde{\phi}_i - k_{2i} \tilde{\phi}_i \phi_i \right) \,.
	\end{equation}
This is decoupled across the different species and only contains linear and quadratic terms in $\phi,\tilde{\phi}$. To indicate that this Hamiltonian (with the linear terms omitted) will be the baseline for our perturbation theory, we use the notation $H_0$. The corresponding baseline action $S_0$ from \cref{supp-Action_discrete} of~\cref{supp-SM:sec:path_integral} is then, in its discrete time form,
\begin{widetext}
	\begin{equation}{\label{action_discrete_baseline}}
		\begin{split}
&\Sfieldsbaseline  =
	\sum_i \Big\lbrace \nzero{i} \tilde{\phi}_i(0) - \phi_i(0) \tilde{\phi}_i(0) +   \Delta t \sum_{\tau=0}^{t}  \left[-\tilde{\phi}_i(\tau) \Delta_\tau \phi_i(\tau) + k_{1i}\tilde{\phi}_i(\tau) - k_{2i} \tilde{\phi}_i(\tau) \phi_i(\tau_-) +  \tilde{\theta}_i(\tau) \tilde{\phi}_i(\tau)  + \theta_i(\tau) \phi_i(\tau)  \right] \Big\rbrace \,.
	\end{split}
	\end{equation}
\end{widetext}
Here $\Delta_\tau$ defines the discrete derivative w.r.t $\tau$, $\Delta_\tau\phi_i(\tau) = \frac{1}{\Delta t}\left( \phi_i(\tau) - \phi_i(\tau_-) \right)$ and the sum over $\tau$ for this term and the $k_{2i} \tilde{\phi}_i(\tau) \phi_i(\tau_-)$ term strictly starts at $\tau = \Delta t$ only, a restriction we will not write explicitly.
Because the path integral is Gaussian, we can calculate the means of the fields by extremizing the action as
	\begin{align}
		\frac{\partial S_0}{\partial \tilde{\phi}_i(\tau)}\Bigg|_{\substack{\phi_i(\tau) = \mu_i(\tau) \\ \tilde{\phi}_i(\tau) = \tilde{\mu}_i(\tau) }} = 0 \quad & \text{and} \quad  \frac{\partial S_0}{\partial \phi_i(\tau) } \Bigg|_{\substack{\phi_i(\tau) = \mu_i(\tau) \\ \tilde{\phi}_i(\tau) = \tilde{\mu}_i(\tau) }} = 0 \,.
\label{S0_stationarity}
	\end{align}
	The derivatives with respect to $\tilde\phi_i(\tau)$ for $0 < \tau \leq t$  give the equations of motion for the means as
	\begin{equation}
		\mu_i(\tau) - \mu_i(\tau-\Delta t) = \Delta t \left( k_{1i} - k_{2i} \mu_i(\tau-\Delta t) + \tilde{\theta}_i(\tau) \right)  \,,
		\label{eq:means_gen}
	\end{equation}
	while for $\tau=0$ one obtains the initial condition
	\begin{equation}
		\mu_i(0) = \nzero{i} + \tilde{\theta}_i(0)\Delta t \,.
	\end{equation}
The $\phi_i(\tau)$ derivatives yield similarly for $0\leq \tau<t$,
	\begin{equation}
		\tilde{\mu}_i(\tau) - \tilde{\mu}_i(\tau+\Delta t) = \Delta t \left( -k_{2i} \tilde{\mu}_i(\tau+\Delta t) + \theta_i(\tau)  \right) \,,
		\label{eq:means_conjugate_gen}
	\end{equation}
	and for $\tau = t$ one obtains a final condition
	\begin{equation}
		\tilde{\mu}_i(t) = \theta_i(t) \Delta t \,.
	\end{equation}
	The equations for the physical means $\mu_i(\tau)$ thus have to be solved forwards in time as expected, while the ones for the conjugate means $\tilde{\mu}_i(\tau)$ are solved backwards starting from the final condition.

	The physical problem is obtained when the generating fields are set to zero, and one then recovers from the relations above that the initial condition for species $i$ is  $\mu_i(0)=\nzero{i}$
and that $\tilde{\mu}_i(\tau) = 0 \text{ } \forall\,i,\tau $ as expected. In the continuous-time limit one also obtains for the physical means the expected equation of motion
	\begin{equation}{\label{mu_eom_baseline}}
		\partial_\tau \mu_i(\tau) = k_{1i} - k_{2i} \mu_i(\tau) \,,
	\end{equation}
	with the two terms on the r.h.s.\ reflecting creation and destruction of particles of species $i$. The same equation is obtained if we generalize to time-dependent creation rates $k_{1i}(\tau)$:
	\begin{equation}
		\partial_\tau \mu_i(\tau) = k_{1i}(\tau) - k_{2i} \mu_i(\tau) \,,
	\end{equation}
	and gives the solution
	\begin{equation}
		\mu_i(\tau) = \nzero{i} e^{-k_{2i}\tau} + \int_0^\tau d\tau' e^{-k_{2i}(\tau-\tau')} k_{1i}(\tau') \,.
		\label{mu0_solution}
	\end{equation}

From the means of the fields we next move on to the response functions. These can be obtained using the relation demonstrated in~\cref{supp-SM:sec:causality_prop}, which incidentally has a direct analogue in Martin-Siggia-Rose-Jansen-de Dominicis (MSRJD) path integrals~\cite{martin_statistical_1973,janssen_lagrangean_1976,dominicis_techniques_1976,hertz_path_2017}:
	\begin{equation}
		R_{0i}(\tau,\tau') = \langle \phi_i(\tau)\tilde\phi_i(\tau')\rangle_0 =
		\frac{1}{\Delta t} \frac{\partial \langle \phi_i(\tau) \rangle_0 }{\partial k_{1i}{(\tau')}}
		=  \frac{1}{\Delta t} \frac{\partial \mu_i(\tau) }{\partial k_{1i}(\tau')} \,.
		\label{R0deriv}
	\end{equation}
Here $R_{0i}$, a shorthand for $R_{0,ii}$, is the bare diagonal response function and all averages are evaluated at zero generating fields, $\tilde\theta=\theta=0$ as indicated by the subscript `0'. Note that all non-diagonal responses $R_{0,ij}$ with $i\neq j$ vanish, because different species are decoupled in the baseline dynamics.

	In the continuous time limit the derivative in~\cref{R0deriv} becomes a functional derivative w.r.t.\ $k_{1i}(\tau')$ and one finds explicitly from \cref{mu0_solution}
	\begin{equation}{\label{R_0}}
		R_{0i}(\tau,\tau') = e^{-k_{2i}(\tau-\tau')} \Theta(\tau-\tau') \,,
	\end{equation}
	with $\Theta(\cdot)$ the Heaviside step function. This response function is causal as expected,
	\begin{equation}{\label{response_causality}}
		R_{0i} (\tau,\tau') = 0 \text{ if } \tau < \tau'  \,,
	\end{equation}
	i.e.\ the real field $\phi$ needs to be ahead in time of the conjugate field $\tilde\phi$, and its equal-time limit is unity, $\lim_{\tau' \rightarrow \tau^-} R_0(\tau,\tau') = 1$. We note for later also the operator inverse of the response function, which is
	\begin{equation}{\label{R_inv_defn}}
		R_{0i}^{-1}(\tau,\tau') = (\partial_{\tau} + k_{2i}) \delta(\tau-\tau') \,.
	\end{equation}
	An alternative calculation of the response function, by inversion of the precision matrix of the discrete time Gaussian path integral, can be found in \cref{supp-matrix_action_plefka}.

We consider finally the correlation functions calculated within our baseline path integral. These are zero because the baseline action only has quadratic terms of the form $\tilde{\phi} \phi$ (see~\cref{supp-SM:Gaussian_action} of~\cref{supp-matrix_action_plefka}), i.e.\ the \textit{bare correlation functions} are
	\begin{equation}{\label{bare_correlation_eqn}}
		C_{0i}(\tau,\tau') = \langle \phi_i(\tau) \phi_i(\tau') \rangle_0 - \langle \phi_i(\tau)\rangle_0 \langle \phi_i(\tau') \rangle_0 = 0 \,.
	\end{equation}

	In the next section, we will employ the Plefka free energy expansions to include the effect of interactions, in the sense of terms not contained in the above baseline Hamiltonian, to calculate the dynamics of CRNs.

\section{Dynamical Plefka free energy}{\label{sec:effective_action_section}}

	We now consider nonlinear reactions. The associated `interacting' Hamiltonian $H_{\text{int}}$ will contain non-quadratic terms and we will treat it as a perturbation to our baseline $H_0$, introducing a parameter $\alpha$ to set up a perturbation theory:
	\begin{equation}
		H_\alpha = H_0 +\alpha H_{\text{int}} \,.
	\end{equation}
	We will call the associated action $S_\alpha=S_0 + \alpha S_{\text{int}}$.

A key quantity in the following will be the Plefka or constrained free energy $\Gamma$,
from which the dynamics can be extracted by a variational principle. This free energy is defined
as the Legendre transform of $\mathcal{W}=\ln\mathcal{Z}$ w.r.t.\ the field means $\mu$ and $\tilde \mu$.
The latter are the derivatives of $\mathcal{W}$ w.r.t.\ $\theta$ and $\tilde{\theta}$, respectively. We already exploited this fact in \cref{sec:observables} by writing physical observables as derivatives of $\mathcal{Z}$; for nonzero generating fields $\theta$ and $\tilde{\theta}$, one needs to use $\mathcal{W}=\ln\mathcal{Z}$ instead to ensure normalization. In the Legendre transform the field means $\mu$ and $\tilde{\mu}$ are multiplied by their generating fields and subtracted from $\mathcal{W}$, leading to the Plefka free energy definition
	\begin{widetext}
	\begin{equation}{\label{plefka_free_energy1}}
		\begin{split}
\Gamma ( \tilde{\mu} ,  \mu )  &= \text{\normalsize extr}_{\tilde{\theta}  ,  \theta } \left\lbrace \ln \int d\tilde{\phi}\, d\phi \,\exp \left[ S_\alpha(  \tilde{\phi} , \phi ) - \sum_i \int_0^t d\tau \left( \tilde{\mu}_i(\tau) \tilde{\theta}_i(\tau) + \mu_i(\tau) \theta_i(\tau)  \right) \right] \right\rbrace  =  \ln \int d\tilde{\phi}\,d\phi \; e^{\Xi_\alpha} \,,
		\end{split}
	\end{equation}
	\end{widetext}
	where $\int d\tilde{\phi}\,d\phi$ is a shorthand for the path integration $\lim_{\Delta t \rightarrow 0} \mathcal{N}^{-1} \int \prod_\tau d \tilde{\bm{\phi}}_\tau d\bm{\phi}_\tau$ from~\cref{partition_function} and
$\Xi_\alpha$ collects all terms in the exponent, evaluated at the extremal values of the generating fields.
Referring back to the form of $S_\alpha\equiv S$ in \cref{action_continuous}, one sees that the combined exponent in square brackets in \cref{plefka_free_energy1} contains the constraint term
\begin{equation}
\sum_i \int_0^t \theta_i(\tau)[\phi_i(\tau)-\mu_i(\tau)] \,,
\end{equation}
and similarly for $\tilde{\phi}$, ensuring that the extremization condition for the generating fields $\theta$ and $\tilde{\phi}$ yields precisely the desired constraint $\langle \phi_i(\tau)\rangle = \mu_i(\tau)$ on the field means.

Conversely, from the constrained free energy we can extract the generating field values by general Legendre transform properties as
	\begin{equation}{\label{fields2}}
		\theta_{i,\alpha}(\tau) = -\frac{\delta \Gamma ( \tilde{\mu} ,  \mu )}{\delta \mu_i(\tau) }, \hspace{20pt}  \tilde{\theta}_{i,\alpha}(\tau) = -\frac{\delta \Gamma ( \tilde{\mu} ,  \mu )}{\delta \tilde{\mu}_i(\tau) }  \,,
	\end{equation}
	where for later use we explicitly indicate the $\alpha$-dependence by $\theta_{i,\alpha}$ and $\tilde{\theta}_{i,\alpha}$.

	The equations for the means of the physical fields are then obtained by asserting that we are considering the physical problem without generating fields,
that is $\tilde{\theta}_i(\tau) = 0 $ and $\theta_i(\tau) =0$ $\forall\,i, \tau$:
\begin{equation}{\label{variational_eqns_mu}}
		\frac{\delta \Gamma ( \tilde{\mu} ,  \mu )}{\delta \mu_i(\tau) } = 0, \hspace{20pt}  \frac{\delta \Gamma ( \tilde{\mu} ,  \mu )}{\delta \tilde{\mu}_i(\tau) } = 0 \,.
	\end{equation}
Formally, these equations constitute the desired variational principle from which the dynamics of $\mu(\tau)$ and $\tilde{\mu}(\tau)$ can be obtained. Note that for the Doi-Peliti field theory, solving the variational equations always gives $\tilde{\mu}_i(\tau) = 0 \ \forall \,i, \tau$ in line with the discussion in~\cref{sec:observables} above. This simplification is the key advantage of the Doi-shift.

	\subsection{The Plefka expansion}{\label{sec:plefka_expansion}}

	The original idea of the Plefka expansion was introduced by Plefka~\cite{plefka_convergence_1982} for the Sherrington-Kirkpatrick (SK) spin glass model~\cite{sherrington_solvable_1975} as a convenient tool to derive the mean-field equations and their refined analogue, the TAP equations~\cite{thouless_solution_1977},
which have since been widely used to understand glassy behaviour in statistical physics models with quenched disorder.

Plefka's original approach for the equilibrium behaviour of systems with discrete (spin) degrees of freedom can be extended to non-equilibrium dynamics~\cite{roudi_dynamical_2011}
and to continuous variables~\cite{bravi_extended_2016}. The key idea consists in calculating the value of the generating fields, here $\theta$ and $\tilde{\theta}$,
for which the non-interacting dynamics (i.e.\ the one without $H_{\text{int}}$) gives the same mean physical fields as for the full Hamiltonian without fields, order by order in $\alpha$.
In~\cite{bravi_extended_2016} this formulation was introduced for pairwise interactions in $H_{\text{int}}$ only, with the fields calculated to second in $\alpha$.
Here we treat the more complex but practically important case of generic non-linear chemical reaction networks and express the fields tractably in terms of the network stoichiometry and rates. Our proposed ansatz in~\cref{sec:ansatz} for resummed corrections beyond the second order extends the applicability of our method to realistic reaction rates, as we demonstrate using a variety of example systems.

To implement the Plefka approach, the Plefka free energy is expanded in powers of $\alpha$ as
	\begin{align}
			\Gamma_\alpha &= \Gamma|_{\alpha=0} + \alpha \frac{\partial \Gamma}{\partial \alpha}\Bigg|_{\alpha=0} + \frac{\alpha^2}{2} \frac{\partial^2 \Gamma}{\partial \alpha^2}\Bigg|_{\alpha=0} + \dots \\
			& = \Gamma^0 + \alpha \Gamma^1 + \frac{\alpha^2}{2} \Gamma^2 + \dots \,.
	\end{align}
One can similarly expand the generating fields in $\alpha$ as
\begin{equation}
			\theta_{i,\alpha} = \theta_{i}^0 + \alpha \theta_{i}^1 + \frac{\alpha^2}{2} \theta_i^2 + \dots \,,
\label{theta_exp}
	\end{equation}
	and
	\begin{equation}
			\tilde{\theta}_{i,\alpha} = \tilde{\theta}_{i}^0 + \alpha \tilde{\theta}_{i}^1 + \frac{\alpha^2}{2} \tilde{\theta}_i^2 + \dots  \,.
		\label{theta_hat_exp}
	\end{equation}
	The field expansion coefficients are then obtained from \cref{fields2} order by order in $\alpha$, giving
\begin{equation}
\theta_i^m(\tau) = -\frac{\delta \Gamma^m}{\delta \mu_i(\tau)} \hspace{10pt} \text{ and } \hspace{10pt}	\tilde{\theta}_i^m(\tau) = -\frac{\delta \Gamma^m}{\delta \tilde{\mu}_i(\tau)} \,,
	\end{equation}
	for $m\in\left\{0,1,2,\dots \right\}$.
The dynamics of the original system, as well as its free energy $\ln \mathcal{Z}$, are retrieved from the Plefka free energy when
	\begin{equation}
		\theta_{i,\alpha}(\tau) = \tilde{\theta}_{i,\alpha}(\tau) = 0 \  \forall \, i,\tau \,,
	\end{equation}
or equivalently, from \cref{theta_exp}, $\theta_{i}^0 = - \alpha \theta_{i}^1 - \frac{\alpha^2}{2} \theta_i^2 + \dots$ and similarly for $\tilde{\theta}_{i}^0$ from~\cref{theta_hat_exp}. By construction these fields at $\alpha=0$ give the same $\mu$ and $\tilde{\mu}$ as in the original system at $\theta=\tilde{\theta}=0$. We therefore call them
\textbf{effective fields}, defined as
	\begin{equation}
		\theta_{i,\alpha}^{\text{eff}}(\tau) = -\alpha\theta_i^1(\tau) -\frac{\alpha^2}{2} \theta_i^2(\tau) + \dots \,,
	\end{equation}
and similarly for $\tilde{\theta}_{i,\alpha}$. The strategy, then, is to obtain the effective fields as functions of $\mu$ and $\tilde{\mu}$, typically by expanding up to $O(\alpha^2)$, and then use them to calculate $\mu$ and $\tilde{\mu}$ in the non-interacting problem ($\alpha=0$). This leads to self-consistency equations that determine the desired time evolution of $\mu_i(\tau)$ and $\tilde{\mu}_i(\tau)$.

The general form of these self-consistency equations has already been calculated in our treatment of the baseline ($\alpha=0$) system in~\cref{baseline_action_section} above. In the later parts of that section we had set the generating fields to zero. We now retain these and replace them by the effective fields. The $\Delta t\to 0$ limits of ~\cref{eq:means_gen} and ~\cref{eq:means_conjugate_gen} then give
\begin{equation}{\label{update_eqns_with_fields}}
		\partial_\tau \mu_i(\tau) = k_{1i} - k_{2i} \mu_i(\tau) + \tilde{\theta}_{i,\alpha}^{\text{eff}}(\tau) \,,
\end{equation}
and
	\begin{equation}{\label{eq:mu_tilde_dynamics}}
		-\partial_\tau \tilde{\mu}_i(\tau) = -k_{2i}\tilde{\mu}_i(\tau) + \theta_{i,\alpha}^{\text{eff}}(\tau) \,,
\end{equation}
with,  as before, initial respectively final condition $\mu_i(0) = \nzero{i}$, $\tilde{\mu}_i(t)=0$. The self-consistent nature of these equations arises from the dependence of the effective fields on $\mu$ and $\tilde{\mu}$.
Note that we will see that $ \theta_{i,\alpha}^{\text{eff}}$ is zero order by order in $\alpha$. This is consistent with $\tilde{\mu}$ being zero in the physical solution, in the Doi-shifted field theory we are using.
We observe finally that the presence of the effective fields does not modify the quadratic part of the baseline action, so that all correlation and response functions remain as worked out in~\cref{baseline_action_section}.

	\subsection{Effective fields}

To obtain the concrete form of the self-consistency equations for the field means, we now calculate the expansion of the Plefka free energy to $O(\alpha^2)$ and obtain from this the effective fields.
In this calculation the full Hamiltonian~\cref{H_expression} is split into $H_0$ containing only the baseline reactions, and the remainder $H_{\rm int}$. In the Plefka expansion it is always the interacting part $H_{\rm int}$ that will appear, which can be expressed exactly as in~\cref{H_expression} but with coefficients $c^{\rm int}_{\bm{m},\bm{n}}$ calculated only from the non-baseline reactions $\beta$. For notational simplicity we will drop the superscript ``int'' again immediately, i.e.\ from now on the $c_{\bm{m},\bm{n}}$ will only contain non-baseline reactions.

\subsubsection{First order in $\alpha$}

One can show in general (\cref{supp-SM:sec:first_order_field}) that the $O(\alpha)$ term in the Plefka free energy is simply given by
\begin{equation}{\label{eq:Gamma1}}
		\Gamma^1  = \left\langle \frac{d \Xi_\alpha}{d\alpha}\right\rangle_{0} = \left\langle S_{\text{int}} \right\rangle_{0} \,,
\end{equation}
where
$S_{\rm int} = \Delta t\sum_\tau H_{\rm int} (\bm{\tilde{\phi}}(\tau), \bm{\phi}(\tau_-))$
and the subscript 0 indicates averaging for $\alpha=0$. Looking at the Hamiltonian in~\cref{H_expression}, one needs to calculate averages of the form $\langle \prod_i \delta\tilde{\phi}_i^{m_i}(\tau)\delta\phi^{n_i}(\tau_-)\rangle$, which using Wick's theorem can be expanded into products of two-point functions $\langle\delta\phi_i(\tau_-)\delta\phi_j(\tau_-)\rangle$, $\langle\delta\tilde{\phi}_i(\tau)\delta\tilde{\phi}_j(\tau)\rangle$ and $\langle\delta\tilde{\phi}_i(\tau)\delta\phi_j(\tau_-)\rangle$. The first two vanish identically at $\alpha=0$; the third is an acausal response function so also vanishes. Only the constant $\bar{m}=\bar{0}$ and $\bar{n}= \bar{0}$ term in~\cref{H_expression} thus survives, giving
	\begin{equation}
		\begin{split}
			\Gamma^1 & = \left\langle S_{\text{int}} \right\rangle_{0} = \Delta t \sum_\tau c_{\bar{0},\bar{0}}(\tau) \;,
\end{split}
	\end{equation}
	The corresponding generating fields to first order in $\alpha$ are then
\begin{equation}{\label{linearOP_theta_o1-2}}
		-\tilde{\theta}_i^1(\tau)  = \frac{1}{\Delta t} \frac{\partial \Gamma^1}{\partial \tilde{\mu}_i(\tau)} = \frac{\partial c_{\bar{0},\bar{0}}(\tau)}{\partial \tilde{\mu}_i(\tau)} =  c_{\bar{e}_i,\bar{0}}(\tau) =
		\parbox{15mm}{
			\begin{fmfgraph*}(50,40)
				\fmfleft{e1}
				\fmfright{r1,r2,r2n,r2nn,r3}
\fmfv{label=\leglabel{i},label.dist=0,label.angle=180}{e1}
				\fmf{end_arrow_dash_two,tension=2}{v1,e1}
				\fmfv{label=$\tau$,label.dist=7,label.angle=90,decor.shape=circle,decor.filled=full, decor.size=4thick}{v1 }
				\fmf{phantom,tension=1}{v1,r1}
				\fmf{phantom,tension=1}{v1,r2}
				\fmf{phantom,tension=1}{v1,r2n}
				\fmf{phantom,tension=1}{v1,r2nn}
				\fmf{phantom,tension=1}{v1,r3}
\end{fmfgraph*}}
	\end{equation}
where we have used the equality
	\begin{equation}{\label{eq:identity_derivative_main_paper}}
		\frac{\partial}{\partial \tilde{\mu}_i(\tau)} \binom{s_i^\beta}{0} \left(1 + \tilde{\mu}_i(\tau) \right)^{s_i^\beta} = \binom{s_i^\beta}{1} \left(1 + \tilde{\mu}_i(\tau) \right)^{s_i^\beta-1}
	\end{equation}
to rewrite the derivative of $c_{\bm{0},\bm{0}}(\tau)$. This is a special case of the following more general identities:
\begin{equation}{\label{eq:c_mn_identity1}}
 	\frac{\partial c_{\bar{m},\bar{n}}(\tau_+)}{\partial \mu_i (\tau)} = c_{\bar{m},\bar{n}+\bar{e}_i}(\tau_+) \,,
 \end{equation}
 and
 \begin{equation}{\label{eq:c_mn_identity2}}
 	\frac{\partial c_{\bar{m},\bar{n}}(\tau)}{\partial \tilde \mu_i (\tau)} = c_{\bar{m}+\bar{e}_i,\bar{n}}(\tau) \,.
 \end{equation}
\Cref{supp-SM:sec:derivative identities} lists further identities and properties of $c_{\bm{m},\bm{n}}$ that will be used below to simplify various expressions.

For the generating field $\theta$ one finds similarly
	\begin{equation}{\label{linearOP_theta_tilde_o1}}
		-\theta_i^1(\tau)  = \frac{1}{\Delta t}\frac{\partial \Gamma^1}{\partial \mu_i(\tau)} = \frac{\partial c_{\bar{0},\bar{0}}}{\partial \mu_i(\tau)} = c_{\bar{0},\bar{e}_i}(\tau_+) \,.
	\end{equation}
Now $c_{\bar{0},\bar{e}_i}(\tau_+)$ vanishes when $\tilde{\mu}=0$. Integrating~\cref{eq:mu_tilde_dynamics} with effective field $
\theta_{i,\alpha}^{\text{eff}}(\tau) = -\alpha\theta_i^1(\tau)$ backwards in time $\tau$ from the final condition $\tilde{\mu}_i(t)=0$ thus gives the expected solution $\tilde{\mu}_i(\tau)=0$ $\forall \ i,\tau$.

Similarly the equation of motion for $\mu_i$ from \cref{update_eqns_with_fields} is to $O(\alpha)$
	\begin{equation}
		\partial_\tau \mu_i(\tau) = k_{1i} - k_{2i}\mu(\tau) -\alpha \tilde{\theta}_i^1(\tau)  \,.
	\end{equation}
Substituting~\cref{linearOP_theta_o1-2} and the explicit expressions for $ c_{\bar{e}_i,\bar{0}}(\tau)$, then setting $\alpha = 1$ leads to \begin{equation}
\partial_\tau \mu_i(\tau) = k_{1i} - k_{2i}\mu_i(\tau) +\sum_\beta k_\beta(\tau) \left(s_i^\beta - r_i^\beta \right) \prod_j \mu_j^{r_j^\beta}(\tau) \,.
	\end{equation}
This is the analogue of the mean field approximation in Plefka's original paper~\cite{plefka_convergence_1982}: we observe that here it reproduces exactly the MAK approximation.

As discussed before, however, MAK predictions can be rather inaccurate in the regime of large intrinsic noise. In the following section we will therefore calculate the $O(\alpha^2)$ corrections to obtain a more accurate description of the dynamics.

	\subsubsection{Second order in $\alpha$}{\label{sec:second_order_linear}}

	We show in~\cref{supp-SM:sec:second_order_field} that the second derivative of the free energy $\Gamma^2$, i.e.~the first non mean-field correction is given by
	\begin{equation}{\label{gamma2_exponent_expansion}}
		\begin{split}
			\Gamma^2 &= \left\langle \left( \delta \frac{d\Xi_\alpha}{d\alpha} \right)^2 \right\rangle_{0} \,.
\end{split}
	\end{equation}
Doing the required averages  (see~\cref{supp-SM:sec:second_order_field}) in a manner analogous to the calculation of $\Gamma^1$ then gives
\begin{equation}{\label{eq:gamma2_linearOP}}
		\Gamma^2 = 2 (\Delta t)^2 \sum_{\tau>\tau'} \sum_{\bar{n}\neq\bar{0},\bar{e}^1} c_{\bar{0},\bar{n}}(\tau) c_{\bar{n},\bar{0}}(\tau') \prod_{i} \cprime{n_i! R_{0i}^{n_i}(\tau_-,\tau')}
{
\frac{R_{0i}^{n_i}(\tau_-,\tau')}{n_i!}
} \,,
\end{equation}
where $\bar{e}^1$ indicates the set of all Cartesian unit vectors $\bar{e}_i$.

From $\Gamma^2$ we can now calculate the $O(\alpha^2)$ terms in the effective fields. As expected one finds $\theta_i^2(\tau) = 0 \ \forall \,i,\tau$; we perform the explicit check  in~\cref{supp-SM:sec:second_order_field}.
The non-trivial field is given by
	\begin{widetext}
	\begin{equation}
		\begin{split}
			-\tilde{\theta}_i^2(\tau) &= \frac{1}{\Delta t} \frac{\partial \Gamma^2}{\partial \tilde{\mu}_i(\tau)}\\
			&= 2\,\Delta t \sum_{\tau'(<\tau)}  \sum_{\bar{n}\neq\bar{0},\bar{e}^1}  \Big[ c_{\bar{e}_i,\bar{n}}(\tau) c_{\bar{n},\bar{0}}(\tau') \prod_j
\cprime{n_j! R_j^{n_j}(\tau_-,\tau')}
{\frac{R_{0j}^{n_j}(\tau_-,\tau')}{n_j!}}
 + c_{\bar{0},\bar{n}}(\tau') c_{\bar{n}+\bar{e}_i,\bar{0}}(\tau) \cprime{(n_i+1)}{} \prod_j \cprime{n_j! R_j^{n_j}(\tau_-,\tau')}
{\frac{R_{0j}^{n_j}(\tau_-,\tau')}{n_j!}}
\Big] \,.
		\end{split}
	\end{equation}
The second term vanishes because $c_{\bar{0},\bar{n}}(\tau')=0$ causality, simplifying the expression to
	\begin{equation}{\label{linearOP_theta_o2}}
		\begin{split}
			-\tilde{\theta}_i^2(\tau) &= 2\,\Delta t \sum_{\tau'(<\tau)}  \sum_{\bar{n}\neq\bar{0},\bar{e}^1}  c_{\bar{e}_i,\bar{n}}(\tau) c_{\bar{n},\bar{0}}(\tau') \prod_j \cprime{n_j! R_j^{n_j}(\tau_-,\tau')}
{\frac{R_{0j}^{n_j}(\tau_-,\tau')}{n_j!}} \,.
\end{split}
	\end{equation}
We now exploit for the first time in the derivation the fact that we are considering reaction networks with at most binary reactions. Recalling that the only non-zero terms have $\sum_i n_i\leq 2$, the condition $\bar{n}\neq \bar{0},\bar{e}^1$ leaves only index vectors of the form $\bar{n} = \bar{e}_k + \bar{e}_l$ for $k\leq l \in \{1,2,\dots,P\}$. For $k\neq l$ the combinatorial factor $\prod_j n_j!$ is unity while for $k=l$ it is two, so symmetrizing the sum over $k\leq l$ into an unrestricted sum one has
\begin{equation}{\label{linearOP_theta_o2}}
		\begin{split}
			-\tilde{\theta}_i^2(\tau) &= \Delta t \sum_{\tau'} \sum_{k,l} c_{\bar{e}_i,\bar{e}_k + \bar{e}_l}(\tau) c_{\bar{e}_k + \bar{e}_l,\bar{0}}(\tau') R_{0k}(\tau_-,\tau') R_{0l}(\tau_-,\tau')
			= \quad \parbox{30mm}{ 	\begin{fmfgraph*}(100,40)
					\fmfleft{e1}
\fmfv{label=\leglabel{i},label.dist=0,label.angle=180}{e1}
					\fmfright{e2,e3,e3n,e3nn,e4}
					\fmf{end_arrow_dash_two,tension=3.}{v1,e1}
					\fmf{phantom}{e2,v2}
					\fmf{phantom}{e3,v2}
					\fmf{phantom}{e4,v2}
					\fmf{phantom}{e3n,v2}
					\fmf{phantom}{e3nn,v2}
\fmf{fermion,left,tension=1.5,label=$R_{0k}$,label.angle=90,label.dist=3}{v2,v1}
					\fmf{fermion,right,tension=1.5,label=$R_{0l}$,label.angle=90,label.dist=3}{v2,v1}
					\fmfv{label=$\tau$,label.dist=7,label.angle=105,decor.shape=circle, decor.filled=full, decor.size= 4thick}{v1}
					\fmfv{label=$\tau'$,label.dist=7, label.angle=70,decor.shape=circle, decor.filled=full, decor.size= 4thick}{v2}
\end{fmfgraph*}} \,.
		\end{split}
	\end{equation}
		\end{widetext}
	The equations of motion for the means to $O(\alpha^2)$ are then given by
	\begin{equation}
		\partial_\tau\mu_i(\tau) = k_{1i} - k_{2i}\mu_i(\tau) -\alpha \tilde{\theta}_i^1(\tau) -\frac{\alpha^2}{2}\tilde{\theta}_i^2(\tau) \,.
	\end{equation}
This are, for our setting, the analogue of the TAP equations~\cite{thouless_solution_1977} for the SK~\cite{sherrington_solvable_1975} model. They contain at $O(\alpha^2)$
the first non-trivial correction to the mean-field mass action dynamics. This (temporally) non-local term can be interpreted as a memory correction~\cite{harsh_accurate_2023}: it integrates the past value of the means -- which appear via the coefficients $c_{\bar{e}_i,\bar{e}_k + \bar{e}_l}(\tau) c_{\bar{e}_k + \bar{e}_l,\bar{0}}(\tau')$ -- weighted by a memory kernel. The latter is constructed from products of the response functions defined in~\cref{R_0}.

	Since the above correction only involves the bare responses,
it turns out that the resulting predictions for the dynamics can become unstable for large interaction strength values, specifically non-baseline reactions with $k_\beta \gtrsim k_2 $ where $\beta$ indexes an interaction or non-baseline reaction and $k_2$ is a typical baseline destruction rate $k_{2i}$.
This has been demonstrated in our previous work~\cite{harsh_accurate_2023}, but can also be easily seen by noting that the response functions in~\cref{R_0} are actually independent of the rate of the interaction reactions $k_\beta$. Note that since the correlation functions also have the bare form, the marginal copy number distributions are constrained to be always Poissonian within the above approximation, independently of the true underlying distribution.

	In the following section, we will extend our analysis by moving to an extended formulation of the Plefka free energy, which incorporates correlation and response functions as dynamical order parameters. This will lead to significantly more accurate and stable predictions for the dynamics that closely reproduce the true correlations and other two-time quantities of interest.

	\section{Extended Dynamical Plefka free energy}{\label{sec:extended_Plefka}}

The idea of the original Plefka expansion was extended in a recent work~\cite{bravi_extended_2016} to not only include continuous degrees of freedom but also two-time correlations and responses for each degree of freedom, as additional order parameters of the system on which the free energy depends. The Plefka expansion procedure then implements the equivalence (order by order in $\alpha$) of these additional order parameters in the unconstrained system with interactions and in the system without interactions but with effective fields.

Concretely, in the action defining the partition  function of \cref{partition_function} one adds a set of second order or \textit{quadratic order parameter generating fields} $\hat{Q} =\lbrace \hat{R}, \hat{B} , \hat C\rbrace$ that couple to the second order statistics of the physical fields, namely $Q = \lbrace \langle\delta \phi \,\delta \tilde{\phi}\rangle, \langle\delta \tilde{\phi}\,\delta \tilde{\phi}\rangle, \langle\delta \phi\,\delta \phi \rangle\rbrace$ respectively.
Explicitly, we add to the action in~\cref{action_continuous} the generating terms
\begin{widetext}
\begin{equation}
\begin{split}
\sum_{ij} \int_{0}^{t} d\tau \int_{0}^{t} d\tau'  \left[ \hat{R}_{ij}(\tau,\tau') \delta \phi_j(\tau') \delta \tilde{\phi}_i(\tau) + \frac{1}{2}\hat{B}_{ij}(\tau,\tau') \delta \tilde{\phi}_i(\tau) \delta \tilde{\phi}_j(\tau') + \frac{1}{2}\hat{C}_{ij}(\tau,\tau') \delta \phi_i(\tau) \delta \phi_j(\tau') \right] \,.
		\end{split}
\label{quadratic_generators}
	\end{equation}
		\end{widetext}
Here the factors of $1/2$ are introduced because of the symmetry of the $\delta \tilde \phi \,\delta \tilde \phi$ and $\delta  \phi \,\delta \phi$ correlators. Note also that we take $\hat{R}_{ij}(\tau,\tau')$ to be conjugate to the response function $R_{ji}(\tau',\tau)$ with inverted species and time indices, which is helpful for later and will in particular make $\hat{R}_{ij}(\tau,\tau')$ causal again for the physical solution. We write the resulting log partition function as $\mathcal{W}(\tilde{\theta},\theta,\hat{Q})=\ln\mathcal{Z}(\tilde{\theta},\theta,\hat{Q})$. Taking derivatives w.r.t.\ the $\hat{Q}$ fields, one then obtains directly the quadratic order parameters $Q = \lbrace R,B,C \rbrace$:
\begin{subequations}
	\begin{align}\label{variational_eq_quad}
R_{ij}(\tau,\tau') &= \langle \delta\phi_i(\tau)\delta\tilde{\phi}_j(\tau')\rangle = \frac{\delta \mathcal{W}( \tilde{\theta}, \theta, \hat{Q})}{\delta \hat{R}_{ji}(\tau',\tau) } \,,\\
\frac{1}{2}B_{ij}(\tau,\tau') &= \frac{1}{2} \langle \delta\tilde{\phi}_i(\tau)\delta\tilde{\phi}_j(\tau')\rangle = \frac{\delta \mathcal{W}(  \tilde{\theta}, \theta, \hat{Q})}{\delta \hat{B}_{ij}(\tau,\tau')} \,, \\ \shortintertext{and}
\frac{1}{2}C_{ij}(\tau,\tau') &= \frac{1}{2} \langle \delta\phi_i(\tau)\delta \phi_j(\tau')\rangle = \frac{\delta \mathcal{W} (  \tilde{\theta}, \theta, \hat{Q})}{\delta \hat{C}_{ij}(\tau,\tau') } \,.
\end{align}
\end{subequations}

The $B$ order parameters can be shown to be zero always in the physical solution, using the causality property or by explicit calculation, see~\cref{supp-matrix_action_plefka}.

One can now similarly define an {\em{extended Plefka free energy}} $\Gamma(\tilde{\mu},\mu,Q)$ with quadratic order parameter constraints: in~\cref{plefka_free_energy1} one uses the action $S_\alpha$ as extended above, and adds the corresponding Legendre transform terms
\begin{widetext}
	\begin{equation}{\label{free_energy_plefka}}
			- \sum_{ij} \int_{0}^{t}  d\tau \int_{0}^{t} d\tau' \left[ \hat{R}_{ij}(\tau,\tau') R_{ji}(\tau',\tau)
			+ \frac{1}{2}\hat{B}_{ij}(\tau,\tau') B_{ij}(\tau,\tau') \right. \left. +  \frac{1}{2}\hat{C}_{ij}(\tau,\tau') C_{ij}(\tau,\tau')  \right] \,.
	\end{equation}
\end{widetext}
The extremum is then taken over all conjugate fields, i.e.\ $\tilde{\theta}$, $\theta$ and $\hat{Q}$.

From $\Gamma(\tilde{\mu},\mu,Q)$ one can then again recover the generating fields by inverse Legendre transform, i.e.
\begin{subequations}
	\begin{align}
		\hat{R}_{ij}(\tau,\tau') &= -\frac{\delta \Gamma(  \tilde{\mu}, \mu , Q)}{\delta R_{ji}(\tau',\tau) } \label{Rhat_from_Gamma1} \,,\\
		\frac{1}{2} \hat{B}_{ij}(\tau,\tau') &= -\frac{\delta \Gamma (  \tilde{\mu}, \mu , Q)}{\delta B_{ij}(\tau,\tau') } \label{Rhat_from_Gamma2} \,, \\ \shortintertext{and}
		\frac{1}{2} \hat{C}_{ij}(\tau,\tau') &= -\frac{\delta \Gamma(  \tilde{\mu}, \mu , Q)}{\delta C_{ij}(\tau,\tau') } \label{Rhat_from_Gamma3} \,.
\end{align}
\end{subequations}
As before, the physical dynamics corresponds to zero generating fields, i.e.\ $\hat{Q}= 0$, so that~\cref{Rhat_from_Gamma1,Rhat_from_Gamma2,Rhat_from_Gamma3}, with zero on the l.h.s., become  the variational equations for the quadratic order parameters of the system.
As a final note on the construction of the extended Plefka free energy we point out that we have chosen to constrain not the second moments of the fields like $\langle\tilde{\phi}\,\phi\rangle$, i.e.\ disconnected correlators as done in~\cite{bravi_extended_2016,bravi_inference_2017},
but the connected correlators such as $\langle\delta\tilde{\phi}\,\delta\phi\rangle$. Given that we are also constraining the field means $\mu$, $\tilde{\mu}$, the two constraints are equivalent as one can check explicitly; we chose the connected correlators as they are much more convenient for calculations. In
the above setup we are building on earlier work with simpler versions that only constrained the (disconnected) single-species response functions~\cite{MovillaMiangolarra2017,Kurjahn2020}.

With the extended Plefka free energy $\Gamma(\tilde{\mu},\mu,Q)$ defined as above, the remaining logic of the calculation follows the earlier case where only the linear order parameters were constrained: one expands the free energy and the generating fields $\hat{Q}$ in $\alpha$, obtaining the expansion coefficients $\hat Q^m$ for $m = 1,2,\dots$; the order parameters of the physical dynamics are then recovered from the non-interacting case ($\alpha=0$) with effective fields $\hat{Q}^{\text{eff}}$
	\begin{equation}
		\hat{Q}^{\text{eff}} = -\alpha \hat{Q}^{1} - \frac{\alpha^2}{2} \hat{Q}^{2} + \dots \,.
	\end{equation}
The resulting equations of motion can then be constructed as in~\cref{sec:plefka_expansion}.
The $\alpha=0$ action now contains the linear generating terms in~\cref{action_continuous} as well as the quadratic ones given by~\cref{quadratic_generators}. It thus remains quadratic overall and the field means can be found from its stationary points defined by~\cref{S0_stationarity}. Because the quadratic generating terms in~\cref{quadratic_generators} are by construction stationary at $\phi=\mu$, $\tilde{\phi}=\tilde{\mu}$, the resulting equations of motion for $\mu$ and $\tilde{\mu}$ then remain as before and are given by~\cref{update_eqns_with_fields} and~\cref{eq:mu_tilde_dynamics} respectively.

To find the corresponding equations for the quadratic order parameters, one can invert the precision matrix of the Gaussian path integral; we defer the details to~\cref{supp-matrix_action_plefka}. The result is
\begin{equation}{\label{FD_eqn_mixed_plefka}}
		\begin{split}
		(\partial_\tau &+ k_{2i}) R_{ij}(\tau,\tau') = \delta(\tau-\tau') \delta_{ij} \\
		&+\int_{\tau'}^{\tau} d\tau''\, \sum_k \hat{R}_{ik}^{\text{eff}}(\tau,\tau'') R_{kj}(\tau'',\tau') \,,
		\end{split}
	\end{equation}
	and
	\begin{align}
(\partial_\tau + k_{2i}) &C_{ij}(\tau,\tau') = \int_{0}^{\tau} d\tau''\, \sum_k \hat{R}_{ik}^{\text{eff}}(\tau,\tau'') C_{kj}(\tau'',\tau') \nonumber \\
		{\label{C_update_multispecies_explicit_plefka}}
		&+ \int_{0}^{\tau} d\tau'' \, \sum_k  \hat{B}_{ik}^{\text{eff}}(\tau,\tau'')  R_{jk}(\tau',\tau'') \,.
\end{align}

	The first of these equations, i.e.~\cref{FD_eqn_mixed_plefka}, will need to be solved numerically on a two-time grid. This aspect is dealt with in detail in~\cref{sec:numerical_solution} below. For evaluation of~\cref{C_update_multispecies_explicit_plefka}, it is more convenient to use its integrated form (see~\cref{supp-matrix_action_plefka}). This gives an efficient method for calculating $C$ once $R$ has been found, which in discrete-time matrix form reads simply
	\begin{equation}{\label{eq:C_update_non_differential}}
		\bm C =  (\Delta t)^2 \bm R \bm{\hat{B}}^{\text{eff}} \bm R^{\rm T} \,.
	\end{equation}
Here the rows and columns of each matrix are labelled by pairs $(i,\tau)$ of a species index and a time.

\subsection{Extended effective fields}

\subsubsection{First order in $\alpha$}

We next calculate the extended Plefka free energy and the resulting effective fields, initially to $O(\alpha)$.	As in the case of linear order parameters, $\Gamma^1$ is given by~\cref{eq:Gamma1}, leading to the expression
	\begin{equation}
		\begin{split}
			\Gamma^1 &= \left\langle S_{\text{int}} \right\rangle_0 \\ &= \Delta t \sum_{\tau}\sum_{\bar{m},\bar{n}}
\cprime{c_{\bar{m},\bar{n}}(\tau)}
{\frac{c_{\bar{m},\bar{n}}(\tau)}{\bm{m}!\bm{n}!}}
\left\langle \prod_{i} \delta \tilde{\phi}_i^{m_i}(\tau) \delta \phi_i^{n_i}(\tau_-) \right\rangle_0 \,.
		\end{split}
	\end{equation}
However, unlike in the linear order parameter case the average is now non-trivial because the correlation functions are themselves order parameters that need to be determined.
Given that the average over the $\alpha=0$ path integral is Gaussian and we have at most binary reactions,
$\Gamma^1$ has contributions from  zeroth, second and fourth order averages. The second order averages are of three types ($\tilde{\phi}\tilde{\phi}$, $\tilde{\phi}\phi$  and $\phi\phi$). The fourth order term always has two factors of $\tilde{\phi}$ and two factors of $\phi$, but gives three distinct Wick pairings. We find in this way
\begin{widetext}
\begin{eqnarray}
\Gamma^1 &=& \Delta t \sum_\tau \left\lbrace
c_{\bm{0},\bm{0}}(\tau) +
\sum_{i,j}c_{\bm{e}_i,\bm{e}_j}(\tau) R_{ji}(\tau_-,\tau)+
\sum_{i\leq j}\left[\frac{c_{\bm{e}_i+\bm{e}_j,\bm{0}}(\tau)}{1+\delta_{ij}} B_{ij}(\tau,\tau) +
\frac{c_{\bm{0},\bm{e}_i+\bm{e}_j}(\tau)}{1+\delta_{ij}} C_{ij}(\tau_-,\tau_-)\right] \right. \nonumber \\
&& \left.+ \sum_{i\leq j, k\leq l}\frac{c_{\bm{e}_i+\bm{e}_j,\bm{e}_k+\bm{e}_l}(\tau)}{
(1+\delta_{ij})(1+\delta_{kl})}\left[
B_{ij}(\tau,\tau)C_{kl}(\tau_-,\tau_-)
+R_{ki}(\tau_-,\tau)R_{lj}(\tau_-,\tau)+
R_{li}(\tau_-,\tau)R_{kj}(\tau_-,\tau)
\right] \right\rbrace \,.
\end{eqnarray}
Here e.g.\ the $R_{ji}$ term in the first line corresponds to $m_i=n_j=1$ and all other entries of $\bm{m}$ and $\bm{n}$ zero, while the $B_{ij}$ term corresponds to $\bm{n}=\bm{0}$ and $\sum_k m_k=2$, which can be realized by either $m_i=m_j=1$ for $i\neq j$, or by $m_i=2$ so that effectively $i=j$. The factor $1+\delta_{ij}$ in the denominator is the corresponding factorial $\bm{m}!$. This factor and its analogues in the other terms are precisely such that the sums can again be symmetrized, giving
\begin{eqnarray}
\Gamma^1 &=& \Delta t \sum_\tau
\left\lbrace c_{\bm{0},\bm{0}}(\tau) +
\sum_{ij}c_{\bm{e}_i,\bm{e}_j}(\tau) R_{ji}(\tau_-,\tau)+
\frac{1}{2}\sum_{ij}\left[c_{\bm{e}_i+\bm{e}_j,\bm{0}}(\tau) B_{ij}(\tau,\tau) +
c_{\bm{0},\bm{e}_i+\bm{e}_j}(\tau) C_{ij}(\tau_-,\tau_-)\right] \right. \nonumber \\
&&  \left.+ \frac{1}{4}\sum_{ijkl}c_{\bm{e}_i+\bm{e}_j,\bm{e}_k+\bm{e}_l}(\tau)\left[
B_{ij}(\tau,\tau)C_{kl}(\tau_-,\tau_-)
+R_{ki}(\tau_-,\tau)R_{lj}(\tau_-,\tau)+
R_{li}(\tau_-,\tau)R_{kj}(\tau_-,\tau)
\right] \right\rbrace \,.
\end{eqnarray}
From this one then finds the following generating fields (for $\hat{R}$ there are four different derivative terms from the two quadratic terms in $R$, but they all turn out to be the same):
\begin{eqnarray}
	-\tilde{\theta}_i^1(\tau) &=& \frac{1}{\Delta t} \frac{\partial \Gamma^1}{\partial \tilde{\mu}_i(\tau)} = \frac{\partial c_{\bar{0},\bar{0}}(\tau)}{\partial \tilde{\mu}_i(\tau)} + \frac{1}{2}\sum_{k,l} \frac{\partial c_{\bar{0},\bar{e}_k + \bar{e}_l}(\tau)}{\partial \tilde{\mu}_i(\tau)} C_{kl}(\tau_-,\tau_-) +   \frac{\partial c_{\bm{e}_k,\bm{e}_l}(\tau)}{\partial \tilde{\mu}_i(\tau) } R_{lk}(\tau_-,\tau) \\
	&=&  c_{\bar{e}_i,\bar{0}}(\tau) + \sum_{k,l} \left[ \frac{1}{2} c_{\bar{e}_i,\bar{e}_k + \bar{e}_l}(\tau) C_{kl}(\tau_-,\tau_-)	+  c_{\bm{e}_k+\bm{e}_i,\bm{e}_l} R_{lk}(\tau_-,\tau) \right] \,, \label{eq:hattheta1_extended} \\
	-\hat{R}^1_{ij}(\tau,\tau')&=& \frac{1}{(\Delta t)^2}\frac{\partial \Gamma^1}{\partial R_{ji}(\tau',\tau)} = \frac{\delta_{\tau',\tau_-}}{\Delta t}\left[
	c_{\bm{e}_i,\bm{e}_j}(\tau)+\sum_{k,l}
	c_{\bm{e}_k+\bm{e}_i,\bm{e}_l+\bm{e}_j}(\tau)R_{lk}(\tau_-,\tau)
	\right] \,, \label{eq:hatR1_extended} \\
	-\frac{1}{2}\hat{B}^1_{ij}(\tau,\tau') &=& \frac{1}{(\Delta t)^2}\frac{\partial \Gamma^1}{\partial B_{ij}(\tau,\tau')} =
	\frac{\delta_{\tau',\tau}}{\Delta t} \left[
	\frac{1}{2}
	c_{\bm{e}_i+\bm{e}_j,\bm{0}}(\tau)
	+\frac{1}{4}
	\sum_{k,l}c_{\bm{e}_i+\bm{e}_j,\bm{e}_k+\bm{e}_l}(\tau)C_{kl}(\tau_-,\tau_-)
	\right] \,, \; \mbox{and} \label{eq:hatB1_extended}
  \\
	-\frac{1}{2}\hat{C}^1_{ij}(\tau,\tau') &=& \frac{1}{(\Delta t)^2}\frac{\partial \Gamma^1}{\partial C_{ij}(\tau,\tau')} =
	\frac{\delta_{\tau',\tau}}{\Delta t}\left[
	\frac{1}{2}
	c_{\bm{0},\bm{e}_i+\bm{e}_j}(\tau_+)
	+\frac{1}{4}
	\sum_{k,l}c_{\bm{e}_k+\bm{e}_l,\bm{e}_i+\bm{e}_j}(\tau_+)B_{kl}(\tau_+,\tau_+)
	\right] \,. \label{eq:hatC1_extended}
\end{eqnarray}

We will need these full expressions later when calculating $\Gamma^2$ and taking its derivatives w.r.t.\ $R$, $B$ and $C$. In the final physical solution, on the other hand, $B$ and the acausal response functions vanish so that the generating fields simplify to
\begin{eqnarray}
		{\label{quadOP_theta_o1}}
		-\tilde{\theta}_i^1(\tau)
		&=&  c_{\bar{e}_i,\bar{0}}(\tau) + \frac{1}{2}\sum_{k,l}c_{\bar{e}_i,\bar{e}_k + \bar{e}_l}(\tau) C_{kl}(\tau_-,\tau_-)
		=	\quad \parbox{15mm}{
			\begin{fmfgraph*}(50,40)
				\fmfleft{e1}
				\fmfright{r1,r2,r2n,r2nn,r3}
\fmfv{label=\leglabel{i},label.dist=0,label.angle=180}{e1}
				\fmf{end_arrow_dash_two,tension=2}{v1,e1}
				\fmfv{label=$\tau$,label.dist=7,label.angle=90,decor.shape=circle,decor.filled=full, decor.size=4thick}{v1 }
				\fmf{phantom,tension=1}{v1,r1}
				\fmf{phantom,tension=1}{v1,r2}
				\fmf{phantom,tension=1}{v1,r2n}
				\fmf{phantom,tension=1}{v1,r2nn}
				\fmf{phantom,tension=1}{v1,r3}
		\end{fmfgraph*}}
		+\frac{1}{2}\quad
		\parbox{30mm}{ 	\begin{fmfgraph*}(40,30)
				\fmfleft{e1}
				\fmfright{v1}
				\fmfv{label=\leglabel{i},label.dist=0,label.angle=180}{e1}
				\fmf{end_arrow_dash,tension=1.}{v1,e1}
				\fmf{plain,tension=1.,label=$C_{kl}$,label.dist=2}{v1,v1}
				\fmfv{label=$\tau$,label.angle=125,decor.shape=circle, decor.filled=full, decor.size= 4thick}{v1}
\end{fmfgraph*}} \,, \\
{\label{eq:hatR1_quad}}
		-\hat{R}^1_{ij}(\tau,\tau') &=&  \frac{\delta_{\tau',\tau_-}}{\Delta t} c_{\bar{e}_i,\bar{e}_j}(\tau)
		=
\quad	\parbox{20mm}{\begin{fmfgraph*}(60,40)
				\fmfleft{e1}
				\fmfright{r1,r2,r2n,r2nn,r3}
\fmfv{label=\leglabel{i},label.dist=0,label.angle=180}{e1}
				\fmf{end_arrow_dash_two,tension=1.5}{v1,e1}
				\fmfv{label=$\tau$,label.dist=7,label.angle=90,decor.shape=circle,decor.filled=full, decor.size=4thick}{v1 }
\fmf{phantom,tension=0.5}{v1,r1}
				\fmf{phantom,tension=0.5}{v1,r2}
				\fmf{dashes,tension=0.5}{v1,r2n}
				\fmf{phantom,tension=0.5}{v1,r2nn}
				\fmf{phantom,tension=0.5}{v1,r3}
\fmfv{label=\leglabel{j},label.dist=5}{r2n}
\end{fmfgraph*}}  \quad \quad \,, \mbox{and}
\\
	{\label{eq:hatB1_quad}}
	-\hat{B}^1_{ij}(\tau,\tau')
	&=& \frac{\delta_{\tau',\tau}}{ \Delta t} \left[ c_{\bar{e}_i+\bar{e}_j,\bar{0}}(\tau) + \frac{1}{2}\sum_{k,l} c_{\bar{e}_i + \bar{e}_j,\bar{e}_k + \bar{e}_l}(\tau) C_{kl}(\tau_-,\tau_-) \right]
= \parbox{13mm}{
		\begin{fmfgraph*}(50,40)
			\fmfleft{e1,e2}
			\fmfright{r1,r2,r2n,r2nn,r3}
\fmfv{label=\leglabel{i},label.dist=0,label.angle=180}{e1}
\fmfv{label=\leglabel{j},label.dist=0,label.angle=180}{e2}
			\fmf{end_arrow_dash_two,tension=1.25}{v1,e1}
			\fmf{end_arrow_dash_two,tension=1.25}{v1,e2}
			\fmfv{label=$\tau$,label.dist=7,label.angle=90,decor.shape=circle,decor.filled=full, decor.size=4thick}{v1 }
\fmf{phantom,tension=1}{v1,r1}
			\fmf{phantom,tension=1}{v1,r2}
			\fmf{phantom,tension=1}{v1,r3}
			\fmf{phantom,tension=1}{v1,r2n}
			\fmf{phantom,tension=1}{v1,r2nn}
\end{fmfgraph*}}
	+ \frac{1}{2} \parbox{32mm}{ 	\begin{fmfgraph*}(40,30)
			\fmfleft{e1,e2}
			\fmfright{v1}
\fmfv{label=\leglabel{i},label.dist=0,label.angle=180}{e1}
\fmfv{label=\leglabel{j},label.dist=0,label.angle=180}{e2}
			\fmf{end_arrow_dash,tension=0.8}{v1,e1}
			\fmf{end_arrow_dash,tension=0.8}{v1,e2}
\fmf{plain,tension=0.8,label=$C_{kl}$,label.dist=2}{v1,v1}
\fmfv{label=$\tau$,label.angle=100,decor.shape=circle, decor.filled=full, decor.size= 4thick}{v1}
\end{fmfgraph*}} \,.
\end{eqnarray}
\end{widetext}
The generating field $\theta_i^1(\tau)$ vanishes (see also~\cref{supp-SM:sec:first_order_extended}) as observed in the simpler version of the theory with only linear order parameters. Similarly, one finds that $\hat C^1_{ij}(\tau,\tau')=0$.
Note that the above expressions, once multiplied by $\alpha$, give the effective fields at this level of the Plefka approximation with linear and quadratic order parameters. These effective fields are easy to interpret: they produce time-local linear and quadratic contributions to the action that are either simply the linear and quadratic terms appearing in $S_{\rm int}$ itself (first terms on the r.h.s.\ of~\cref{quadOP_theta_o1,eq:hatB1_quad,eq:hatR1_quad})
or cubic and quartic terms from $S_{\rm int}$ in which occurrences of $\delta\phi\,\delta\phi$ have been replaced by their average, giving the $C_{kl}$-contributions.

To improve the approximation further, we now proceed to calculate in an analogous fashion the effective fields to the next order, $O(\alpha^2)$.

	\subsubsection{Second order in $\alpha$: Mode Coupling Approximation (MCA)}

The $O(\alpha^2)$ correction to the extended Plefka free energy $\Gamma^2$ can again be found using~\cref{gamma2_exponent_expansion} but now including the quadratic order parameters, as detailed in~\cref{supp-SM:sec:second_order_extended}. From this one can then obtain the $O(\alpha^2)$ corrections to the effective fields.

	In~\cref{supp-SM:sec:second_order_extended} we show generally that the $O(\alpha^2)$ contribution $\theta_i^2$ vanishes, consistent with the expectation that $\theta_i$ should be zero to all orders in $\alpha$. We also demonstrate there that $\tilde{\theta}^2_i$ is zero for our case of reaction networks with at most binary reactions.

The $\hat R ^2$ fields can be calculated with more work (\cref{supp-SM:sec:second_order_extended}). They contain products $RR$ of two response functions, and products $RC$ of one response and one correlation function. Since $C$ is $O(\alpha)$  while $R$ is nonzero even at $O(\alpha^0)$, we will drop the $RC$ contributions in the following. The $RR$ terms that we retain are
\begin{align}
			-\frac{1}{2} \hat{R}^2_{ij}(\tau,\tau')
			&= \sum_{klpq}  c_{\bar{e}_i,\bar{e}_k+\bm{e}_l}(\tau) c_{\bm{e}_p+\bm{e}_q,\bar{e}_j}(\tau'_+)  \Lambda^{kl,pq}(\tau_-,\tau'_+) \nonumber \\
			{\label{hatR2}}
			&= \quad \parbox{40mm}{ 	\begin{fmfgraph*}(100,40)
					\fmfleft{e1}
					\fmfright{e2,e3,e4}
\fmfv{label=\leglabel{i},label.dist=0,label.angle=180}{e1}
					\fmfv{label=\leglabel{j},label.dist=2}{e3}
					\fmf{end_arrow_dash_two,tension=3.}{v1,e1}
\fmf{phantom}{e2,v2}
					\fmf{dashes}{e3,v2}
					\fmf{phantom}{e4,v2}
					\fmf{fermion,left,tension=1.5,label=$\Lambda$,label.side=right,label.dist=12}{v2,v1}
					\fmf{fermion,right,tension=1.5}{v2,v1}
					\fmfv{label=$\tau$,label.dist=7,label.angle=105,decor.shape=circle, decor.filled=full, decor.size= 4thick}{v1}
					\fmfv{label=$\tau'$,label.dist=7, label.angle=70,decor.shape=circle, decor.filled=full, decor.size= 4thick}{v2}
\end{fmfgraph*}} \,,
	\end{align}
	where  $\Lambda^{kl,pq}(\tau_-,\tau'_+) = \frac{1}{2} R_{kp}(\tau_-,\tau'_+)  R_{lq}(\tau_-,\tau'_+)$
For example, for the $A+B\rightarrow C$ system, the $c$-coefficients are nonzero only if $k=A$ and $l=B$ or vice versa, and similarly for $p$ and $q$, so that the sum becomes, omitting the common time arguments $(\tau_-,\tau'_+)$ of the response functions,
\begin{eqnarray}
c_{\bm{e}_i,\bm{e}_A+\bm{e}_B}(\tau)
c_{\bm{e}_A+\bm{e}_B,\bm{e}_j}(\tau'_+)[R_{AA}R_{BB}+R_{AB}R_{BA}] \,.
\end{eqnarray}

Moving on to the other two effective fields on the quadratic order parameters, $\hat C^2$ again vanishes in general. For systems with at most binary reactions, $\hat B^2$ involves $RC$ and $CC$ terms (see~\cref{supp-SM:sec:second_order_extended}), which we neglect for the same reason as the $RC$-term in $\hat{R}^2$.

Numerically, we find with the above expression for the $\hat R^2$ effective field that solutions can still develop  instabilities~\cite{harsh_accurate_2023}. As in the case of only linear order parameters, this occurs if the response functions do not decay fast enough, which is usually the case for large nonlinear reaction rates. This is, in the end, not surprising since we are only keeping terms up to $O(\alpha^2)$ in the perturbation theory in $\alpha$. We note that keeping all the $O(\alpha^2)$ terms with the extended set of order parameter fields would lead to the well known Mode Coupling Approximation (MCA)~\cite{janssen_mode-coupling_2018,bouchaud_mode-coupling_1996,szamel_mode-coupling_2013} in its schematic or field theory form. Our systematic derivation of this approximation from a second order expansion of the extended dynamical Plefka free energy complements other approaches such as self-consistent replacement of propagators~\cite{bouchaud_mode-coupling_1996,reichman_mode-coupling_2005}. In the study of the glass transition, where MCA is widely employed, time translational invariance and equilibrium assumptions are further imposed on the solution of the dynamical equations, which leads to a tractable form of the memory functions. Since we are interested in the full non-equilibrium problem, however, and since even the full MCA including the $RC$ and $CC$ terms does not cure the instabilities,
we shall proceed differently.

\subsection{Beyond the Mode Coupling Approximation: generalized Self-consistent Bubble Resummation (gSBR)}{\label{sec:ansatz}}

In this section, we propose an ansatz to conveniently include the effect of higher order terms in the response functions, by summing an infinite subseries of diagrams. A specific case of this ansatz was previously considered in the literature~\cite{falkovich_john_2008} in the study of the asymptotic vertex renormalization of a binary reaction. The ansatz was also derived diagrammatically in our own work~\cite{harsh_accurate_2023}.
Finally, in the study of density fluctuations in systems of interacting Brownian particles~\cite{szamel_dynamics_2007} and mode coupling theory for supercooled liquids~\cite{szamel_mode-coupling_2013}, related diagrammatic series were motivated to derive memory function corrections, while in solid state theory the Bethe-Salpeter equation~\cite{nambu_force_1950,salpeter_relativistic_1951,rohlfing_electron-hole_2000} is also closely related.

We call our ansatz the generalized Self-consistent Bubble Resummation (gSBR). It is constructed from a chain of bubble like diagrams of~\cref{hatR2} joined together by $c_{\bar{m},\bar{n}}$, in a way that makes it applicable to an arbitrary reaction network with at most binary reactions. We can write this ansatz series for $\hat{R}^2$ as
\begin{widetext}
\begin{equation}{\label{eq_chain_hatR}}
	\begin{split}
&-\frac{1}{2}\hat{R}^{2,\text{gSBR}}_{ij}(\tau,\tau'_-) = \sum_{k l p q} c_{\bar{e}_i,\bar{e}_{k}+\bar{e}_{l}}(\tau) \Big[ \Lambda^{k l,p q}(\tau_-,\tau'_+) c_{\bar{e}_{p}+\bar{e}_{q},\bar{e}_j}(\tau'_+)  \\
	&+ (\Delta t) \sum_{k_1 l_1 p_1 q_1 \tau^1} \Lambda^{k l,p_1 q_1}(\tau_-,\tau^1_+) c_{\bar{e}_{p_1} + \bar{e}_{q_1},\bar{e}_{k_1} + \bar{e}_{l_1}}\!(\tau^1)  \Big\lbrace \Lambda^{k_1 l_1,p q}(\tau^1_-,\tau'_+)c_{\bar{e}_{p} + \bar{e}_{q},\bar{e}_j}(\tau'_+)\\
	&+ (\Delta t) \sum_{k_2 l_2 p_2 q_2 \tau^2} \Lambda^{k_1 l_1,p_2 q_2}(\tau^1_-,\tau^2_+) c_{\bar{e}_{p_2} + \bar{e}_{q_2},\bar{e}_{k_2} + \bar{e}_{l_2}}(\tau^2) \Lambda^{k_2 l_2,p q}(\tau^2_-,\tau'_+) c_{\bar{e}_{p} + \bar{e}_{q},\bar{e}_j}(\tau')	+\dots \Big\rbrace \Big] \,,
	\end{split}
\end{equation}
or diagrammatically,
\begin{equation}{\label{eq:bubble_sum_diag}}
		\begin{split}
			-\frac{1}{2}\hat{R}^{2,\text{gSBR}}_{ij} = \;
			\parbox{33mm}{
				\begin{fmfgraph*}(90,40)
					\fmfleft{e1}
					\fmfright{e2,e3,e4}
\fmfv{label=\leglabel{i},label.dist=0,label.angle=180}{e1}
					\fmfv{label=\leglabel{j},label.dist=2}{e3}
					\fmf{end_arrow_dash_two,tension=3.}{v1,e1}
\fmf{phantom}{e2,v2}
					\fmf{dashes}{e3,v2}
					\fmf{phantom}{e4,v2}
					\fmf{fermion,left,tension=1.5,label=$\Lambda$,label.side=right,label.dist=12}{v2,v1}
					\fmf{fermion,right,tension=1.5}{v2,v1}
					\fmfv{decor.shape=circle, decor.filled=full, decor.size= 4thick}{v1}
					\fmfv{label.angle=70,decor.shape=circle, decor.filled=full, decor.size= 4thick}{v2}
\fmffreeze
					\fmfshift{
						vloc(__v2)-vloc(__e3)
						+abs(vloc(__e1)-vloc(__v1))
						*unitvector(vloc(__e3)-vloc(__v2))
					}{e3}
				\end{fmfgraph*}
			}
			+\; \parbox{43mm}{
				\begin{fmfgraph*}(120,40)
					\fmfleft{e1}
					\fmfright{e2,e3,e4}
\fmfv{label=\leglabel{i},label.dist=0,label.angle=180}{e1}
					\fmfv{label=\leglabel{j},label.dist=2}{e3}
					\fmf{end_arrow_dash_two,tension=3.}{v0,e1}
\fmf{phantom}{e2,v2}
					\fmf{dashes}{e3,v2}
					\fmf{phantom}{e4,v2}
					\fmf{fermion,left,tension=1.5,label=$\Lambda$,label.side=right,label.dist=12}{v1,v0}
					\fmf{fermion,right,tension=1.5}{v1,v0}
					\fmf{fermion,left,tension=1.5,label=$\Lambda$,label.side=right,label.dist=12}{v2,v1}
					\fmf{fermion,right,tension=1.5}{v2,v1}
					\fmfv{decor.shape=circle, decor.filled=full, decor.size= 4thick}{v1}
					\fmfv{decor.shape=circle, decor.filled=full, decor.size= 4thick}{v2}
					\fmfv{decor.shape=circle, decor.filled=full, decor.size= 4thick}{v0}
\fmffreeze
					\fmfshift{
						vloc(__v2)-vloc(__e3)
						+abs(vloc(__e1)-vloc(__v0))
						*unitvector(vloc(__e3)-vloc(__v2))
					}{e3}
				\end{fmfgraph*}
			}
			+ \; \parbox{52mm}{
				\begin{fmfgraph*}(150,40)
					\fmfleft{e1}
					\fmfright{e2,e3,e4}
\fmfv{label=\leglabel{i},label.dist=0,label.angle=180}{e1}
					\fmfv{label=\leglabel{j},label.dist=2}{e3}
					\fmf{end_arrow_dash_two,tension=3.}{v0n,e1}
\fmf{phantom}{e2,v2}
					\fmf{dashes}{e3,v2}
					\fmf{phantom}{e4,v2}
					\fmf{fermion,left,tension=1.5,label=$\Lambda$,label.side=right,label.dist=12}{v0,v0n}
					\fmf{fermion,right,tension=1.5}{v0,v0n}
					\fmf{fermion,left,tension=1.5,label=$\Lambda$,label.side=right,label.dist=12}{v1,v0}
					\fmf{fermion,right,tension=1.5}{v1,v0}
					\fmf{fermion,left,tension=1.5,label=$\Lambda$,label.side=right,label.dist=12}{v2,v1}
					\fmf{fermion,right,tension=1.5}{v2,v1}
					\fmfv{decor.shape=circle, decor.filled=full, decor.size= 4thick}{v1}
					\fmfv{decor.shape=circle, decor.filled=full, decor.size= 4thick}{v2}
					\fmfv{decor.shape=circle, decor.filled=full, decor.size= 4thick}{v0}
					\fmfv{decor.shape=circle, decor.filled=full, decor.size= 4thick}{v0n}
\fmffreeze
					\fmfshift{
						vloc(__v2)-vloc(__e3)
						+abs(vloc(__e1)-vloc(__v0n))
						*unitvector(vloc(__e3)-vloc(__v2))
					}{e3}
				\end{fmfgraph*}
			}
			\; + \dots \,.
		\end{split}
	\end{equation}
Note that because of the way that we have defined the $c$ and $\Lambda$ factors, all reactions and all combinations of response functions are considered in this series.

Since every term of the bubble series above contains an extra factor of the form
$$
\chi^{kl,k'l'}(\tau,\tau') = \sum_{pq}\Lambda^{kl,pq}(\tau_-,\tau'_+) c_{\bar{e}_p+\bar{e}_q, \bar{e}_{k'} +\bar{e}_{l'}}(\tau')\ ,$$
the series can be summed as a geometric series in the space of species pair ($kl$) and time ($\tau$) indices.
In the continuous time limit $\Delta t \rightarrow 0$ one obtains the following expression for the field, \begin{equation}{\label{eq_hatR2_contnuous}}
	\begin{split}
		-\frac{1}{2}\hat{R}^{2,\text{gSBR}}_{ij}(\tau,\tau')
		= \sum_{klk'l'} c_{\bar{e}_i,\bar{e}_k + \bar{e}_l}(\tau)  \int_0^{\tau} d\tau'' \left( \delta_{kk'}\delta_{ll'}\delta(\tau-\tau'') -  \chi^{kl,k'l'}(\tau,\tau'') \right)^{-1} \sum_{pq}\Lambda^{k' l', pq}(\tau'',\tau') c_{\bar{e}_{p} + \bar{e}_{q},\bar{e}_j}(\tau') \,.
	\end{split}
\end{equation}
Here the operator in the brackets of the form $A^{kl,k'l'}\!(\tau,\tau'')$ has discrete indices $(kl,k'l')$ and continuous time indices $\tau,\tau''$. The inverse $A^{-1}$ of $A$ is accordingly defined by
\begin{equation}
	\int_{\tau'}^{\tau} d\tau'' \sum_{k'l'} (A^{-1})^{kl,k'l'}(\tau,\tau'')A^{k' l', p q}(\tau'',\tau') = \delta_{kp}\delta_{lq}\delta(\tau-\tau') \,.
\end{equation}
\end{widetext}
	The bubble series ansatz has a convenient regularising property in that it prevents response functions from diverging (as long as the time steps in the numerical integrations are small enough) because of the functional/matrix inverse that appears. For systems with at most binary reactions, it can be shown~\cite{harsh_accurate_2023} that this gSBR series includes the effects of all $O(\alpha^3)$ corrections and an infinite series of higher order corrections in the equation for the copy number means $\mu$. For response functions, the bubble series similarly includes all $O(\alpha^2)$ corrections and an infinite series of corrections beyond $O(\alpha^2)$~\cite{harsh_accurate_2023}.

Once $\hat{R}^{2,\text{gSBR}}$ has been calculated, the final effective field $\hat{R}^{\text{eff}}$ in the gSBR approximation is
	\begin{equation}{\label{eq:R_eff_gSBR}}
		\hat{R}^{\text{eff}} = -\hat{R}^1 -\frac{1}{2}\hat{R}^{2,\text{gSBR}} \,,
	\end{equation}
with $\hat R^1$ as given by~\cref{eq:hatR1_quad}.
	In the following section, we will outline a numerical recipe for the solution of the entire set of equations to obtain consistent time evolution equations for the all the order parameters considered and thus the one-time and two-time quantities of the system.

\end{fmffile}

\subsection{Connection with standard field theory terminology}{\label{sec:field-theory-term}}

We pause briefly at this point to connect the Plefka approach to standard field theory language, for readers more familiar with the latter. In field theory, $\mathcal{W} = \ln \mathcal{Z}$ with $\mathcal{Z}$ as defined in~\cref{partition_function} is the generating function of the connected $n$-point functions. $\Gamma$ in~\cref{plefka_free_energy1} is generally referred to as the one-particle-irreducible (1PI) effective action~\cite{berges_introduction_2004,bellac_quantum_1991}, which is derived as the Legendre transform of $\mathcal{W}$ w.r.t.\ the so-called background fields means $\mu$ and $\tilde \mu$. The extended Plefka free energy of~\cref{sec:extended_Plefka} is conceptually the two-particle-irreducible (2PI) effective action~\cite{cornwall_effective_1974,berges_introduction_2004} with $R,B$ and $C$ as additional background fields. The effective action has this name because it is stationary at the physical value of the background fields.

The two-time integro-differential equations for the response and correlation functions, i.e.~\cref{FD_eqn_mixed_plefka} and~\cref{C_update_multispecies_explicit_plefka} respectively, are Dyson equations~\cite{hertz_path_2017,bellac_quantum_1991} with $\hat{R}^{\text{eff}}$ and $\hat{B}^{\text{eff}}$ being the respective self-energy contributions. These integro-differential equations are of Kadanoff-Baym type~\cite{baym_conservation_1961} and occur widely in quantum field theory, in what would be imaginary time there.

The connection to 2PI field theory can be seen more transparently by writing the extended Plefka expansion in more generic notation, as explained in~\cref{supp-app:structure}.

\section{Numerical solution}{\label{sec:numerical_solution}}

Since we are interested in the full non-equilibrium solution of the reaction network dynamics without any further approximations beyond those explained above, we need to solve the full Kadanoff-Baym (KB)~\cite{baym_conservation_1961} type equation~\cref{FD_eqn_mixed_plefka} for the two-time response function $R$, along with the corresponding~\cref{update_eqns_with_fields} for the means $\mu$ and~\cref{eq:C_update_non_differential} for the correlators $C$.
The simplest way to obtain numerical results is use the simple time discretization scheme that we have also used in the derivations, i.e.\ time grids with constant time steps $\Delta t$.
However, this is numerically expensive (see~\cref{supp-sec:computational_complexity_gSBR}), leading to a computational effort scaling as $O(\zeta^3)$ where $\zeta\propto 1/(\Delta t)$ is the number of time steps.

Fortunately, KB-type integro-differential equations for the response and correlation propagators appear in a variety of fields, especially in the study of strongly correlated solid-state systems. Accordingly, their numerical solutions has received much attention~\cite{kohler_numerical_2000,Talarico2018,meirinhos_adaptive_2022}. We therefore employ a recently developed~\cite{meirinhos_adaptive_2022}, out-of-the-box adaptive time step KB equation solver.
This is based on a two-time generalization of the Variable Adams method~\cite{norsett_solving_1993}. Other numerical schemes for the solution of the KB equations could certainly also be used, and exploring their benefits for the application to reaction networks will be an interesting avenue for future work.

The adaptive time-step solver allows us to access much longer integration times than would be possible with a fixed time-step scheme. For example, for the case of non-driven chemical reactions that we consider in the next section, small time-steps with high sensitivity are desired during the initial phase of the dynamics, when the order parameters evolve rapidly. Later on, larger time-steps then allow one to efficiently study the slow relaxation to equilibrium.

In the examples below we will compare the different levels of constraints and approximations that we have developed above. We summarize them here for ease of  reference:
\begin{enumerate}

	\item \textbf{gSBR}: Within this approximation, the order parameter $\mu$ is given by~\cref{update_eqns_with_fields}, $R$ by~\cref{FD_eqn_mixed_plefka} and $C$ by~\cref{eq:C_update_non_differential}. The effective field $\hat R^{\text{eff}}$ is given by~\cref{eq:R_eff_gSBR},
$\hat B^{\text{eff}}$ by~\cref{eq:hatB1_quad} and $\tilde \theta^{\text{eff}}$ by~\cref{quadOP_theta_o1}.

	\item \textbf{SBR}: This version differs from the gSBR version, only in the expression of the $\hat R^2$ field. Instead of a summation in the extended space of species pairs and time indices, we consider the sum over the different reactions separately; see~\cref{supp-sec:simpleSBR_sum} for details.

	\item \textbf{MCA}: This is the mode coupling approximation or the $O(\alpha^2)$ expansion. $\mu$ is given by~\cref{update_eqns_with_fields}, $R$ by~\cref{FD_eqn_mixed_plefka} and $C$ by~\cref{eq:C_update_non_differential}. For the calculation of $\hat R^{\text{eff}}$, $\hat R^1$ is given by~\cref{eq:hatR1_quad} and $\hat R^2$ by~\cref{hatR2}; $\hat B^{\text{eff}}$ by~\cref{eq:hatB1_quad} and $\tilde \theta^{\text{eff}}$ by~\cref{quadOP_theta_o1}.

	\item \textbf{MAK}:
In the mass action kinetics approximation the order parameter $\mu$ is given by~\cref{update_eqns_with_fields}, $R = R_0$ by~\cref{R_0} and $C = 0$. The only effective field used is $\tilde \theta^{\text{eff}}$ from~\cref{linearOP_theta_o1-2}.

\end{enumerate}
These options are all implemented in our software package \texttt{Achedmy.jl}. We also provide there the option of only constraining the \textit{single-species quadratic order parameter fields} $Q_{ii}$, rather than the full set of $Q_{ij}$. This is less accurate but only requires the calculation of $P$ responses, instead of the $P^2$ responses that are needed when constraining also cross-species correlators and responses as described above. The reduction to single-species quadratic order parameters can therefore signifantly reduce the computational effort when the number of species $P$ is large.

In \texttt{Achedmy.jl}, the reaction network or the stochastic population dynamics system can be defined using \texttt{Catalyst.jl}, which is the standard \texttt{Julia} ecosystem for chemical reaction networks. Within this ecosystem, implementations of other popular dynamical approximations like stochastic state simulations (SSA) or Gillespie simulations~\cite{gillespie_exact_1977}, MAK, linear noise approximation (LNA)~\cite{van_kampen_expansion_1976}, moment closure~\cite{sukys_momentclosurejl_2021} are also available. We compare our results to these methods as appropriate below, in the cases where they methods are stable and convergent. We also compare our results to our own implementation of the state-of-the-art effective mesoscopic reaction equations (EMRE)~\cite{grima_effective_2010}, which include higher order corrections in the system size expansion~\cite{van_kampen_expansion_1976} of the master equation.

For each of the above approximations, the \texttt{Achedmy.jl} package outputs a list of the following order parameters and physical quantities as a function of time and for all species: (1) the copy number means $\bm \mu$, (2) the copy number variances as defined in~\cref{num_correlator_relation_1}
, (2) the response functions $\bm R$, (3) the correlation functions $\bm C$, (4) the two-time number correlator $\bm N$ as defined in~\cref{num_correlator_relation_2,num_correlator_relation_3}, (5) the effective fields $\tilde{\bm \theta}^{\text{eff}}$, (6) $\hat{\bm B}^{\text{eff}}$, (7) $\hat{\bm R}^{\text{eff}}$, and (8) the adaptive time-step sizes.

Accumulating reliable ground truth statistics from stochastic simulation algorithms (SSA) in the regime of large intrinsic noise requires running a very large number of such simulations~\cite{harsh_accurate_2023}, which becomes computationally extremely expensive, especially for multi-species dynamics. In order to reliably determine the ground truth statistics, we resort instead to a numerical solution of the master equation using our implementation and extension of the finite space projection (FSP) method~\cite{munsky_finite_2006}, which truncates the state-space of the master equation from above, see~\cref{supp-sec:app_ME}. This method,
simply called ``Master'' below, is still computationally intensive, but in our context provides a reliable and accurate baseline to compare the approximation results to.

In the subsequent section we discuss three chosen application examples: gene regulation, enzyme kinetics and infection dynamics. We explain the phenomenology arising from small copy numbers or correspondingly small population size, and discuss the advantage of our method over existing approximations. Further details and supplementary plots concerning these examples can be found in~\cref{supp-sec:examples_SM}
and the \texttt{Jupyter} notebooks provided with the code.

\section{Applications}{\label{sec:applications}}

\subsection{Gene regulation}{\label{sec:gene_regulation}}

\begin{figure}[!t]
	\includegraphics[width=\linewidth]{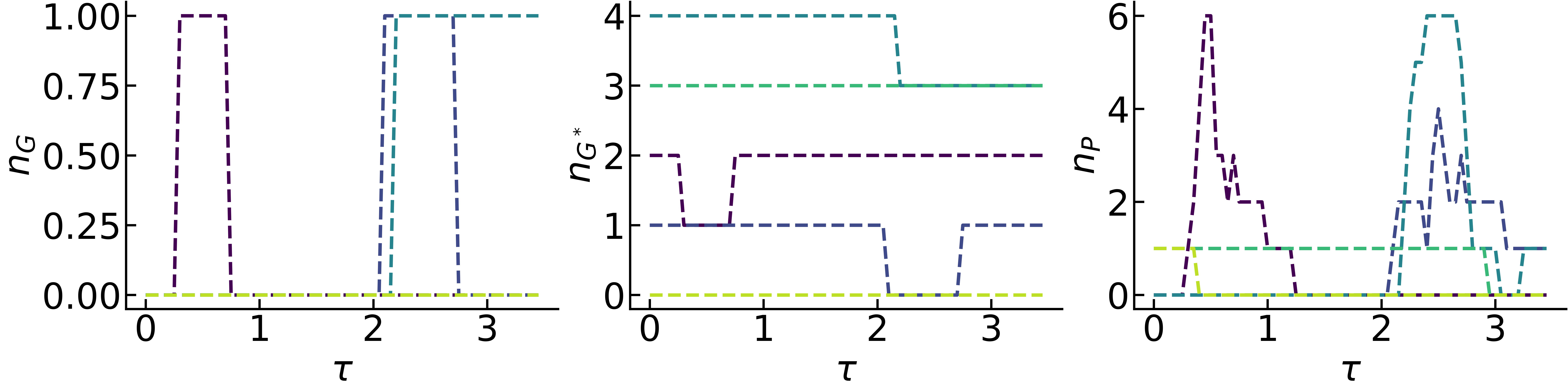}
	\caption[Stochastic time traces of the gene regulation system ]{Stochastic time traces of the gene regulation system using 5 stochastic realizations
using the Gillespie algorithm simulated at the standard gene regulation parameter set values. Each realization is plotted in a different colour. From left to right, we plot the copy numbers of active gene \textit{G}, inactive gene $\text{\textit{G}}^*$ and the protein \textit{\textit{P}}. The switching dynamics between \textit{G} and $\text{\textit{G}}^*$ and the large fluctuations of \textit{P} in this regime are apparent here.}
	\label{fig_gene_1}
\end{figure}

A minimal stochastic description of gene expression is provided by the {\em telegraph model}~\cite{peccoud_markovian_1995,iyer-biswas_stochasticity_2009,raj_stochastic_2006}, in which a gene switches stochastically between an active and an inactive state, with gene products (RNA, proteins) being synthesized only while the gene is active. This model can be regarded as the stochastic building block of the incoherent feedforward loop~\cite{mangan_structure_2003,mangan_incoherent_2006}, in which the switching rates of the target promoter are modulated in opposite directions by the upstream activator and the intermediate repressor. The noise properties of this loop have also been studied using the van Kampen system size expansion~\cite{grigolon_noise_2016}.

Here, we use a modification of the telegraphic model~\cite{schnoerr_approximation_2017,ocal_model_2023} with an autorepressive loop such that the protein can bind to the active gene in order to inactivate it. This presents a more realistic model of gene regulation since many genes are known to be regulated by an inhibition mechanism that is driven by the protein that they produce. We do not explicitly model mRNA as an intermediate molecule in protein production.

Finite copy number stochastic effects in gene regulation systems are quite pronounced since genes coding for a particular protein are usually present in low copy numbers of the order of one or two in each cell~\cite{alberts_molecular_2002}. This gives rise to intrinsic noise in cells and leads to cell to cell variability or protein burst production, a topic much studied in the last two decades~\cite{elgart_connecting_2011}.

The model we consider here consists of three species: Active gene (\textit{G}), inactive gene ($\text{\textit{G}}^*$) and protein (\textit{P}). \textit{G} can spontaneously produce \textit{P} at the translation rate $k_t$. \textit{P} binds to \textit{G} at binding rate $k_b$, inactivating it into the state $\text{\textit{G}}^*$, thus forming the autorepressive loop. \textit{P} is degraded at rate $k_{2P}$ such that it does not accumulate in the system. This system can be represented schematically as
\begin{equation}
	\begin{split}
G \xrightarrow{k_t} G+P \xrightleftharpoons[k_u]{k_b} G^* \,, \\
		P \xrightarrow{k_{2P}} \emptyset \,.
	\end{split}
\end{equation}
The total gene copy number $n_G + n_{G^*}$ is conserved in this system. Note that there are no baseline reactions in this system other than the degradation of \textit{P}.

We choose the following parameters: total time $T=3.5$, time step $\Delta t = 0.05$ (for the fixed time step methods) with Poisson initial conditions with means $\overline{n}_{0G}= 0.01$, $\overline{n}_{0G^*}=1$, $\overline{n}_{0P}=0.1$ such that most of \textit{G} is in the inactive state $\text{\textit{G}}^*$ in the beginning. The rates are chosen as $k_{2P}=2,\ k_t=10,\ k_b=1,\ k_u=0.1$.
We call this parameter set the {\em standard gene regulation parameter set} and unless specifically mentioned all the simulations for this system are carried out using this parameter set.

\begin{figure}[!t]
	\includegraphics[width=\linewidth]{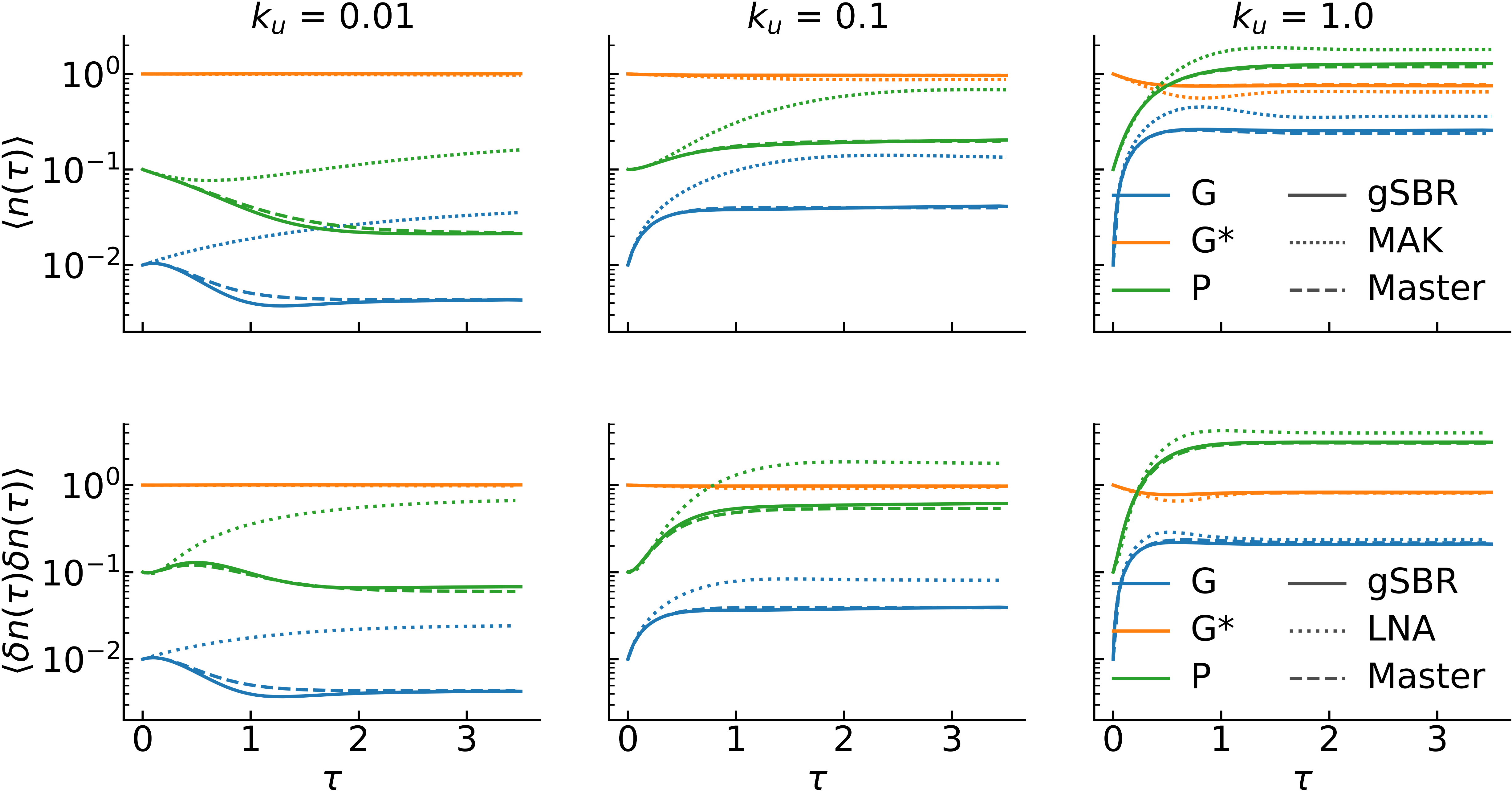}
	\caption[Means and variances of gene regulation system for three values of unbinding rates]{Mean copy numbers $\langle n(\tau) \rangle$ (top) and copy number variance $\langle \delta n(\tau) \delta n(\tau) \rangle$ (bottom) as function of time for three different $k_u = 0.01$ (left), $k_u = 0.1$ (middle) and $k_u = 1$ (right) for \textit{G}, $\text{\textit{G}}^*$ and \textit{P}. In the top plot we show the gSBR, Master and MAK methods and in the bottom plot the results from gSBR, Master and LNA. For small $k_u = 0.01$, small average numbers of \textit{P} are produced and most of the gene is in state $\text{\textit{G}}^*$. At larger $k_u$, however, more \textit{P} is produced since more \textit{G} is available and correspondingly the variance of the process also increases. The gSBR method reproduces the means and variances in close agreement with the master equation solution for all the three species, while the other methods show significant deviations, especially at smaller $k_u$ values. }
	\label{fig_gene_2}
\end{figure}

In~\cref{fig_gene_1} we plot a few stochastic time traces of the copy numbers of  \textit{G}, $\text{\textit{G}}^*$ and \textit{P}, which demonstrate the switching dynamics of the gene from \textit{G} to $\text{\textit{G}}^*$. Since $k_b \gg k_u$, \textit{G} spends most time in the inactive state $\text{\textit{G}}^*$ and any stochastic switch to the active state is quickly reverted. But since \textit{P} is produced at a large rate $k_t$ when \textit{G} is in the active state, this results in `bursty' production and intermittent, long-lived large fluctuations in the copy numbers of \textit{P}.  This limit is in fact the natural regime of operation for many of these networks. For example, there are only a few copies of each gene coding for a protein in each cell, which transcribe a few copies of mRNA that are later translated into proteins~\cite{larson_single_2009,chen_stochastic_2006}.

\begin{figure}[!t]
	\includegraphics[width=\linewidth]{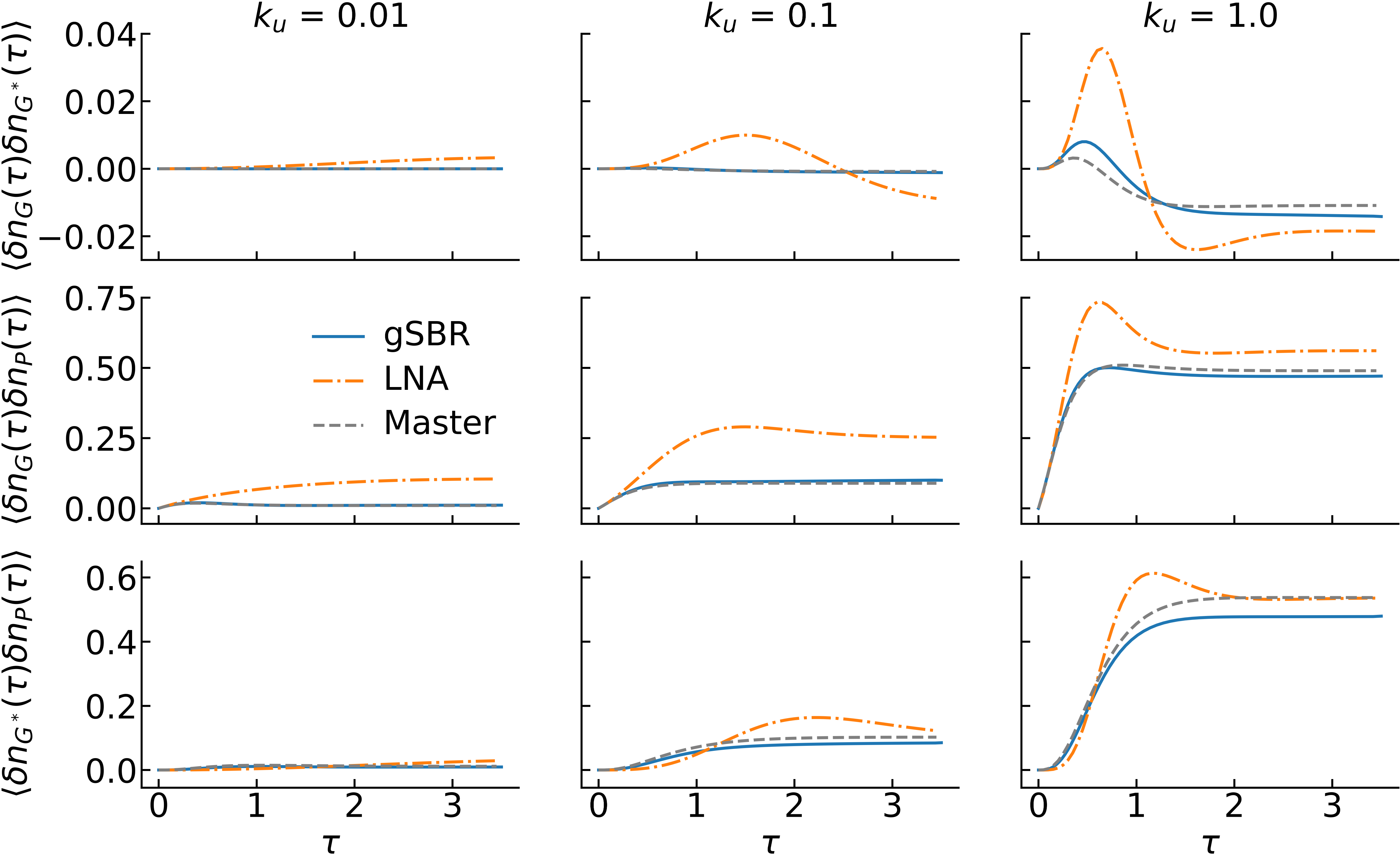}
	\caption[Cross correlators for gene regulation system for three unbinding rates]{The equal time cross species correlators $\langle \delta n_G(\tau) \delta n_{G^*}(\tau) \rangle$ (top),  $\langle \delta n_G(\tau) \delta n_P(\tau) \rangle$ (middle) and  $\langle \delta n_{G^*}(\tau) \delta n_{P}(\tau) \rangle$ (bottom) as a function of time $\tau$ for $k_u = 0.01$ (left), $k_u = 0.1$ (middle) and $k_u = 1$ (right). The correlators are plotted using the gSBR, LNA and the ground truth master equation. sSBR is much better at capturing these correlators as compared to the LNA for all times and parameter values, except the $\text{\textit{G}}^*P$ correlator at large $k_u$ values where the LNA performs better. The magnitude of all the correlators increases with $k_u$ because of increased unbinding and subsequent production of \textit{P}. }
	\label{fig_gene_3}
\end{figure}

We now vary the unbinding rate $k_u$ from $10^{-2} \text{ to } 10^1$ such that more and more of \textit{G} is available to produce \textit{P} as $k_u$ is increased. This shifts the steady state numbers of \textit{P} to larger and larger values. In~\cref{fig_gene_2} we plot the average copy numbers $\mu(\tau)=\langle n(\tau) \rangle$ as well as the equal time variance $\langle \delta n(\tau)\delta n(\tau) \rangle$ as obtained using the gSBR method and compare it with the ground truth solution of the master equation.
The MAK results for the means and the LNA results for the variance are also shown. All quantities are plotted at three different values of the unbinding rate $k_u = 0.01,\ 0.1\ \text{and}\ 1$.

For $k_u = 0.01$, most of the gene is in the state $\text{\textit{G}}^*$, so \textit{P} production is balanced with its degradation at small \textit{P}. The variance of \textit{P} first increases because of its production, but then quickly decreases because of the spontaneous degradation. For $k_u = 0.1$ and $k_u =1$, both \textit{P} and \textit{G} increase from their initial values and then settle at some intermediate value in both mean and variance.
For all the parameter values, gSBR has predictions very close to the ground truth Master equation while the MAK and LNA have significant corrections to the mean and the variance respectively. These deviations, however, become small as $k_u$ is increased, slowly approaching a mean field limit.
This figure demonstrates the utility of using gSBR method since it is much more reliable for the dynamics as compared to other methods, across a large parameter range. The (adaptive) time-step sizes from the gSBR solver are plotted in~\cref{supp-fig_gene_time}
in~\cref{supp-sec:example_SM_gene}.

We compare the results for other dynamical methods like MCA, single species response function gSBR, SBR in~\cref{supp-fig_gene_5} in~\cref{supp-sec:example_SM_gene}. Normal Moment Closure~\cite{sukys_momentclosurejl_2021} is widely known to be unstable and quickly diverges for this system~\cite{ocal_model_2023}. The EMRE~\cite{grima_effective_2010} yields negligible inverse system size corrections and its predictions are in very close agreement with the MAK, hence not repeated. The gSBR methods again outperforms all the other methods, followed by the SBR and the MCA.

In addition to the variance, we can also measure the equal time cross correlators $\langle \delta n_i \delta n_j \rangle$ from the different methods and compare them to the ground truth Master equation. In~\cref{fig_gene_3} we plot the three cross correlators $GG^*$, \textit{GP} and $G^*P$ as a function of time, for the three values of $k_u$ also considered in~\cref{fig_gene_2}. For small $k_u=0.01$, all the correlators are small since the reversible reaction is dominated by $k_b$ and no appreciable reaction takes place to create correlations from an uncorrelated initial condition. For intermediate $k_u$ there are some non-trivial correlations since small amounts of \textit{P} are produced. For both of these cases gSBR performs much better when compared to the conventional LNA. For $k_u = 1$, the steady state $GG^*$-correlator is negative as these two species must be anti-correlated on account of the conservation law.  \textit{G }and \textit{P}, on the other hand, are positively correlated because of the $k_b$ and $k_t$ reactions. gSBR captures the correct trend, while LNA overestimates these correlators. $\text{\textit{G}}^*$ and \textit{P }are also positively correlated because of the $k_u$ reaction; the gSBR approximation somewhat underestimates this correlation.

\begin{figure}[!t]
	\includegraphics[width=\linewidth]{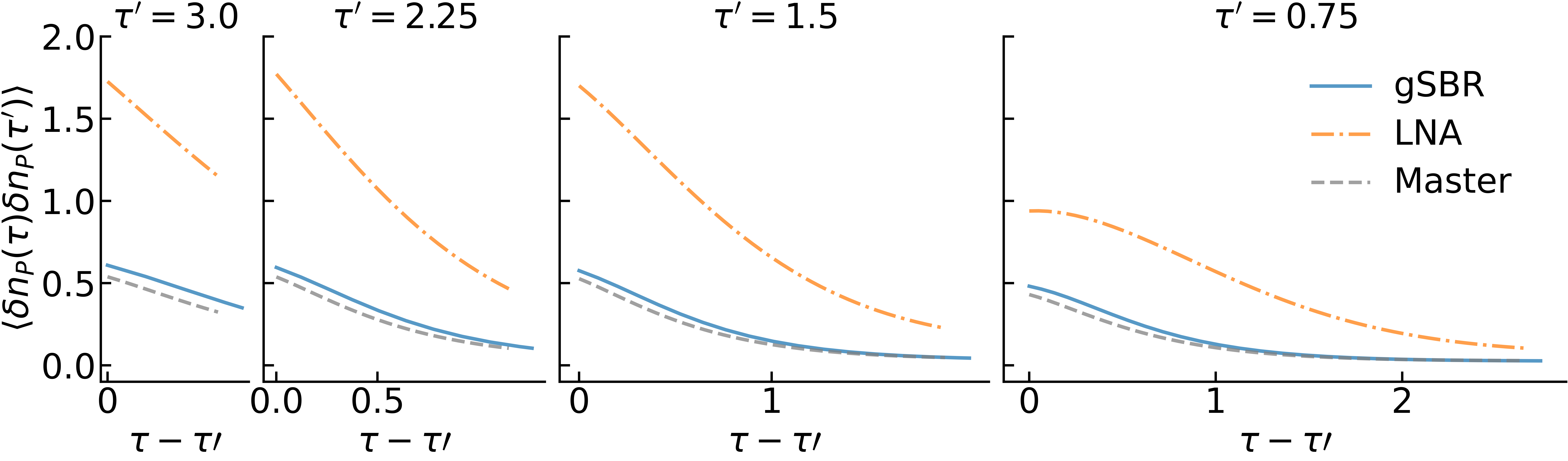}
	\caption[Two time Protein-Protein number correlator in the gene regulation system]{The two time number correlator for \textit{P}, i.e.~$\langle \delta n_P(\tau) \delta n_P(\tau')\rangle$ as a function of the lag time $\tau-\tau'$ for $\tau'=0.75,\ 1.5,\ 2.25,\ 3.0$ from right to left. These simulations are carried out using the standard gene regulation parameter set. The gSBR method here predicts the true correlations closely while the LNA has substantial deviations, especially at small lag times. gSBR consistently captures the decay time scale of the order of $\sim 1/k_{2P}$, which is overestimated by the LNA.}
\label{fig_gene_4}
\end{figure}

Of great interest in stochastic processes such as reaction networks are the two time quantities; in~\cref{fig_gene_4}, we plot the time lagged number-number correlator for the protein $\langle \delta n_P(\tau) \delta n_P(\tau')\rangle$ as a function of lag time $\tau-\tau'$ for different values of $\tau'$. Such a quantity is experimentally accessible as it can be calculated directly from time traces of \textit{P}. The latter could be measured using, for example, Fluorescence microscopy or Fluorescence Correlation Spectroscopy (FCS)~\cite{magde_fluorescence_1974,magde_thermodynamic_1972}. The zero lag time values
are the equal-time correlations shown in~\cref{fig_gene_2}-bottom and are largely independent of $\tau'$ because the system quickly reaches a steady state. At nonzero lag times, the correlator decays on a time scale of $\sim 1$ consistent with the protein decay rate of $k_{2P}$ of $O(1)$.
The gSBR method closely reproduces the master equation ground truth, while the LNA shows substantial deviations especially at small lag times. The other lagged time correlators behave similarly and are thus not shown here.

\subsection{Enzyme Kinetics}{\label{sec:enzyme}}

\begin{figure}
	\includegraphics[width=\linewidth]{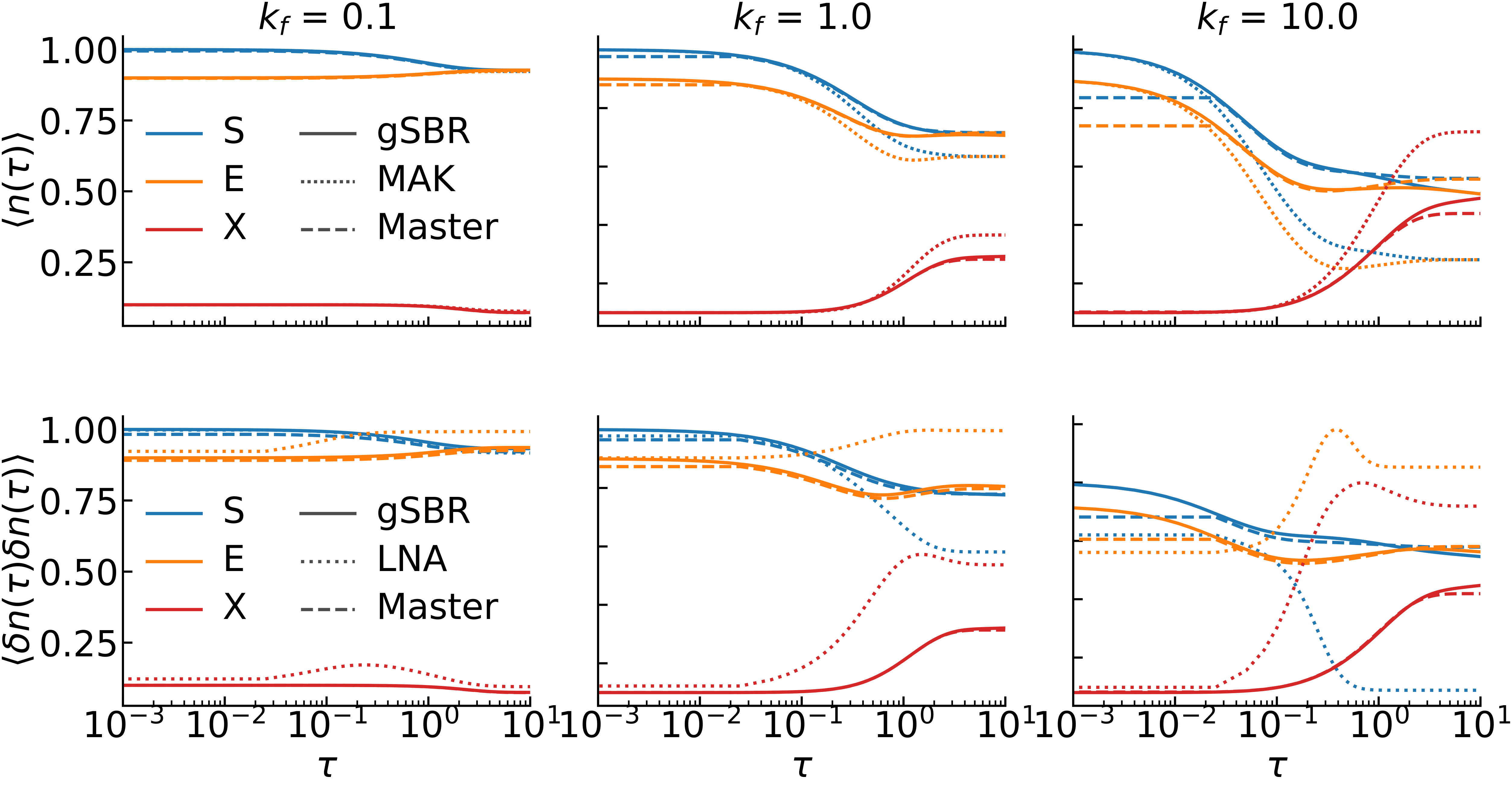}
	\caption[Means and variances of enzyme kinetics system for three values of the forward rates]{Analog of~\cref{fig_gene_2} for Michaelis-Menten enzyme kinetics. Mean copy numbers (top) and variance (bottom) for substrate \textit{S}, enzyme \textit{E} and product \textit{X} for $k_f = 0.1,\ 1,\ 10$ from left to right. Adaptive time-step gSBR solutions can resolve the dynamics across the varied time scales (here four
orders of magnitude from left to right, x-axis in log-scale) over the course of the dynamics.
The MAK predictions for the means and the LNA for the variances are unreliable in this challenging regime, while gSBR performs much better. In particular, the MAK predicts much greater production of \textit{X} and saturation of \textit{E}, which is not seen in the master equation solution. The apparently initial discrepancies in the means and variances are artefacts arising from limitations of the FSP~\cite{munsky_finite_2006} algorithm for solving the master equation.}
	\label{fig_enz_1}
\end{figure}

\noindent We next turn our attention to another ubiquitous building block of typical biochemical reaction networks: substrate enzyme kinetics, also called irreversible Michaelis-Menten kinetics~\cite{MichaelisMenten2007}, where the substrate (\textit{S}) reacts reversibly with the enzyme (\textit{E}) with forward rate $k_f$ and backward rate $k_b$ to form the complex (\textit{C}), which disintegrates irreversibly into product (\textit{X}) and the enzyme (\textit{E}) at disintegration rate $k_d$. We further assume that the product is degraded at rate $k_{2X}$ and the substrate spontaneously created at rate $k_{1S}$ and degraded at rate $k_{2S}$, so that both \textit{S} and \textit{X} approach a steady state.
This reaction network can be schematically represented as
\begin{equation}
	\begin{split}
S + E \xrightleftharpoons[k_b]{k_f} C \xrightarrow{k_d} E + X \,,\\
		S  \xleftrightharpoons[k_{2S}]{k_{1S}} \emptyset, \quad \quad X \xrightarrow{k_{2X}}  \emptyset \,.
	\end{split}
\end{equation}

To investigate the dynamics in this system in the regime of small copy numbers, we consider that on average we only have one enzyme molecule in the system: we start with initial Poisson copy number distributions with means
$\overline{n}_{0S}=1$, $\overline{n}_{0E}=0.9$, $\overline{n}_{0C}=0.1$,  $\overline{n}_{0X}=0.1$. We use the following parameters: $k_b=0.1$, $k_d=1$, $k_f= 1$, $k_{2X} = 1$, $k_{1S} = 1$, $k_{2S} = 1$, total simulation time $T=10$, time step $\Delta t=0.025$ (for the fixed time step methods).
We call this parameter set the {\em standard enzyme parameter set} and carry out simulations using this unless otherwise specified. Since we consider the dynamics in the low copy number regime, it exhibits large fluctuations. Note that $n_E+n_C$ is conserved in this system, so we do not show $n_C$ explicitly.

\begin{figure}[!t]
	\includegraphics[width=\linewidth]{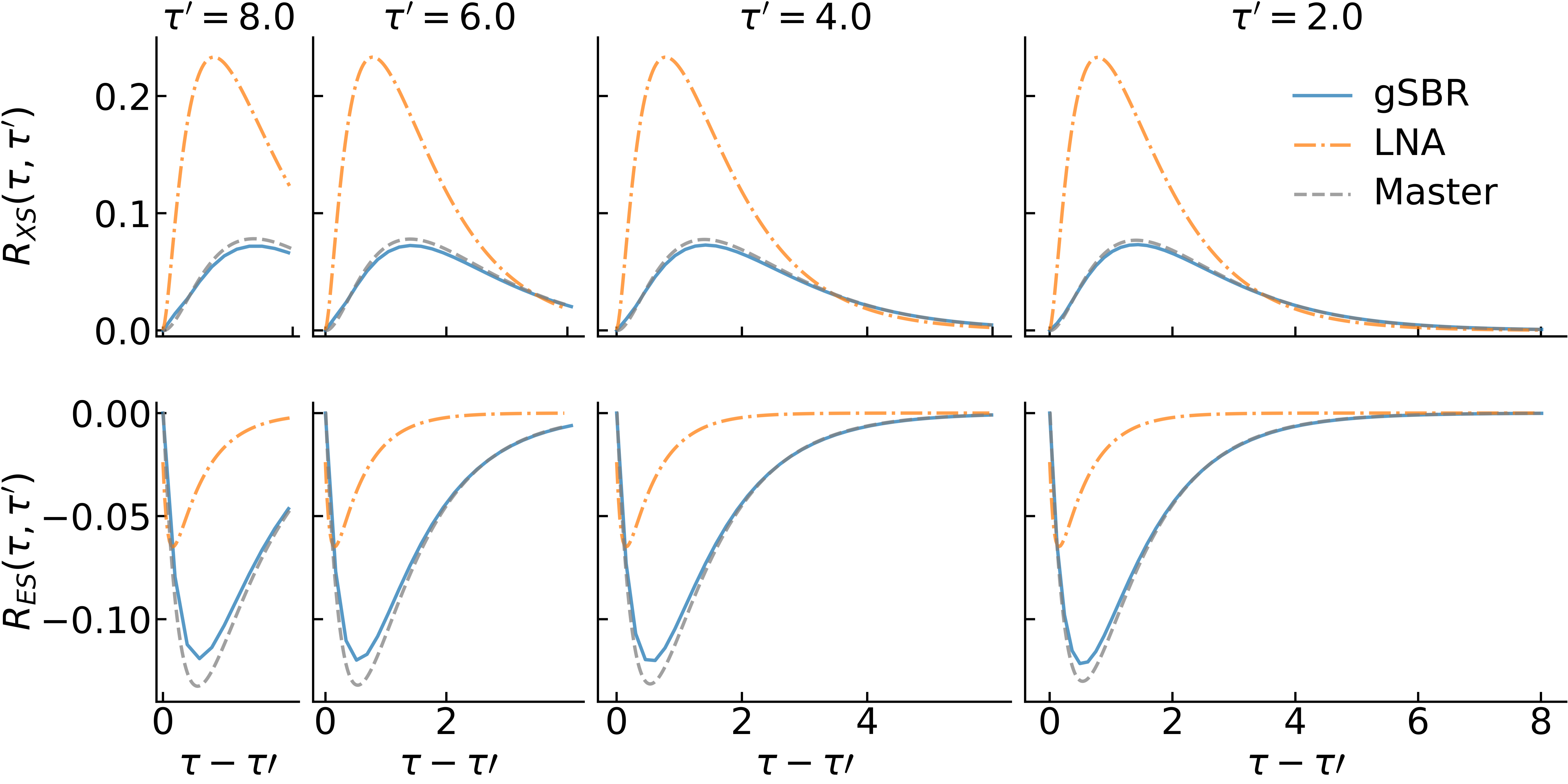}\\
	\caption[Time lagged response of product and enzyme to perturbations of the substrate as a function of the lag time]{Two-time cross response functions $R_{XS}(\tau,\tau')$ and $R_{ES}(\tau,\tau')$, i.e.~the response of \textit{X }and \textit{E }at time $\tau$ as functions of the lag times $\tau-\tau'$ for impulse perturbations of $S$ at $\tau' = 2,\ 4,\ 6,\ 8$ from right to left, simulated using the standard enzyme kinetics parameter set. \textit{X }responds positively while \textit{E }responds negatively to increasing \textit{S}, both as expected. At $k_f=1$, the dynamics takes place on a time scale of $\sim 1$ and hence the response functions appear approximately time translation invariant for $\tau>2$.
The gSBR is significantly more faithful in reproducing these response functions at small to intermediate lag times when compared to the LNA. }
	\label{fig_enz_3}
\end{figure}

We will study the dynamics in this system at different values of the forward rate $k_f$  from $0.1 \text{ to } 10$, which is the rate for the binary reaction. Similarly to~\cref{fig_gene_2} we start by considering the average copy numbers in~\cref{fig_enz_1} at $k_f = 0.1,\ 1,\ 10$. The gSBR method predicts the means and variances of all species in reasonable agreement with the ground truth solution of the master equation in this range, while the MAK for mean values overestimates the saturation of \textit{S} and \textit{E}. The LNA for the variances spuriously suppresses fluctuations of \textit{S}, while enhancing those of \textit{E }and \textit{X}.

We compare the results for other dynamical methods like MCA, single species response function gSBR, EMRE~\cite{grima_effective_2010}, Normal Moment Closure~\cite{sukys_momentclosurejl_2021} and the Michaelis Menten~\cite{MichaelisMenten2007} Quasi Steady State approach (MM QSS,~\cref{supp-sec:example_SM_enzyme}) in~\cref{supp-fig_enzyme_3}. The gSBR methods again outperforms all the other methods, followed by Normal Moment Closure. Since the MM QSS is a coarse grained approach, which assumes that \textit{E} and \textit{C} follow the dynamics of \textit{S} in a quasi steady state, this method performs worse than the MAK as expected.

As the forward rate is increased, the time scale of the dynamics becomes correspondingly shorter, more enzyme and substrate get saturated and more product is formed. Eventually all the species settle into their corresponding steady states on a time scale of the order of $\sim 1/k_f$. This highlights the utility of the adaptive time step solver (see ~\cref{sec:numerical_solution}), which successfully captures the dynamics at such variable time scales.

In the steady state, production of new substrate is balanced by its spontaneous degradation and conversion to product by the enzyme. In the absence of the enzymatic reaction, the average substrate copy number would thus remain at its starting value of unity, which is already in equilibrium with its creation and destruction reactions. In~\cref{fig_enz_1} the change in the dynamics of \textit{S} is thus primarily driven by the enzymatic reactions and not the baseline reactions of the substrate.

The equal-time cross correlators between \textit{S} and \textit{X}, and \textit{E} and \textit{X} are also interesting to consider and, in analogy to~\cref{fig_gene_3}, are plotted in~\cref{supp-fig_enz_2} in~\cref{supp-sec:example_SM_enzyme}, with gSBR capturing the correlations much better than LNA across the entire parameter range.

A further quantity of interest in this system, which could also in principle be probed experimentally, is the response of the system to perturbation of \textit{S}. In~\cref{fig_enz_3} we plot for $k_f =1$ the two-time response functions $R_{XS}(\tau,\tau')$ and $R_{ES}(\tau,\tau')$, which measure the change in \textit{X} and \textit{E} respectively at time $\tau$ to an impulse perturbation in the creation rate of \textit{S}, and thus a corresponding change in $n_S$, at time $\tau$; see the response function discussion in and around~\cref{R0deriv}.
We plot this as a function of the lag time $\tau-\tau'$ for several values of $\tau'$. $R_{XS}$ is positive for small lag times because increasing \textit{S }increases \textit{X}, while $R_{ES}$ is negative since increasing \textit{S }saturates more \textit{E }into \textit{C}. The gSBR approximation is much more accurate in this regime than the LNA, which predicts much stronger responses. For small $\tau'$, since \textit{E} is already partly saturated, the response is slightly lower at its peak as compared to the peak response at larger $\tau'$ values. At longer lag times both of the responses decay to zero as expected, but the gSBR response decays slower than the LNA, in good agreement with the master equation. The other response functions show similar trends and hence are not shown explicitly here.

\begin{figure}
	\includegraphics[width=\linewidth]{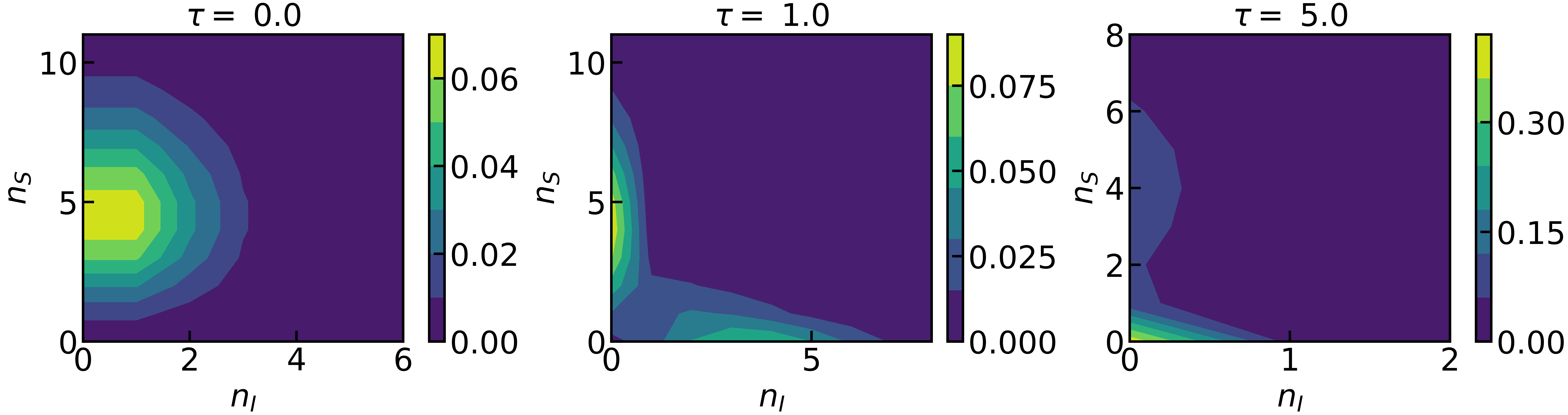}
	\caption[Joint probability of infectious and susceptible individuals in the SIR system for different times ]{The joint probability distribution $P(n_I,n_S,\tau)$ for three different times $\tau = 0,\ 1,\ 5$ from left to right. The dynamics proceeds from the initial product Poisson state with means $\mu_I = 1,\ \mu_S = 5$ to the intermediate state where \textit{S }and \textit{I }are anti-correlated to the final state with the bi-modal marginal distribution of \textit{S}.}
	\label{SIR_1}
\end{figure}

\subsection{SIR infection dynamics}{\label{sec:SIR}}

\noindent
A further interesting problem of significant consequence is that of infection or epidemic spreading in small populations, which can have large intrinsic noise fluctuations leading to vastly different outcomes. The Susceptible-Infected-Recovered (SIR) model~\cite{kermack_contribution_1927,neymann_deterministic_1956} is one of the most popular models for the study of infection dynamics or epidemic modelling. The SIR system is defined such that an infectious individual (\textit{I}) infects a susceptible individual (\textit{S}) at an infection rate $k_i$. Species \textit{I}
can then either recover (or, for the purposes of the model equivalently, die) at rate $k_{2I}$; reinfection is not allowed. The reaction system can thus be written as
\begin{equation}
	\begin{split}
		S + I &\xrightarrow{k_i} 2I \,,\\
		I &\xrightarrow{k_{2I}} R \,,\\
	\end{split}
\end{equation}
where we do not explicitly consider the dynamics of recovered individuals \textit{R} because their number can be deduced from the others by conservation of $n_S+n_I+n_R$.

We use the parameter values $k_i=0.75$ for the rate constant of the binary reaction, and $k_{2I} = 1$ for the recovery rate. Since we are not explicitly tracking the dynamics of \textit{R}, we can equivalently replace the reaction $I \xrightarrow {}R$ by $I \xrightarrow{k_{2I}} \emptyset$, so that $k_{2I}$ becomes a baseline rate for the destruction of \textit{I}.
We start with Poisson initial conditions with means
$\overline{n}_{0I} = 1$ and $\overline{n}_{0S} = 5$.
With our count-based infection propensity, the deterministic initial growth ratio is then \( R_0 = k_i \overline{n}_{0S}/k_{2I}=3.75\) which in a large system would always ensure an epidemic outbreak. However, individual stochastic realizations in small systems can exhibit no outbreak or early extinction. In particular, given the initial Poisson statistics there is a probability of $\approx 0.736$ that any realization of the system starts with none or only one infected individual. This is relevant in the situation of infection dynamics at the scale of different households, where the initial number of infected people can be Poisson distributed at the beginning of an epidemic. In realizations where there are no initially infected individuals, i.e.~$n_I(0) = 0$, it follows that both $n_I(\tau)$ and $n_S(\tau)$ are constant in time.

\begin{figure}
	\includegraphics[width=\linewidth]{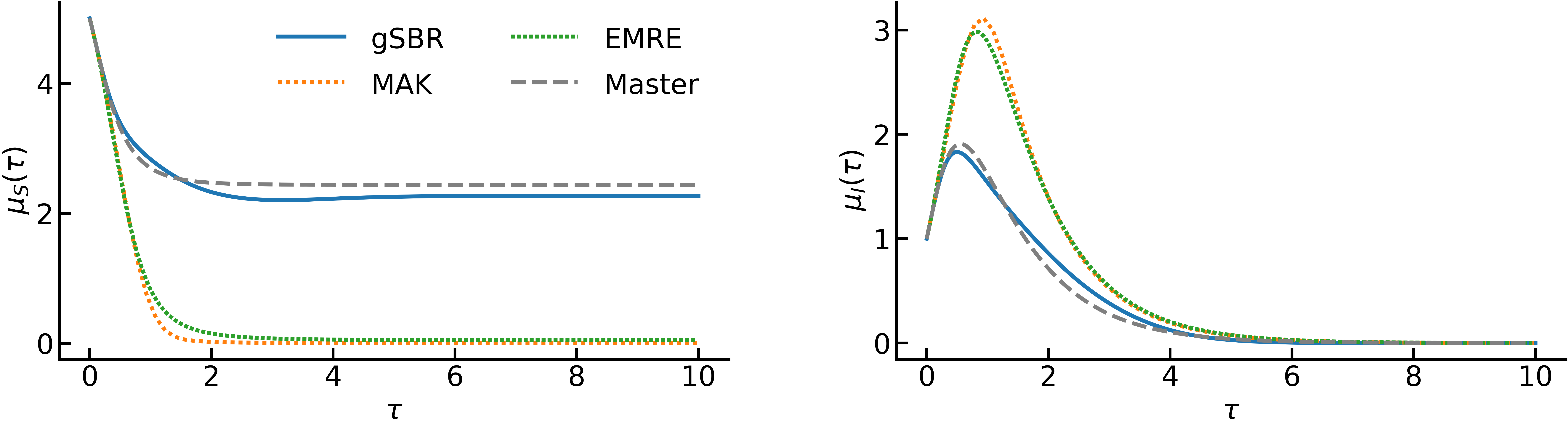}\\
	\caption[Mean copy number of susceptible and infectious individuals in the SIR system]{ The average copy number of \textit{S }(left) and \textit{I }(right) as a function of time obtained using the gSBR, MAK, EMRE and Master equation methods. The average number of \textit{S }decreases from the starting value while the number of \textit{I }first increases because of infection and then decreases because of recovery. The size of the infection is exaggerated by the MAK and the EMRE, which predict that essentially all susceptible individuals (\textit{S}) get infected. In comparison, gSBR predicts the mean copy numbers of \textit{S }and \textit{I }much more accurately.}
	\label{SIR_2}
\end{figure}

By way of orientation we show in~\cref{SIR_1} the joint probability distribution of \textit{I }and \textit{S }at three times: at the beginning of the dynamics, at an intermediate time and in the steady state, in each case obtained by numerical solution of the master equation. The initial state is a product of Poisson distributions by assumption.
From this the system evolves to a bi-modal distribution with negligible probability across the diagonal since the numbers of \textit{S }and \textit{I }become anti-correlated.
In the steady state, on average, there is almost no population of \textit{I}, but we see a clear bi-modality in the marginal distribution of \textit{S }with one peak around $S=0$ and another one around $S=4$. The latter, which in the joint distribution corresponds to a peak around $S=4$ and $I=0$, is caused by histories without any outbreak. To capture the moments of the marginal distribution of \textit{S }is, then, a challenging problem for any approximation method because of the bi-modality that appears spontaneously with time.

In~\cref{SIR_2} we test the performance of the different methods in capturing the correct mean values as a function of time. The MAK predicts that on average, almost no \textit{S }remains in the steady state and essentially the entire population gets infected, with $\mu_S^{\text{MAK}}$
approaching $0.058$ for $\tau\to\infty$. However, in the true underlying dynamics as given by the solution of the master equation, the mean copy number of \textit{S} in the steady state has a much larger finite value, indicating that a significant fraction of the population never gets infected. The gSBR method is able to capture this effect with small deviation from the true steady state value while the EMRE, like MAK, predicts a very small steady state population of \textit{S}. As a corollary, the peak infection numbers are predicted accurately by gSBR while MAK and EMRE exaggerate this quantity.

We consider the variances and cross correlators in~\cref{supp-SIR_3}
in~\cref{supp-sec:example_SM_SIR}, which demonstrates the poor performance of LNA as compared to gSBR in predicting these quantities. In particular, the LNA predicts that the variance of \textit{S} falls sharply to zero
while in the ground truth solution the variance saturates to a large constant value.

\section{Conclusions and Discussions}{\label{sec:conclusions}}

We have introduced a robust analytical scheme for approximating the dynamics in the large intrinsic noise limit in stochastic discrete population dynamics systems, and have demonstrated it to outperform all other approximation schemes for a range of fundamental chemical reaction networks across the wide range of parameter values considered. Our formulation is also directly applicable to a broad range of models from ecological or evolutionary population dynamics, collective behaviour in small agent systems or opinion dynamics. More broadly, our method enables an analytical approach to discovering new phenomenology driven by intrinsic noise, such as noise-induced Turing  patterns~\cite{karig_stochastic_2018}, and a unifying framework to address elusive observations like noise-induced persistence via quasi-stationary states and metapopulation rescue, e.g.\ long-lived non-extinction even when deterministic models predict collapse~\cite{block_population_2000, schreiber_extinction_2021, constable_demographic_2016}.

Starting from Doi-Peliti field theory, which maps a discrete jump process to continuous fields, we construct a path integral for the probability of any given trajectory of the system starting from some initial conditions. We set up a baseline system of unary and first order reactions for which the path integral is Gaussian, and treat higher order reactions as interactions in the sense of a perturbation theory. We then effectively replace these interactions by effective fields using the extended Plefka expansion, where we constrain the path integral to reproduce the means, responses and correlations in the presence of interactions. Dynamical equations then follow from a variational principle for the correspondingly constrained free energy, the Plefka free energy.
At first order and with constraints only on the means, these equations give the analogue of the mean field approximation in equilibrium systems, which for reaction networks we found to be simply mass action kinetics.

For the more sophisticated approximations the time evolution equations for the means involve memory, via integrals over the past weighted by appropriate memory functions.
Of course the underlying microscopic dynamics we consider is Markovian, with the rate for each chemical reaction depending only on the current state of the system via the propensity function. The fact that the equations for copy number means (as well as correlations and responses) contain memory terms can be seen as a consequence of the fact that a representation in terms of these low-order moments is effectively a projection of the full microscopic dynamics of the system, which would be specified by the joint distribution of the copy numbers for all species and times.

For the more sophisticated approximation schemes we focused on networks of at most binary reactions. We showed that for these the Plefka expansion to second order in interactions, with linear and quadratic order parameters included, recovers the mode coupling approximation. To improve the approximation further we then proposed the gSBR ansatz, which includes an infinite subseries of higher order terms in the interaction. This effectively results in a regularized memory kernel, making the approximation stable and accurate even for very large interacting reaction rates. The resulting ansatz also allows for a relatively efficient computational solution in tensorial form, along with the other approximations in \texttt{Achedmy.jl}.

Our approach is more versatile still than we have been able to present within the scope of this paper. For example, higher order correlations like $n$-time quantities, though experimentally difficult to probe, could be easily approximated by expansion in terms of second order correlators and using the gSBR Ansatz. Similarly, non-Poisson initial condition distributions could be relatively easily incorporated into our approach, with delta distributions for deterministic initial conditions an useful special case. This change would introduce different $t=0$ boundary terms in the action. Finally, since our method does not rely on the existence of a steady state, it will also be interesting in future to explore examples where a steady state does not exist.

We have implicitly assumed that our reaction systems are well mixed rather than diffusion limited, so that only overall copy numbers of each species affect the reaction rates. This assumption does not always hold, e.g.\ the binding and unbinding of transcription factors to DNA in gene regulation networks can be significantly influenced by the spatial arrangement of the DNA~\cite{pai_spatial_2010,hafner_spatial_2023}. The challenges arising from strong fluctuations thus become more acute for true spatially resolved reaction-diffusion dynamics.
It is, however, straightforward to include space in our analysis by considering a compartmentalized spatial grid, with molecules in different compartments treated effectively as different species. Diffusion can then be modelled~\cite{schnoerr_cox_2016} as a unary conversion reaction between species in neighbouring compartments. This only introduces quadratic terms in the Hamiltonian and thus does not give rise to memory corrections. In the continuous time and space limit, the Hamiltonian simply acquires an additional term $D \tilde{\phi} \nabla^2 \phi $ with $D$ the diffusion coefficient and the fields explicitly a function of space.  We have tested this approach in a numerical analysis of a spatial gene regulation model considering nucleus and cytoplasm, with promising initial results.

Our approach further allows for an arbitrary time-dependence of the reaction rates. We have not explored this in the application examples presented here, but this flexibility may be useful in the study of e.g.\  time-dependent branching processes~\cite{assaf_population_2008}. Here new physical phenomena may arise
from the competition between the relaxation time scale of a system and the timescale on which reaction rates vary, e.g.\ under periodic driving,

To manage the computational complexity associated with our method, in addition to the adaptive time step solver, additional simplifications could be explored including appropriate memory function truncation~\cite{stahl_memory_2022}, low rank compression~\cite{kaye_low_2021} or imposition of time translation invariance.
An interesting extension of our method in this direction would be to directly determine the statistical correlations in the time translation invariant steady state, without the need to integrate the dynamics through potentially long transients. Many initial results already exist in this regard: Feinberg~\cite{feinberg_chemical_1987} classified reaction networks based on a topological property called deficiency and showed the existence of a unique MAK steady state for zero deficiency networks, a result later extended~\cite{anderson_product-form_2010} to product Poisson steady states for arbitrary stochastic dynamics. In our approach, for example, a reaction network with only baseline reactions corresponds to a zero deficiency network and results in a product Poisson steady state. The Doi-Peliti path integral approach gives us a powerful tool to address the dependence of the steady state distribution on the network topology, contrasting with recently developed steady state moment based approaches~\cite{krishnamurthy_solving_2017}.

In the future it would be interesting to develop an accurate dynamical inference method for reconstructing the topology and rate constants of a reaction network from its time traces in the regime of small copy numbers. This should take into account the corrections from copy number fluctuations that we have focussed on in this work. In this challenging regime of small signal to noise ratio, inference may require long time traces or many repeats, which may be obtainable from emerging high-throughput technologies~\cite{la_manno_rna_2018, izeddin_single_molecule_2014, tang_mrna_seq_2009}. Either way, inference would always have to build on approaches -- like the one in this paper -- for solving the ``forward problem'' of making accurate predictions of the dynamics for given parameters.

Though limited methods~\cite{craciun_statistical_2013} exist for inference in biochemical networks, this field has been receiving progressively more attention~\cite{loskot_comprehensive_2019} as more biological data is becoming available. The existing approaches are predominantly based on expensive numerical simulations~\cite{diamataris_statistical_2023}. The few existing analytical  methods~\cite{loskot_comprehensive_2019,schnoerr_approximation_2017} would be generally ineffective in the regime of small copy numbers since they assume underlying mean field models.

Using the dynamical variational and Plefka free energy expansion approach as presented here, we have the effective Gaussian path integral with the interaction terms replaced by the effective fields, which can be employed for likelihood based inference. Indeed, the extended Plefka expansion has already been used~\cite{bravi_inference_2017} for inferring the hidden bulk nodes for networks with linear couplings. Related dynamical mean field approaches~\cite{dunn_learning_2013} such as Belief Propagation~\cite{battistin_belief_2015} have also been used for inference from non-equilibrium Ising model dynamics. Another approach~\cite{schnoerr_cox_2016} based on the Cox-process representation tackles the inference problem in non-linear binary reaction networks, but effectively linearises the binary reactions and is thus likely not to capture the full stochasticity. As a more long-term goal, finally, one would like to tackle the inference problem in the yet more challenging situation when not all species of the reaction network can be observed, because of the additional memory effects~\cite{harsh_physics-inspired_2024,herrera-delgado_memory_2018} that emerge from this.

\section*{Acknowledgements}
We thank Ramon Grima for pointing us to the gene regulation example studied above, including the failure of standard approximation and closure schemes for this system, and Grzegorz Szamel for discussions regarding the mode coupling approximations. We also thank Ander Movilla~\cite{MovillaMiangolarra2017} and Maximilian Kurjahn~\cite{Kurjahn2020} for important contributions in the early stages of the development of the approach presented here.

\section*{Data and code availability}

The results reported in this work were obtained using numerical
simulations. The source code implementing the methods, together
with the parameters and scripts used to reproduce the numerical
results and figures, is publicly available at \href{https://github.com/btemoshir/Achedmy.jl}{\texttt{Achedmy.jl}} (\href{https://github.com/btemoshir/Achedmy.jl}{https://github.com/btemoshir/Achedmy.jl}).

\clearpage
\bibliography{paper}

\end{document}


\maketitle

\begin{fmffile}{diagrams_SM/fields_SM}

	\fmfcmd{
		style_def end_arrow expr p =
		cdraw p;
		cfill (harrow (p, 0.5));
		enddef;}

	\fmfcmd{
		style_def end_arrow_two expr p =
		cdraw p;
		cfill (harrow (p, 0.6));
		enddef;}

	\fmfcmd{
		style_def end_arrow_dash expr p =
		draw_dashes p;
		cfill (harrow (p, 0.4));
		enddef;}

	\fmfcmd{
		style_def end_arrow_dash_two expr p =
		draw_dashes p;
		cfill (harrow (p, 0.5));
		enddef;}

	\fmfcmd{
		style_def double_arrow expr p =
		draw_double p;
		cfill (arrow p);
		enddef;
	}

	\fmfcmd{
		style_def mixed_response_GB expr p =
		save oldpen;
		pen oldpen;
		oldpen := currentpen;
		pickup oldpen scaled 3;
		draw (subpath( (0 ,0.5)*length(p)) of p) withcolor green;
		draw (subpath( (0.5,1)*length(p)) of p) withcolor blue;
		pickup oldpen;
		cullit ; undraw p; cullit;
		cfill (marrow (p, 0.5));
		enddef;}

	\fmfcmd{
		style_def mixed_response_BG expr p =
		save oldpen;
		pen oldpen;
		oldpen := currentpen;
		pickup oldpen scaled 3;
		draw (subpath( (0 ,0.5)*length(p)) of p) withcolor blue;
		draw (subpath( (0.5,1)*length(p)) of p) withcolor green;
		pickup oldpen;
		cullit ; undraw p; cullit;
		cfill (marrow (p, 0.5));
		enddef;}

	\section{Propensity function}{\label{SM:sec:propensity_function}}

	This section follows the analysis originally presented in~\cite{harsh_accurate_2023}. The propensity function in \cref{main-propensity_1} of the main paper uses rate constants $k_\beta$ of physical dimension of inverse time, $1/t$, since we are working with absolute molecule numbers. In the MAK propensities however, i.e.\ \cref{main-eq:MAK_propersity} of the main paper, the rate constants $j_\beta$ have additional dimensions of volume depending on the order of the reaction considered.

	Considering the binary reaction $A+B \rightarrow C$ as an example, the mass action description in terms of densities $\bm x=\bm n/V$ would be $\partial_\tau x_A = -j_\beta x_A x_B$.
	Converting this to an equation for the change in particle number $\bm n= \bm x V$ we get $\partial_\tau n_A = -\frac{j_\beta}{V} n_A n_B$. In the general case one would accordingly write the propensity function as originally defined in~\cref{main-propensity_1} of the main paper as
	\begin{equation}
		f_\beta(\bm{n}) = j_\beta V \prod_i \frac{n_i !}{(n_i - r_i^\beta)! V^{r_i^\beta}}
	\end{equation}
	with the rate constant $j_\beta$ having dimension $(V^{\sum_i r_i^\beta-1}/t)$.
	This is again easiest to understand from the examples:
	\begin{enumerate}
		\item For a creation
		reaction such as $\emptyset \rightarrow A$, $f_\beta(\bm{n}) = j_\beta V$
		\item For a unary destruction
		reaction such as $A \rightarrow \emptyset$, $f_\beta(\bm{n}) = j_\beta n_A$
		\item For a
		binary reaction such as $A+B \rightarrow C$, $f_\beta(\bm{n}) =\frac{j_\beta}{V} n_A n_B$
	\end{enumerate}

	In our work, we are primarily interested in the standard case where the reactions are at most binary ($\sum_i s_{ir}\leq 2$, $\sum_i s'_{ir}\leq 2$); higher order reactions can be represented as sequences of such binary ones.~\Cref{table:different_reactions} collects the types of reactions that can occur and gives the mass action reaction fluxes and stoichiometries.

	\begin{table}[h!]
		\centering
		\begin{tabular}{| c|c|c|c|}
			\hline
			Reaction & Type & Fluxes, $f_i$ & Stoichiometry\\
			\hline
			$X_j + X_l \xrightarrow{k_{jl,i}} X_i $ & Complex formation & $ j_{jl,i}x_j x_l $ & $S_j=S_k=-1$, $S_i=+1$ \\
			\hline
			$X_j + X_l \xleftarrow{k_{i,jl}} X_i $ & Complex dissociation & $ j_{i,jl} x_i $ & $S_j=S_k=+1$, $S_i=-1$ \\
			\hline
			$X_i \xrightarrow{\lambda_{i,j}} X_j $  & Unary conversion & $ j_{i,j}x_i $ & $S_i=-1$, $S_j=+1$ \\
			\hline
$X_i \xrightarrow{\omega_{i}} \emptyset $ & Destruction & $ j_i x_i $ & $S_i=-1$\\
			\hline
			$X_i \xleftarrow{\kappa_{i}} \emptyset $ & Creation & $ j_i $ & $S_i=+1$\\
			\hline
		\end{tabular}
		\caption[Possible reaction types with fluxes and stoichiometry]{Possible reactions in a network with at most bi-molecular reactions, with the corresponding reaction fluxes and stoichiometries.}
		\label{table:different_reactions}
	\end{table}

MAK has severe deviation in the large fluctuation regime as outlined in the main text. Retaining the discrete nature of particle number, in the next section we treat the CRN dynamics using the so called Chemical Master equation, which will lead to a path integral for the CRN dynamics.

	\section{Second quantization of the Chemical Master Equation}{\label{SM:sec:second_quant}}
The CME of~\cref{main-Master_equation} of the main paper can, following the seminal work by Doi and Peliti~\cite{doi_second_1976,peliti_path_1985} and as the first step towards a path integral formulation, be cast in a quantum mechanical ``second quantized'' form~\cite{tauber_applications_2005} by introducing annihilation and creation operators for species $i$, $\hat{a}_i$ and $\hat{a}_i^\dagger$, respectively, which obey the following commutation relations
	\begin{eqnarray}
		\left[\hat{a}_i^\dagger,\hat{a}_j^\dagger \right] = \left[ \hat{a}_i,\hat{a}_j \right] =0 \quad & \left[ \hat{a}_i,\hat{a}_j^\dagger \right] = \delta_{ij}
	\end{eqnarray}
	We introduce a ket $| \bm{n} \rangle = |n_1,n_2,\dots, n_i,\dots,n_P\rangle$ that defines the state of the system. The creation and annihilation operators act on the state ket $| \bm{n} \rangle$ as (notice that the normalization differs from the standard choice used in quantum mechanics)
	\begin{eqnarray}
		\hat{a}_i |\bm{n} \rangle &= \ n_i & \! \! \!\! |n_1,n_2,\dots, n_i-1,\dots,n_P\rangle \\
		\hat{a}_i^\dagger |\bm{n}\rangle &\!= \ \ \ {}
		& \!\!\!\!  |n_1,n_2,\dots, n_i+1,\dots,n_P\rangle
	\end{eqnarray}
	The state $ |\bm{n}\rangle$ can therefore be obtained by acting on the zero or ``vacuum'' state $| \bm{0} \rangle$ with the appropriate product of creation operators $\hat{a}_i^\dagger$,
	\begin{equation}
		|\bm{n}\rangle = \prod_i (\hat{a}_i^\dagger)^{n_i}|\bm{0}\rangle
		\label{n_from_0}
	\end{equation}
	The operator $\hat{n}_i$ that counts the number of molecules of species $i$ is given by
	\begin{equation}
		\hat{n}_i = \hat{a}_i^\dagger \hat{a}_i
	\end{equation}
	We define the associated bra states of the kets defined earlier, namely  $\langle \bm{n}|  = \langle n_1, \dots, n_P |$ which is generated by the action of $a_i$ on the state $\langle \bm{0}|$:
	\begin{equation}
		\langle \bm{n} | =	\langle \bm{0} | \prod_i \frac{a_i^{n_i}}{n_i!}
		\label{bra}
	\end{equation}
	The difference from the standard quantum mechanics normalization is explicit here and comes from demanding that the bra states should be (bi-)orthogonal to the ket states, $\langle \bm{n} | \bm{n}'\rangle = \delta_{\bm{n},\bm{n}'}$.

	We also define a uniform state or  ``sum state'',
	\begin{equation}
		\langle \bm{1} | = \sum_{\bm{n}} \langle \bm{n}| = \langle \bm{0} | e^{\sum_i \hat{a}_i }
	\end{equation}
	where the second expression follows from \cref{bra}.

	To obtain the quantum mechanical form of the CME, one identifies the probability distribution $P(\bm{n},\tau)$ across states with the vector $|P(\tau)\rangle = \sum_{\bm{n}} P(\bm{n},\tau)|\bm{n}\rangle$. As the CME is a linear equation for the $P(\bm{n},\tau)$, it can then equivalently be written in the form of
	an imaginary time Schr\"odinger equation,
	\begin{equation}
		\partial_\tau | P(\tau) \rangle = \hat{H} | P(\tau) \rangle
	\end{equation}
	with an effective Hamilton operator $\hat{H}$. Comparing with the original CME, one can read off $\hat{H}$ (see  \cite{tauber_applications_2005,falkovich_john_2008}) as
	\begin{equation}{\label{Hamiltonian_operator}}
		\hat{H} = \sum_\beta k_\beta (\tau) \left[ \prod_i (\hat{a}_i^\dagger)^{s_i^\beta} (\hat{a}_i)^{r_i^\beta} - \prod_i (\hat{a}_i^\dagger)^{r_i^\beta} (\hat{a}_i)^{r_i^\beta} \right]
	\end{equation}
The two terms in \cref{Hamiltonian_operator}
	correspond to those in the CME \cref{main-Master_equation} of the main paper, with the positive one representing ``incoming'' probability fluxes due to the reactions and the negative the corresponding ``outgoing'' fluxes that ensure probability conservation.

	The formal solution for the time evolution of the system to time $t$ is then generally given by
	\begin{equation}
		| P(t) \rangle = e^{\hat{H}t} | P(0) \rangle
	\end{equation}
Because of $\langle \bm{n}' | \bm{n}\rangle = \delta_{\bm{n}',\bm{n}}$ one has $\langle \bm{1}| \bm{n}\rangle = 1$, and so
	\begin{equation}{\label{prob_cons_proj}}
		\langle \bm{1} | P(\tau) \rangle = \sum_{\bm{n}} P(\bm{n},\tau) = 1 \qquad \qquad \forall \text{ } 0<\tau<t,
	\end{equation}
	which expresses conservation of probability. More generally, the mean value of any observable $A$ at time $\tau$, denoted by $\bar{A}(\tau)$, is given by
	\begin{equation}
		\begin{split}
			\bar{A}(\tau) &= \sum_{\bm{n}} P(\bm{n},\tau) A(\bm{n}) \\
			&=  \sum_{\bm{n}} P(\bm{n},\tau)  \langle \bm{1} | \hat{A}
			|\bm{n} \rangle \\
			&= \langle  \bm{1} |\hat{A}| P(\tau) \rangle
		\end{split}
	\end{equation}
	where $\hat{A}$ is obtained by replacing $n_i \rightarrow \hat{a}_i^\dagger \hat{a}_i$ in $A(\bm{n})$.
	Using the explicit expression for $|P(\tau)\rangle$ we can also write
	\begin{equation}
		\bar{A}(\tau) = \langle \bm{1} | \hat{A} e^{\hat{H}\tau} | P(0) \rangle
	\end{equation}
	where the initial state $|P(0)\rangle$ is the Poisson mixture of species such that
the copy number of each molecular species $i$ has independent Poisson fluctuations around a mean $\nzero{i}$.
Such states can be written in the simple form
\begin{equation}{\label{Poisson_initial_state}}
	| P(0) \rangle = \sum_{\bm{n}} \prod_i \left(
	e^{-\nzero{i}}\frac{\nzero{i}^{n_i}}{n_i!}
	\right)	|\bm{n}\rangle = e^{\sum_i \nzero{i}(\hat{a}_i^\dagger - 1)} | \bm{0} \rangle
\end{equation}
	where the second version follows from \cref{n_from_0}.

	A coherent state path integral can now be constructed for this stochastic process which enables the calculation of the ensemble averages over the dynamic trajectories of the system.

	\section{Coherent state path integral}{\label{SM:sec:path_integral}}
	To construct the path integral, one can split the time evolution by $e^{\hat{H}t}$ into small discrete time intervals $\Delta t$, and insert at each time step resolutions of the identity operator, expressed as integrals over an (over-complete) set of coherent states~\cite{doi_second_1976,peliti_path_1985,tauber_applications_2005}. Our description of the construction of this path integral closely follows \cite{harsh_accurate_2023,tauber_applications_2005}.

One begins by introducing coherent states $| \psi \rangle, \langle \psi|$ as eigenstates of the creation and annihilation operators $\hat{a}$ and $\hat{a}^\dagger$ with eigenvalues $\psi$ and $\psi^*$, respectively:
	\begin{eqnarray}
		\hat{a} | \psi \rangle = \psi | \psi \rangle  \quad & \langle \psi | \hat{a}^\dagger = \langle \psi | \psi^*
	\end{eqnarray}
	(For simplicity we treat a single species for now and so omit the species index $i$.) Explicitly, these eigenstates can be written as
	\begin{equation}
		| \psi \rangle = \exp \left( -\frac{1}{2} |\psi|^2 + \psi \hat{a}^\dagger \right) |0 \rangle, \qquad
		\langle \psi | = \langle 0| \exp \left( -\frac{1}{2} |\psi|^2 + \psi^* \hat{a} \right)
	\end{equation}
	with overlap given by
	\begin{equation}
		\langle \psi_1| \psi_2 \rangle = \exp \left( -\frac{1}{2} |\psi_1|^2 -\frac{1}{2} |\psi_2|^2 + \psi_1^* \psi_2 \right)
	\end{equation}
	From these states one can construct an over-complete resolution of the identity,
	\begin{equation}
		\mathbb{I} = \int \frac{d\phi^* d\phi}{\pi} | \phi \rangle \langle \phi | \quad \text{with} \quad d\phi^* d\phi =  d(\text{Re} \phi) d(\text{Im} \phi)
	\end{equation}
	We now insert such a resolution of the identity after every time step $\Delta t$, using one field for each species, i.e.\ $\bm{\phi}(\tau) \equiv \bm{\phi}_\tau = (\phi_1(\tau), \phi_2(\tau), \dots, \phi_P(\tau))$.
	Doing this in the normalization \cref{prob_cons_proj} at time $t$ gives
	\begin{equation}
		1 = \mathcal{N}^{-1} \lim_{\Delta t \rightarrow 0} \int \prod_{\tau=0}^t d \bm{\phi}^*_\tau  d \bm{\phi}_\tau \langle \bm{1} |
		\bm{\phi}_t \rangle \langle \bm{\phi}_t | e^{\hat{H} \Delta t} | \bm{\phi}_{t - \Delta t}\rangle \dots \langle \bm{\phi}_\tau | e^{\hat{H} \Delta t} | \bm{\phi}_{\tau - \Delta t}\rangle  \dots \langle \bm{\phi}_0 | P(0) \rangle
		\label{normalization_one}
	\end{equation}
	with a normalization factor $\mathcal{N}=\pi^{(t/\Delta t)+1}$ that will not play a role in the following. The matrix element factors can be evaluated to $O(\Delta t)$ as
	\begin{equation}{\label{matrix_element}}
		\langle \bm{\phi}_\tau | e^{\hat{H} \Delta t} | \bm{\phi}_{\tau-\Delta t} \rangle =
		\langle \bm{\phi}_\tau | \bm{\phi}_{\tau-\Delta t} 	\rangle
		[1+H(\bm{\phi}^*_\tau, \bm{\phi}_{\tau-\Delta t})\Delta t]
			=\langle \bm{\phi}_\tau | \bm{\phi}_{\tau-\Delta t} \rangle  e^{H(\bm{\phi}^*_\tau, \bm{\phi}_{\tau-\Delta t})\Delta t}
		\end{equation}
		because the Hamiltonian in \cref{Hamiltonian_operator} is \textit{normal ordered}, i.e.\ with creation operators $\hat{a}_i^\dagger$ always to the left of annihilation operators $\hat{a}_i$; these operators then simply act on their eigenvectors to the left and right, respectively, giving the relevant factors of $\phi^*_i(\tau)$ or $\phi_i(\tau)$:
		\begin{equation}
			H(\bm{\phi}^*_\tau, \bm{\phi}_{\tau-\Delta t}) = \langle \bm{\phi}_\tau | \hat{H}| \bm{\phi}_{\tau - \Delta t} \rangle
		\end{equation}
		Note that the conjugate field $\phi^*$ is always a time step $\Delta t$ ahead of $\phi$. The remaining overlap factor in \cref{matrix_element} is
		\begin{equation}{\label{overlap_app1}}
			\langle \bm{\phi}_\tau | \bm{\phi}_{\tau_-}\rangle = \exp\left[ -(\bm{\phi}^*_\tau)^{\rm T}( \bm{\phi}_\tau - \bm{\phi}_{\tau_-}) \right] \exp\left(\frac{1}{2} | \bm{\phi}_\tau|^2 - \frac{1}{2} |\bm{\phi}_{\tau_-} |^2 \right)
		\end{equation}
		After multiplication across $\tau=0,\Delta t, \ldots t$ the second exponential cancels for all times other than the initial and the final ones. The first exponential can be expressed in terms of the discrete derivative $\Delta_\tau \phi_i(\tau) = \frac{1}{\Delta t} (\phi_i(\tau) - \phi(\tau_-))$:
		\begin{equation}
			\exp\left[ -(\bm{\phi}^*_\tau)^{\rm T} ( \bm{\phi}_\tau - \bm{\phi}_{\tau_-}) \right]
			= \exp\left[ -(\bm{\phi}^*_\tau)^{\rm T} \Delta_\tau \bm{\phi}(\tau) \Delta t \right]
		\end{equation}
		The last factor in \cref{normalization_one} including the final surviving squared exponential can be evaluated by noticing that the sum state $\langle \bm{1}|$ is, up to normalization, a coherent state with eigenvalue 1, such that
		\begin{equation}{\label{overlap_app2}}
			\langle \bm{1} | \bm{\phi}_t \rangle \exp\left(\frac{1}{2}|\bm{\phi}_t|^2\right)  =  \exp\left(\sum_i \phi_i(t)\right)
		\end{equation}
		For the initial term one has similarly, using that $|P(0)\rangle$ is proportional to an eigenvector of $a_i$ with eigenvalue $\nzero{i}$, \begin{equation}{\label{initial_overlap_path_integral}}
	\langle \bm{\phi}_0 | P(0) \rangle \exp\left(-\frac{1}{2}|\bm{\phi}_0|^2\right) = \exp\left(\sum_i 		\left[\nzero{i} (\phi_i^*(0) - 1) - |\bm{\phi}_i(0)|^2\right]\right)
	\end{equation}

		We now generalize the above construction by inserting, instead of an identity $\mathbb{I}$, a factor
		\begin{equation}
			e^{ \sum_i \theta_i(\tau)\Delta t\,\hat{a}_i} \, \mathbb{I} \, 	e^{ \sum_i \tilde\theta_i(\tau)\Delta t\,(\hat{a}^\dagger_i -1 )}
		\end{equation}
		at each discretized time $\tau$. This allows one to generate averages such as means, correlation functions etc.\ by taking derivatives w.r.t.\ the generating fields ${\theta}$ and ${\tilde{\theta}}$ defined for each species at each time step. Defining here the $\tilde{\theta}$ factor with $(\hat{a}^\dagger - 1)$ rather than $\hat{a}^\dagger$ will simplify a number of later formulas. This leads to the following \textit{path integral} representation of the \textit{generating function} or the \textit{partition function}:
		\begin{equation}{\label{path_integral1}}
			\mathcal{Z}( \tilde{\theta}  , \theta ) = \lim_{\Delta t \rightarrow 0} \mathcal {N}^{-1} \int \prod_{\tau} d \bm{\phi}_\tau^* d\bm{\phi}_\tau  \langle \bm{1} |\bm{\phi}_t \rangle \prod_{\tau=\Delta t}^{t} 	\langle \bm{\phi}_\tau | e^{\sum_i (\hat{a}_i^\dagger - 1 )\tilde{\theta}_i(\tau) \Delta t} e^{\hat{H} \Delta t} e^{\sum_i \hat{a}_i \theta_i(\tau-\Delta t) \Delta t} | \bm{\phi}_{\tau - \Delta t} \rangle \langle \bm{\phi}_0 | P(0) \rangle
		\end{equation}
		The matrix element for the propagation is then given by
		\begin{equation}
			\begin{split}
				\langle \bm{\phi}_\tau | e^{\sum_i (\hat{a}_i^\dagger-1) \tilde{\theta}_i(\tau) \Delta t} e^{\hat{H} \Delta t} e^{\sum_i \hat{a}_i \theta_i(\tau-\Delta t) \Delta t} | \bm{\phi}_{\tau - \Delta t} \rangle = \\
				e^{H(\bm{\phi}^*_\tau, \bm{\phi}_{\tau-\Delta t})\Delta t} \langle \bm{\phi}_\tau | \bm{\phi}_{\tau-\Delta t} \rangle  \prod_i \left\lbrace e^{\tilde{\theta}_i(\tau) (\phi_i^*(\tau)-1)  \Delta t + \theta_i(\tau-\Delta t) \phi_i(\tau-\Delta t) \Delta t} \right\rbrace
			\end{split}
		\end{equation}
		Putting all the elements together with the initial and final overlap, that is \cref{overlap_app1,overlap_app2,initial_overlap_path_integral}, the partition function can be written in terms of an action, $S$, as
\begin{equation}{\label{partition_function}}
			\mathcal{Z}( \tilde{\theta}  , \theta ) = \lim_{\Delta t \rightarrow 0} \mathcal{N}^{-1} \int \prod_\tau d \bm{\phi}_\tau^* d\bm{\phi}_\tau e^{S\left(  \phi^* ,  \phi  \right)}
		\end{equation}
where the normalization constant $\mathcal{N}$ ensures that $\mathcal{Z}(0,0)=1$.	The integrations are over the (time-discretized) paths of the complex fields $\phi_i(\tau)$ defining the coherent states. The action $S$ depends on these fields and their complex conjugates $\phi_i^*(\tau)$ and reads
\begin{equation}{\label{discrete_action}}
	\begin{split}
		S\left( \phi^* ,  \phi  \right) &= \Delta t \sum_{\tau=\Delta t}^{t} H(\bm{\phi}^*(\tau), \bm{\phi}(\tau_-)) + \sum_i \Big\lbrace \phi_i(t) + \nzero{i} (\phi_i^*(0)-1) - \phi_i(0)\phi_i^*(0) \\
		&- \Delta t \sum_{\tau=\Delta t}^{t} \phi_i^*(\tau) \Delta_\tau \phi_i(\tau)
		+ \Delta t \sum_{\tau=0}^{t} \left[ \tilde{\theta}_i(\tau) \left( \phi_i^*(\tau)-1 \right)  + \theta_i(\tau) \phi_i(\tau)  \right] \Big\rbrace
	\end{split}
\end{equation}
		where $\tau_-\equiv \tau-\Delta t$, $t$ is the total time of the dynamics considered, and as before $\Delta_\tau \phi_i(\tau) = \frac{1}{\Delta t} (\phi_i(\tau) - \phi(\tau_-))$ is a shorthand for the discrete time derivative. $ H(\bm{\phi}^*(\tau), \bm{\phi}(\tau_-))$ is obtained from the Hamiltonian $\hat{H}$ in \cref{Hamiltonian_operator} by replacing $\hat{a}_i^\dagger$ by $\phi^*_i(\tau)$ and $\hat{a}$ by $\phi_i(\tau_-)$.
For definiteness in the discretized time representation we also take the reaction rate $k_\beta$ to be evaluated at $\tau_-$.

		One can now apply a \textit{Doi-shift}~\cite{tauber_applications_2005,falkovich_john_2008}  by replacing $\phi_i^*(\tau) = 1 + \tilde{\phi}_i(\tau)$. This turns out to make the average of $\bm{\tilde{\phi}} = 0$ and can be justified by appropriate rearrangements in the partition function before the coherent states are introduced~\cite{tauber_applications_2005}. If we continue to use the same label for the function $H$ evaluated at $\bm{\phi}^* = 1 + \bm{\tilde{\phi}}$, then in terms of the new variables the action reads
\begin{equation}{\label{Action_discrete}}
	\begin{split}
		\Sfields &= \Delta t \sum_{\tau=\Delta t}^{t} H(\bm{\tilde{\phi}}(\tau), \bm{\phi}(\tau_-)) + \sum_i \Big\lbrace  \nzero{i} \tilde{\phi}_i(0) - \phi_i(0) \tilde{\phi}_i(0)  \\
		&- \Delta t \sum_{\tau=\Delta t}^{t} \tilde{\phi}_i(\tau) \Delta_\tau \phi_i(\tau)
		+ \Delta t \sum_{\tau=0}^{t} \left[ \tilde{\theta}_i(\tau) \tilde{\phi}_i(\tau)  + \theta_i(\tau) \phi_i(\tau) \right] \Big\rbrace
	\end{split}
\end{equation}
		We will generally work in discrete time and only take the continuous time limit in the final equations of motion, but one can also take this limit in the expression for the action to obtain
\begin{equation}{\label{action_continuous}}
	\begin{split}
		\Sfields =  \int_0^{t} d\tau\, H(\bm{\tilde{\phi}}(\tau), \bm{\phi}(\tau_-))  + \sum_i \Big( & \nzero{i} \tilde{\phi}_i(0) - \phi_i(0) \tilde{\phi}_i(0) \\
		&+ \int_0^{t} d\tau \left[ -\tilde{\phi}_i(\tau) \partial_\tau \phi_i(\tau) + \tilde{\theta}_i(\tau) \tilde{\phi}_i(\tau)  + \theta_i(\tau) \phi_i(\tau)  \right] \Big)
	\end{split}
\end{equation}

	With this powerful Doi-Peliti path integral in hand, we next demonstrate an important causality property.

	\section{Causality property of the path integral}{\label{SM:sec:causality_prop}}

	To see the causality property, consider a reaction where a molecule of species $i$ is created spontaneously with rate $k_{1i}$. From \cref{Hamiltonian_operator}, this corresponds to a term $k_{1i}(a_i^\dagger-1)$ in $\hat{H}$, and hence, for the more general case of a time-dependent creation rate, to a contribution $\Delta t \sum_\tau k_{1i}(\tau)\tilde\phi_i(\tau)$ in the action. Comparing this with the generating term
	$\Delta t \sum_\tau \tilde\theta_i(\tau) \tilde\phi_i(\tau)$
	from the $\tilde\theta_i(\tau)$-field shows that, whenever we take derivatives of the partition function, we have the identity
	\begin{equation}
		\frac{1}{\Delta t}\frac{\partial}{\partial k_{1i}(\tau)} = \frac{1}{\Delta t}\frac{\partial}{\partial \tilde\theta_{i}(\tau)}
	\end{equation}
	Now any product $f_{<\tau}$ of fields evaluated at times before $\tau$ can be generated by a sequence of appropriate derivatives of $\mathcal{Z}$. Taking then an additional derivative as in our last identity and setting the generating fields to zero afterwards shows that
	\begin{equation}
		\frac{1}{\Delta t}\frac{\partial}{\partial k_{1i}(\tau)} \langle f_{<\tau}\rangle = \langle \tilde\phi_i(\tau)f_{<\tau}\rangle
		\label{response_field_relation}
	\end{equation}
	However, the l.h.s.\ vanishes by causality -- the average of $f_{<\tau}$ cannot depend on a creation rate at a later time -- and therefore so does the r.h.s., as claimed.

	Summarizing so far, then, and introducing the symbol $\mu$ for the means of our fields, we have
	\begin{equation}
		\begin{split}
			\mu_i(\tau) &= \langle \phi_i(\tau)\rangle = \langle n_i(\tau) \rangle \\
			\tilde{\mu}_i(\tau) &= \langle \tilde{\phi}_i(\tau)\rangle = 0
		\end{split}
	\end{equation}
	The second line follows again from the causality property.

	\section{Marginal copy number distribution}{\label{SM:sec:marginal_copy_dist}}
	Following the same line of arguments as in~\cref{main-sec:observables} of the main paper we will consider the marginal copy number distribution as described by our path integral, defined in \cref{path_integral1}. We will consider the two cases of correlations functions being zero or not.

	In the first case we get a marginal Poisson copy number distributions for each species if the path integral is Gaussian, i.e.\ the action only has has linear and quadratic terms, and if the field correlation functions $C_i$ vanish. We start by taking the $k$ derivatives w.r.t.\ $\tilde{\theta}$ and $\theta$ such that the operators are normal ordered,
\begin{equation}
		\begin{split}{\label{copy_number_dist_1}}
			\frac{1}{(\Delta t)^{2k}} \left( \frac{\partial }{\partial \tilde\theta_i(\tau_+)} + 1 \right)^k \frac{\partial^k }{\partial \theta_i^k(\tau)} \mathcal{Z} \Bigg|_{\theta,\tilde{\theta}=0} &= \langle \bm{1}| (\hat{a}_i^\dagger)^k (\hat{a}_i)^k| \bm{\phi}_{\tau} \rangle \\
			&= \langle n_i(\tau) (n_i(\tau)-1) \dots (n_i(\tau)-k+1) \rangle \\
			&=\langle \lbrace n_i(\tau) \rbrace_k\rangle
		\end{split}
	\end{equation}
	where we acted on the state by the annihilation operator to lower it, then used the creation operator and finally exploited probability conservation as defined in \cref{prob_cons_proj}. $\langle \lbrace n_i \rbrace_k \rangle \equiv \langle n_i!/(n_i-k)! \rangle $ is the $k^{\text{th}}$ \textit{factorial moment}.

	Calculating the same derivatives from the path integral we have
	\begin{equation}{\label{eq_appendix_n_correlator_expansion_field}}
		\begin{split}
			\frac{1}{(\Delta t)^{2k}} \left( \frac{\partial }{\partial \tilde\theta_i(\tau_+)} + 1 \right)^k  \frac{\partial^k }{\partial \theta_i^k(\tau)} \mathcal{Z} \Bigg|_{\theta,\tilde{\theta}=0} &= \langle (\tilde\phi_i(\tau_+)+1)^k \phi_i^k(\tau) \rangle \\
			&= \langle \phi_i^k(\tau)\rangle \\
			& = \langle [\delta \phi_i(\tau) + \langle n_i(\tau)\rangle ]^k\rangle
		\end{split}
	\end{equation}
	where we have used the causality property discussed in \cref{SM:sec:causality_prop} in going from the first to the second line.

	The final average can now be expanded.
If the path integral is Gaussian, the averages can be evaluated using Wick's theorem.
All terms involving $\delta\phi_i$ then reduce to powers of $C_i(\tau,\tau) = \langle \delta \phi_i(\tau) \delta \phi_i(\tau)\rangle$, so if the latter vanishes
	one has simply
	\begin{equation}
		\begin{split}
			\frac{1}{(\Delta t)^{2k}} \left( \frac{\partial }{\partial \tilde\theta_i(\tau_+)} + 1 \right)^k  \frac{\partial^k }{\partial \theta_i^k(\tau)} \mathcal{Z} \Bigg|_{\theta,\tilde{\theta}=0} = \langle n_i(\tau)\rangle^k
		\end{split}
	\end{equation}
	Comparing with \cref{copy_number_dist_1} we obtain that the $k^{\text{th}}$ factorial moment of the copy number distribution is equal to the mean to the power $k$, that is $\langle \lbrace n_i \rbrace_k\rangle = \langle n_i\rangle^k $ for all times, which uniquely characterizes a Poisson distribution~\cite{potts_note_1953} as claimed. Note that even though the marginal distributions are Poisson, there can still be correlations between species.

In the second case, if $C_i(\tau,\tau)$ is not generally zero, then using the Wick's expansion we have from~\cref{copy_number_dist_1,eq_appendix_n_correlator_expansion_field}
	\begin{equation}
		\begin{split}
			\langle \lbrace n_i(\tau) \rbrace_k\rangle &= \langle [\delta \phi_i(\tau) + \langle n_i(\tau)\rangle ]^k\rangle \\
			&= \sum_{j=0}^{k} \binom{k}{j}  \langle (\delta \phi_i(\tau))^j \rangle \langle n_i(\tau)\rangle^{k-j} \\
&= \sum_{l=0}^{[k/2]} \frac{k!}{2^l l! (k-2l)!} C_i^l(\tau,\tau) \langle n_i(\tau)\rangle^{k-2l}
		\end{split}
	\end{equation}
	where the last line simplifies because the odd moments of Gaussian fields are zero. These factorial moments can be summarized into the generating function $G(z)=\sum_n P(n_i(\tau)=n)z^n$ for the probability distribution of $n_i(\tau)$ as
\begin{equation}
		G(z) = \exp\left[ \langle n_i\rangle(z-1) + \frac{C_i(\tau,\tau)}{2}(z-1)^2 \right]
	\end{equation}
Although this uniquely defines the probability distribution in terms of two time-dependent parameters, namely $\langle n_i(\tau)\rangle$ and $C_i(\tau,\tau)$, it is of limited practical utility as it does not immediately imply a closed form for this distribution.

	\section{Update equations with effective fields}{\label{matrix_action_plefka}}

	In this section we cast the baseline action, with the effective fields $\theta$, $\tilde{\theta}$, $\hat{R}$ and $\hat{B}$ included, in a matrix form for a multivariate Gaussian distribution and obtain the resulting response and correlation functions by inverting the precision matrix of the Gaussian. We start by defining the following vectors,
\begin{subequations}
	\begin{align}
			\boldsymbol{\phi} &= (\delta \phi_1(0),\delta \phi_1(\Delta t),\dots,\delta \phi_1(t),\delta \phi_2(0), \delta \phi_2(\Delta t),\dots,\delta \phi_2(t),\dots,\delta \phi_P(0), \delta \phi_P(\Delta t),\dots,\delta \phi_P(t)) \\
			\boldsymbol{\tilde{\phi}} &=(\delta \tilde \phi_1(0),\delta \tilde \phi_1(\Delta t),\dots,\delta \tilde \phi_1(t),\delta \tilde \phi_2(0), \delta \tilde \phi_2(\Delta t),\dots, \delta \tilde \phi_2(t),\dots,\delta \tilde \phi_P(0),\delta \tilde \phi_P(\Delta t),\dots,\delta \tilde \phi_P(t))  \\
			\boldsymbol{b} &= (\theta_1(0),\theta_1(\Delta t),\dots,\theta_1(t),\theta_2(0),\theta_2(\Delta t),\dots,\theta_2(t),\dots,\theta_P(0),\theta_P(\Delta t),\dots,\theta_P(t)) \Delta t \\
			\boldsymbol{\tilde{b}} &=(\tilde \theta_1(0)+k_{11},\tilde \theta_1(\Delta t) +k_{11} ,\dots,\tilde \theta_1(t)+k_{11},\tilde \theta_2(0)+k_{12},\tilde \theta_2(\Delta t) +k_{12},\dots,\tilde \theta_2(t)+k_{12},\dots, \nonumber\\
			&\dots,\tilde \theta_P(0)+k_{1P},\tilde \theta_P(\Delta t)+k_{1P},\dots,\tilde \theta_P(t)+k_{1P}) \Delta t
	\end{align}
	\end{subequations}
Then one has the quadratic action for the Gaussian path integral
	\begin{equation}{\label{SM:Gaussian_action}}
		S_0\left( \delta \bm{\tilde{\phi}} , \delta \bm \phi  \right)  =  \delta  \boldsymbol{\tilde{\phi}}^{\rm T} \boldsymbol{\tilde{b}} + \delta \boldsymbol{\phi}^{\rm T} \boldsymbol{b} - \delta \boldsymbol{\phi}^{\rm T} \bm{J}^{\rm T} \delta \boldsymbol{\tilde{\phi}}
		+ \frac{1}{2}(\Delta t)^2 \delta \boldsymbol{\tilde{\phi}} ^T  \bm{\hat{B}} \delta \boldsymbol{\tilde{\phi}} + \mbox{const.}
\end{equation}
	where $\bm{J}$ is a lower triangular matrix and the constant term is independent of $\phi$ and $\tilde\phi$. We have already used the fact that $\hat C = 0$ in the physical solution. The relationship between $\bm{J}$ and $\hat{\bm{R}}$ is easiest to see by dividing $\bm{J}$ into blocks with species indices so that block $J_{ij}$ is defined as $J_{ij} = $
	\begin{equation}
\begin{pmatrix}
			\delta_{ij} - \tilde{R}_{ij}(0,0) & 0 &  \dots & 0\\
			\delta_{ij}(k_{2i}\Delta t - 1) - \tilde{R}_{ij}(\Delta t ,0) & \delta_{ij} - \tilde{R}_{ij}(\Delta t,\Delta t) &  \dots & 0 \\
			\vdots & \vdots &  \dots & 0 \\
			-\tilde{R}_{ij}(t ,0) & \dots & \delta_{ij}(k_{2i}\Delta t -1) - \tilde{R}_{ij}(t,t-\Delta t) & \delta_{ij} - \tilde{R}_{ij}(t,t)
		\end{pmatrix}
\end{equation}
where $\tilde{R}_{ij}(\tau,\tau') = (\Delta t)^2  \hat{R}_{ij}(\tau,\tau') $.

	To find the response and correlation functions one writes the quadratic part of the action as
\begin{equation}
		-\frac{1}{2}
		\begin{pmatrix}
			\delta \bm{\phi} \\
			\delta \bm{ \tilde{\phi}}
		\end{pmatrix}^{\rm T}
		\begin{pmatrix}
			0 & \bm{J}^{\rm T} \\
			\bm{J} & - (\Delta t)^2 \bm{\hat{B}}
		\end{pmatrix}
		\begin{pmatrix}
			\delta \bm{\phi} \\
			\delta \bm{\tilde{\phi}}
		\end{pmatrix}
	\end{equation}
	Because the fields have a Gaussian distribution, the matrix that appears here is the inverse covariance or precision matrix. Denoting the covariance matrix in a $2\times 2$ block form by
	\begin{equation}
\begin{pmatrix}
			\langle \delta\bm{\phi} \delta\bm{\phi}^{\rm T}\rangle & \langle \delta\bm{\phi} \delta\bm{\tilde{\phi}}^{\rm T}\rangle \\
			\langle \delta\bm{\tilde{\phi}} \delta\bm{\phi}^{\rm T}\rangle & \langle \delta\bm{\tilde{\phi}} \delta\bm{\tilde{\phi}}^{\rm T}\rangle
		\end{pmatrix}
		=
		\begin{pmatrix}
			\bm C & \bm R \\
			\bm R^{\rm T}& \bm B
		\end{pmatrix}
	\end{equation}
we must then have
	\begin{equation}{\label{eq:SM_matrix_RCB}}
		\begin{pmatrix}
			\bm C & \bm R \\
			\bm R^{\rm T}& \bm B
		\end{pmatrix}
		\begin{pmatrix}
			0 & \bm{J}^{\rm T} \\
			\bm{J} & -(\Delta t)^2 \bm{\hat{B}}
		\end{pmatrix} =
		\begin{pmatrix}
			\boldsymbol{\mathbb{I}} & 0 \\
			0 &  \boldsymbol{\mathbb{I}}
		\end{pmatrix}
	\end{equation}
Multiplying this out one has for the top left block
	\begin{equation}{\label{eq_RJ_1}}
		\bm R \bm J =  \boldsymbol{\mathbb{I}}
	\end{equation}
which written out explicitly reads
	\begin{equation}
		R_{ij}(\tau,\tau') + \sum_{m} \delta_{im}(k_{2i}\Delta t - 1)R_{mj}(\tau_-,\tau') - (\Delta t)^2 \sum_{\tau''=\tau'}^{\tau} \sum_k \hat R_{ik}(\tau,\tau'')R_{kj}(\tau'',\tau') = \delta_{ij}\delta_{\tau,\tau'}
	\end{equation}
This is the discrete time update equation for the response function. Taking the limit $\Delta t \rightarrow 0$ and substituting $\hat R$ with the effective field $\hat{R}^{\text{eff}}$
then leads to~\cref{main-FD_eqn_mixed_plefka} of the main paper.

From the matrix multiplication from~\cref{eq:SM_matrix_RCB}
we also have
	\begin{equation}{\label{eq_CJ_1}}
		\bm C \bm J^{\rm T} -  (\Delta t)^2 \bm R \bm{\hat{B}} = 0
	\end{equation}
Expanding this, substituting $\hat B$ with $\hat{B}^{\text{eff}}$ and taking the limit $\Delta t \rightarrow 0$ leads to~\cref{main-C_update_multispecies_explicit_plefka} of the main text. We actually use the following more convenient form to update $\bm C$ once $\bm R$ has been calculated:
	\begin{equation}
		\bm C =  (\Delta t)^2 \bm R \bm{\hat{B}} \bm R^{\rm T}
	\end{equation}
	Finally we have from~\cref{eq:SM_matrix_RCB}
\begin{equation}
		\bm B \bm J = 0
	\end{equation}
	which implies as expected that
	\begin{equation}{\label{eq:B0}}
		\bm B = 0
	\end{equation}

We return finally to the response function and its causality property. From~\cref{eq_RJ_1}
the $ij$-block of the response function matrix is
	$\langle \delta\bm{\phi}_i \delta\bm{\tilde{\phi}}_j^{\rm T}\rangle = (\bm{J}^{-1})_{ij}$,
	with entries $R_{ij}(\tau,\tau') = \langle \delta \phi_i(\tau) \delta \tilde{\phi}_j(\tau') \rangle$.
	Now $\bm{J}_{ij}$ is lower triangular with unit diagonal entries if $i = j$, and with zeros on the diagonal if $i \neq j$ since $\hat{R}_{ij}(\tau,\tau) = 0\ \forall\ i,j,\tau$. It then follows directly that the response function matrix has the same structure:
	the equal-time response (diagonal elements in each block) is
	\begin{equation}
		R_{ij}(\tau,\tau) =
		\begin{cases}
			1 &\text{ if } i = j \\
			0 &\text{ if } i \neq j
		\end{cases}
\end{equation}
As expected the response is also causal in that the elements above the diagonal vanish,
	\begin{equation}
		R_{ij} (\tau,\tau') = 0 \text{ if } \tau < \tau'\ \forall \ i,j\ .
	\end{equation}

	\section{Calculation of Plefka free energy expansion and effective fields}

\subsection{Identities for coefficients $c_{m,n}$}
{\label{SM:sec:derivative identities}}

For ease of reference we collect here some useful identities that are used frequently in later calculations, which are also stated in the main text as~\cref{main-eq:c_mn_identity1,main-eq:c_mn_identity2}.

We start by noticing that for the physical solution where $\tilde\mu_i(\tau)=0$, one has
\begin{equation}{\label{c_on_property}}
	c_{\bm 0, \bm n} = 0 \quad \forall \; \bm n
\end{equation}
The identities of~\cref{main-eq:identity_derivative_main_paper} of the main paper are a special case of the following general relations, which follow by direct calculation from~\cref{main-c_mn_defn}:

		\begin{equation}
	\frac{\partial c_{\bar{m},\bar{n}}(\tau_+)}{\partial \mu_i (\tau)} = c_{\bar{m},\bar{n}+\bar{e}_i}(\tau_+)
\end{equation}
and
\begin{equation}
	\frac{\partial c_{\bar{m},\bar{n}}(\tau)}{\partial \tilde \mu_i (\tau)} = c_{\bar{m}+\bar{e}_i,\bar{n}}(\tau_+)
	\label{eq:cderivative_mutilde}
\end{equation}

\subsection{First order in $\alpha$}{\label{SM:sec:first_order_field}}

We give here some details of the calculation of the Plefka free energy $\Gamma$ as defined in~\cref{main-plefka_free_energy1} of the main paper, i.e.\ for the case where the constrained order parameters are the field averages $\mu$ and $\tilde{\mu}$. The Plefka free energy is then given by
\begin{equation}
				\Gamma ( \tilde{\mu} ,  \mu )  = \text{\normalsize extr}_{\tilde{\theta}  ,  \theta } \left\lbrace \ln \int d\tilde{\phi}\, d\phi \,\exp \left[ S_\alpha(  \tilde{\phi} , \phi ) - \sum_i \int_0^t d\tau \left( \tilde{\mu}_i(\tau) \tilde{\theta}_i(\tau) + \mu_i(\tau) \theta_i(\tau)  \right) \right] \right\rbrace  =  \ln \int d\tilde{\phi}\,d\phi \; e^{\Xi_\alpha}
\end{equation}
so that its derivative $\Gamma^1$ at $\alpha = 0$ can be expressed as
		\begin{equation}
			\begin{split}
				\Gamma^1 = \frac{\partial \Gamma_\alpha}{\partial \alpha}\Bigg|_{\alpha=0}  = \left\langle \frac{d \Xi_\alpha}{d \alpha} \right\rangle_0
			\end{split}
		\end{equation}
		 The total derivative $d/d\alpha$ here indicates that one in principle needs to include the derivatives of the fields $\tilde{\theta}_{i,\alpha},\theta_{i,\alpha}$ with respect to $\alpha$. However, since these are multiplied by factors of $\delta \tilde\phi_i,\delta \phi_i$ respectively, they make a vanishing contribution after averaging.

As explained in the main text, this leads to the following expression for $\Gamma^1$,
	\begin{equation}{\label{linear_OP_theta_tilde_o1}}
	\begin{split}
		\Gamma^1 & = \left\langle S_{\text{int}} \right\rangle_{0} = \Delta t \sum_\tau c_{\bar{0},\bar{0}}(\tau) \end{split}
\end{equation}
and from this one obtains the effective fields to $O(\alpha)$ as
\begin{equation}{\label{linearOP_theta_o1-2}}
	-\tilde{\theta}_i^1(\tau)  = \frac{1}{\Delta t} \frac{\partial \Gamma^1}{\partial \tilde{\mu}_i(\tau)} = \frac{\partial c_{\bar{0},\bar{0}}}{\partial \tilde{\mu}_i(\tau)} =  c_{\bar{e}_i,\bar{0}}(\tau) =
	\parbox{15mm}{
		\begin{fmfgraph*}(50,40)
			\fmfleft{e1}
			\fmfright{r1,r2,r2n,r2nn,r3}
\fmfv{label=\leglabel{i},label.dist=0,label.angle=180}{e1}
			\fmf{end_arrow_dash_two,tension=2}{v1,e1}
			\fmfv{label=$\tau$,label.dist=7,label.angle=90,decor.shape=circle,decor.filled=full, decor.size=4thick}{v1 }
			\fmf{phantom,tension=1}{v1,r1}
			\fmf{phantom,tension=1}{v1,r2}
			\fmf{phantom,tension=1}{v1,r2n}
			\fmf{phantom,tension=1}{v1,r2nn}
			\fmf{phantom,tension=1}{v1,r3}
\end{fmfgraph*}}
\end{equation}
and
\begin{equation}{\label{linearOP_theta_tilde_o1}}
	-\theta_i^1(\tau)  = \frac{1}{\Delta t}\frac{\partial \Gamma^1}{\partial \mu_i(\tau)} = \frac{\partial c_{\bar{0},\bar{0}}}{\partial \mu_i(\tau)} = c_{\bar{0},\bar{e}_i}(\tau_+)
\end{equation}
which vanishes for the physical solution but will be needed later.

\subsection{Second order in $\alpha$}{\label{SM:sec:second_order_field}}

		The $O(\alpha^2)$ contribution to the Plefka free energy is given quite generally by the following calculating (as discussed in~\cite{roudi_dynamical_2011,bravi_extended_2016}): \begin{equation}{\label{gamma_2_exponent_calculation}}
			\begin{split}
				\Gamma^2 & = \frac{\partial^2 \Gamma_\alpha}{\partial \alpha^2}\Bigg|_{\alpha=0} = \frac{\partial}{\partial \alpha} \left( \frac{ \int d\tilde{\phi}\,d\phi\ \frac{\partial \Xi_\alpha}{\partial \alpha} e^{\Xi_\alpha} }{ \int d\tilde{\phi}\,d\phi\ e^{\Xi_\alpha}} \right)\Bigg|_{\alpha=0}  \\
				& = \frac{\int d\tilde{\phi}\,d\phi\ \frac{\partial^2 \Xi_\alpha}{\partial \alpha^2} e^{\Xi_\alpha}  + \int d\tilde{\phi}\,d\phi\ \left( \frac{\partial \Xi_\alpha}{\partial \alpha}\right)^2 e^{\Xi_\alpha}}{ \int d\tilde{\phi}\,d\phi\ e^{\Xi_\alpha}} -\frac{\left( d\tilde{\phi}\,d\phi\ \frac{\partial \Xi_\alpha}{\partial \alpha} e^{\Xi_\alpha}  \right)^2}{\left( d\tilde{\phi}\,d\phi\ e^{\Xi_\alpha}\right)^2} \\
				&= \left\langle \frac{d^2 \Xi_\alpha}{d \alpha^2}\right\rangle_0 +  \left\langle \left( \frac{d \Xi_\alpha}{d \alpha}\right)^2\right\rangle_0 -  \left\langle \frac{d \Xi_\alpha}{d \alpha}\right\rangle^2_0 \\
				&= \left\langle \left( \frac{d \Xi_\alpha}{d \alpha} - \left\langle \frac{d \Xi_\alpha}{d \alpha }\right\rangle_0 \right)^2\right\rangle + \left\langle \frac{d^2 \Xi_\alpha}{d \alpha^2}\right\rangle_0 \\
				&= \left\langle \left( \delta \frac{d \Xi_\alpha}{d \alpha}\right)^2 \right\rangle_0
			\end{split}
		\end{equation}
		where the last term in the penultimate line vanishes because $\partial^2 H_\alpha/\partial \alpha^2 = 0$ and because $\partial^2 h_{i,\alpha} /\partial \alpha^2$ is multiplied by a factor that is zero on average; here $h$ represents any of the fields $\tilde \theta$ or $\theta$.

Evaluating the above expression for $\Gamma^2$ explicitly and, for notational simplicity, denoting the average w.r.t.\ the baseline $\langle \dots \rangle_0$ simply by $\langle \dots \rangle$, leads to
		\begin{align}
			\Gamma^2  &= \left\langle \left[\Delta t \sum_\tau H_{\text{int}} + \sum_{i,\tau} \Delta t \left( \tilde{\theta}_i^1(\tau)\delta\tilde{\phi}_i(\tau) +\theta_i^1(\tau_-)\delta\phi_i(\tau_-) \right) - \left\langle \Delta t \sum_\tau H_{\text{int}} \right\rangle \right]^2 \right\rangle \\
			\label{gamma2_simpleOP}
			&= (\Delta t)^2 \left[ \left\langle \left( \sum_\tau H_{\text{int}} \right)^2 \right\rangle - \left\langle \sum_\tau H_{\text{int}} \right\rangle^2   + 2\sum_{i,\tau,\tau'}\left\langle H_{\text{int}}\left(  \tilde{\theta}_i^1(\tau')\delta \tilde{\phi}_i(\tau') + \theta_i^1(\tau_-')\delta\phi_i(\tau_-')\right) \right\rangle \right. \nonumber\\
			&\left. + \sum_{i,j,\tau,\tau'}  \left \langle \left(  \tilde{\theta}_i^1(\tau')\delta \tilde{\phi}_i(\tau') + \theta_i^1(\tau_-')\delta\phi_i(\tau_-')\right) \left(  \tilde{\theta}_j^1(\tau)\delta \tilde{\phi}_j(\tau) + \theta_j^1(\tau_-)\delta\phi_j(\tau_-)\right)  \right \rangle
			\right]
		\end{align}
Consider now the first term: \begin{align}
	\left\langle \left( \sum_\tau H_{\text{int}} \right)^2 \right\rangle
	= \sum_{\tau,\tau'} \sum_{\bar{m},\bar{n}}\sum_{\bar{m}',\bar{n}'} \frac{c_{\bar{m},\bar{n}}(\tau)}{\bm m! \bm n!} \frac{c_{\bar{m}',\bar{n}'}(\tau')}{\bm m'! \bm n'!}  \left\langle \prod_{i} \delta\tilde{\phi}_i^{m_i}(\tau) \delta\tilde{\phi}_i^{m_i'}(\tau') \delta\phi_i^{n_i}(\tau_-) \delta\phi_i^{n_i'}(\tau'_-) \right\rangle
\end{align}
	or equivalently, by ordering the time arguments,
\begin{equation}{\label{hint_expansion_gamma1}}
		\begin{split}
			\left\langle \left( \sum_\tau H_{\text{int}} \right)^2 \right\rangle  - \left\langle \sum_\tau H_{\text{int}}\right\rangle^2 =
			2 &\sum_{{\tau>\tau'}} \sum_{\bar{m},\bar{n}} \sum_{\bar{m}',\bar{n}' }^{(\bm m, \bm n, \bm m', \bm n') \neq (\bm 0, \bm 0, \bm 0, \bm 0)}  \frac{c_{\bar{m},\bar{n}}(\tau)}{\bm m! \bm n!} \frac{c_{\bar{m}',\bar{n}'}(\tau')}{\bm m'! \bm n'!}  \\
			&\left\langle \prod_{i}  \delta\tilde{\phi}_i^{m_i}(\tau) \delta\tilde{\phi}_i^{m_i'}(\tau') \delta\phi_i^{n_i}(\tau_-) \delta\phi_i^{n_i'}(\tau'_-) \right\rangle
		\end{split}
	\end{equation}
	where the constraint is that the term where {\em all} index vectors are zero is omitted.

The average over the baseline dynamics can now be evaluated using Wick's theorem.
Since we only have $\phi_i \tilde\phi_i$ terms in the baseline action, the only non-vanishing second order averages are the response functions
$R_{0i}(\tau,\tau') = $ $\left\langle \delta\phi_i(\tau')\delta\tilde{\phi}_i(\tau) \right\rangle $.
From this observation follow two simple rules: the average~\cref{hint_expansion_gamma1} vanishes (by straightforward causality) if the factor at the latest time is a $\delta\tilde{\phi}$; but it also vanishes if the factor at the {\em earliest} time is a $\delta\phi$ (which in a Wick pairing would have to be paired with a factor $\delta\tilde{\phi}$ at a later time, giving an acausal response that vanishes).
Applying these rules, the only average that will survive in the above equation is the one of $\delta\tilde{\phi}_i^{m_i'}(\tau') \delta\phi_i^{n_i}(\tau_-)$, for $\bar{m}' =\bar{n}$ and $\bar{m} =\bar{n}'=\bar{0}$. Thus we get\begin{align}
		\left\langle \left( \sum_\tau H_{\text{int}} \right)^2 \right\rangle  - \left\langle \sum_\tau H_{\text{int}}\right\rangle^2 &=
		2 \sum_{{\tau>\tau'}} \sum_{\bar{n}\neq \bar{0}} \frac{c_{\bar{0},\bar{n}}(\tau)}{\bm n!} \frac{c_{\bar{n},\bar{0}}(\tau')}{\bm n!} \prod_i \left\langle \delta\tilde{\phi}_i^{n_i}(\tau') \delta\phi_i^{n_i}(\tau_-) \right\rangle \\
		& = 2  \sum_{{\tau>\tau'}} \sum_{\bar{n} \neq \bar{0}} \frac{c_{\bar{0},\bar{n}}(\tau)}{\bm n!} \frac{c_{\bar{n},\bar{0}}(\tau')}{\bm n!} \prod_i n_i! R_{0i}^{n_i}(\tau_-,\tau')
	\end{align}
Here the factor of $n_i!$ arises since there are $n_i$ possible ways to pair the fields $\delta \tilde \phi_i $ with $ \delta \phi_i$.

	We proceed to calculate the cross term in~\cref{gamma2_simpleOP}. Using $\tilde{\theta}_i^1(\tau) = -c_{\bar{e}_i,\bar{0}}(\tau)$ and $\theta_i^1(\tau) = -c_{\bar{0},\bar{e}_i}(\tau_+)$ from~\cref{linearOP_theta_o1-2,linear_OP_theta_tilde_o1}
we have
	\begin{equation}
		\begin{split}
			&2 \sum_{i,\tau,\tau'} \left\langle H_{\text{int}} \left\lbrace \tilde{\theta}_i^1(\tau)\delta\tilde{\phi}_i(\tau) +  \theta_i^1(\tau_-) \delta\phi_i(\tau_-)\right\rbrace \right\rangle	\\
			& = -2 \sum_i\sum_{\tau,\tau'} \sum_{\bar{m},\bar{n}} c_{\bar{m},\bar{n}}(\tau) c_{\bar{e}_i,\bar{0}}(\tau') \left\langle \delta \tilde{\phi}_i(\tau') \prod_{j} \delta\tilde{\phi}_j^{m_j}(\tau) \delta \phi_j^{n_j}(\tau_-) \right\rangle\\
			& - 2 \sum_i \sum_{\tau,\tau'} \sum_{\bar{m},\bar{n}} c_{\bar{m},\bar{n}}(\tau) c_{\bar{0},\bar{e}_i}(\tau') \left\langle \delta \phi_i(\tau_-') \prod_{j} \delta\tilde{\phi}_j^{m_j}(\tau) \delta \phi_j^{n_j}(\tau_-) \right\rangle \\
			& = -4\sum_{i} \sum_{{\tau>\tau'}} c_{\bar{e}_i,\bar{0}}(\tau') c_{\bar{0},\bar{e}_i}(\tau) R_{0i}(\tau_-,\tau')
		\end{split}
	\end{equation}
Here we have used that the first term is non-vanishing only if $\tau'<\tau$ because $\delta\tilde{\phi}_i(\tau')$ is otherwise the last factor; the $m_j$ need to vanish for the same reason, and one needs $n_j=\delta_{ij}$ to get a response function without further (vanishing) factors of $\langle \delta\phi\,\delta\phi\rangle$. In the second term one similarly needs $n_j=0$, $m_j=\delta_{ij}$ and $\tau'>\tau$.

	Calculating the last term~\cref{gamma2_simpleOP} using the same arguments gives
	\begin{equation}
		\begin{split}
			\sum_{i,j,\tau,\tau'} & \left \langle \left(  \tilde{\theta}_i^1(\tau')\delta \tilde{\phi}_i(\tau') + \theta_i^1(\tau_-')\delta\phi_i(\tau_-')\right) \left(  \tilde{\theta}_j^1(\tau)\delta \tilde{\phi}_j(\tau) + \theta_j^1(\tau_-)\delta\phi_j(\tau_-)\right)  \right \rangle  \\
			&=  2 \sum_{i} \sum_{{\tau>\tau'}} \theta_i^1(\tau_-) \tilde{\theta}_i^1(\tau') R_{0i}(\tau_-,\tau') \\
			& = 2 \sum_{i} \sum_{{\tau>\tau'}} c_{\bar{e}_i,\bar{0}}(\tau') c_{\bar{0},\bar{e}_i^1}(\tau) R_{0i}(\tau_-,\tau')
		\end{split}
	\end{equation}
Combining the above terms gives for $\Gamma^2$, which is the $O(\alpha^2)$ contribution to the Plefka free energy, i.e.\ the first non-trivial correction to MAK,
\begin{equation}
	\Gamma^2 = 2 (\Delta t)^2 \sum_{{\tau>\tau'}} \sum_{\substack{\bar{n}\\\bar{n}\neq\bar{0},\bar{n}\neq\bar{e}^1}} \frac{1}{\bm n!} c_{\bar{0},\bar{n}}(\tau) c_{\bar{n},\bar{0}}(\tau') \prod_{i} R_{0i}^{n_i}(\tau_-,\tau')
\end{equation}
	where $\bar{e}^1$ indicates the collection of all the (Cartesian) unit vectors $\bar{e}_i$. This is the result given in the main text in~\cref{main-eq:gamma2_linearOP}.

	Calculating the second order corrections to the fields from $\Gamma^2$, one finds $\tilde{\theta}_i^2(\tau)$ as given in the main text:
		\begin{equation}
		\begin{split}
			-\tilde{\theta}_i^2(\tau) &= \frac{1}{\Delta t} \frac{\partial \Gamma^2}{\partial \tilde{\mu}_i(\tau)}\\
			&= 2\,\Delta t \sum_{\tau'(<\tau)}  \sum_{\bar{n}\neq\bar{0},\bar{e}^1}  \Big[ c_{\bar{e}_i,\bar{n}}(\tau) c_{\bar{n},\bar{0}}(\tau') \prod_j
			\cprime{n_j! R_j^{n_j}(\tau_-,\tau')}
			{\frac{R_{0j}^{n_j}(\tau_-,\tau')}{n_j!}}
			+ c_{\bar{0},\bar{n}}(\tau') c_{\bar{n}+\bar{e}_i,\bar{0}}(\tau) \cprime{(n_i+1)}{} \prod_j \cprime{n_j! R_j^{n_j}(\tau_-,\tau')}
			{\frac{R_{0j}^{n_j}(\tau_-,\tau')}{n_j!}}
			\Big]
		\end{split}
	\end{equation}
As shown in the main text, this simplifies to
		\begin{equation}{\label{linearOP_theta_o2}}
		\begin{split}
			-\tilde{\theta}_i^2(\tau) &= \Delta t \sum_{\tau'(<\tau)} \sum_{k,l} c_{\bar{e}_i,\bar{e}_k + \bar{e}_l}(\tau) c_{\bar{e}_k + \bar{e}_l,\bar{0}}(\tau') R_{0k}(\tau_-,\tau') R_{0l}(\tau_-,\tau')
			= \quad \parbox{40mm}{ 	\begin{fmfgraph*}(100,40)
					\fmfleft{e1}
\fmfv{label=\leglabel{i},label.dist=0,label.angle=180}{e1}
					\fmfright{e2,e3,e3n,e3nn,e4}
					\fmf{end_arrow_dash_two,tension=3.}{v1,e1}
					\fmf{phantom}{e2,v2}
					\fmf{phantom}{e3,v2}
					\fmf{phantom}{e4,v2}
					\fmf{phantom}{e3n,v2}
					\fmf{phantom}{e3nn,v2}
\fmf{fermion,left,tension=1.5,label=$R_{0k}$,label.angle=90,label.dist=3}{v2,v1}
					\fmf{fermion,right,tension=1.5,label=$R_{0l}$,label.angle=90,label.dist=3}{v2,v1}
					\fmfv{label=$\tau$,label.dist=7,label.angle=105,decor.shape=circle, decor.filled=full, decor.size= 4thick}{v1}
					\fmfv{label=$\tau'$,label.dist=7, label.angle=70,decor.shape=circle, decor.filled=full, decor.size= 4thick}{v2}
\end{fmfgraph*}}
		\end{split}
	\end{equation}

For the other generating field one obtains
\begin{equation}
		\begin{split}
			-\theta_i^2(\tau) &= \frac{1}{\Delta t} \frac{\partial \Gamma^2}{\partial \mu_i(\tau)}\\
			&= 2\Delta t \sum_{\tau'(<\tau)}  \sum_{\substack{\bar{n}\\{\bar{n}\neq\bar{0},\bar{e}^1}}}  \Big[ c_{\bar{0},\bar{n}+\bar{e}_i}(\tau_+) c_{\bar{n},\bar{0}}(\tau') \prod_j \frac{R_{0j}^{n_j}(\tau,\tau')}{n_j!} + c_{\bar{0},\bar{n}}(\tau') c_{\bar{n},\bar{e}_i}(\tau_+) \prod_j \frac{R_{0j}^{n_j}(\tau_-',\tau_+)}{n_j!} \Big]
		\end{split}
	\end{equation}
	and using~\cref{c_on_property} leads to
\begin{equation}
		\begin{split}
			\theta_i^2(\tau) &= 0
		\end{split}
	\end{equation}
	as expected on general grounds.

\section{Calculation of extended effective fields}

In this section we provide details of the calculations for the Plefka approximation with extended effective fields, i.e.\ where correlation and response functions are retained as additional order parameters beyond $\mu$ and $\tilde{\mu}$.

\subsection{First order in $\alpha$}{\label{SM:sec:first_order_extended}}

$\Gamma^1$ as given in the main text is given by
	\begin{equation}
	\begin{split}
		\Gamma^1 = \left\langle S_{\text{int}} \right\rangle_0 = \Delta t \sum_{\tau}\sum_{\bar{m},\bar{n}}
		\cprime{c_{\bar{m},\bar{n}}(\tau)}
		{\frac{c_{\bar{m},\bar{n}}(\tau)}{\bm{m}!\bm{n}!}}
		\left\langle \prod_{i} \delta \tilde{\phi}_i^{m_i}(\tau) \delta \phi_i^{n_i}(\tau_-) \right\rangle
	\end{split}
\end{equation}
and as explained there can be simplified for binary reaction networks to
\begin{eqnarray}
	&&\Gamma^1 = \Delta t \sum_\tau
	\left\lbrace c_{\bm{0},\bm{0}}(\tau) +
	\sum_{i,j}c_{\bm{e}_i,\bm{e}_j}(\tau) R_{ji}(\tau_-,\tau)+
	\frac{1}{2}\sum_{i,j}\left[c_{\bm{e}_i+\bm{e}_j,\bm{0}}(\tau) B_{ij}(\tau,\tau) +
	c_{\bm{0},\bm{e}_i+\bm{e}_j}(\tau) C_{ij}(\tau_-,\tau_-)\right] \right. \nonumber \\
	&&  \left.+ \frac{1}{4}\sum_{i,j,k,l}c_{\bm{e}_i+\bm{e}_j,\bm{e}_k+\bm{e}_l}(\tau)\left[
	B_{ij}(\tau,\tau)C_{kl}(\tau_-,\tau_-)
	+R_{ki}(\tau_-,\tau)R_{lj}(\tau_-,\tau)+
	R_{li}(\tau_-,\tau)R_{kj}(\tau_-,\tau)
	\right] \right\rbrace
\end{eqnarray}
The resulting first order fields were shown to be in the main text
\begin{eqnarray}
	-\tilde{\theta}_i^1(\tau) &=&  c_{\bar{e}_i,\bar{0}}(\tau) + \sum_{k,l} \left[ \frac{1}{2} c_{\bar{e}_i,\bar{e}_k + \bar{e}_l}(\tau) C_{kl}(\tau_-,\tau_-)	+  c_{\bm{e}_k+\bm{e}_i,\bm{e}_l} R_{lk}(\tau_-,\tau) \right] \label{eq:hattheta1_extended} \\
	-\hat{R}^1_{ij}(\tau,\tau')&=& \frac{\delta_{\tau',\tau_-}}{\Delta t}\left[
	c_{\bm{e}_i,\bm{e}_j}(\tau)+\sum_{k,l}
	c_{\bm{e}_k+\bm{e}_i,\bm{e}_l+\bm{e}_j}(\tau)R_{lk}(\tau_-,\tau)
	\right]\label{eq:hatR1_extended} \\
	-\frac{1}{2}\hat{B}^1_{ij}(\tau,\tau') &=&	\frac{\delta_{\tau',\tau}}{\Delta t} \left[
	\frac{1}{2}
	c_{\bm{e}_i+\bm{e}_j,\bm{0}}(\tau)
	+\frac{1}{4}
	\sum_{k,l}c_{\bm{e}_i+\bm{e}_j,\bm{e}_k+\bm{e}_l}(\tau)C_{kl}(\tau_-,\tau_-)
	\right] \label{eq:hatB1_extended}
	\\
	-\frac{1}{2}\hat{C}^1_{ij}(\tau,\tau') &=& 	\frac{\delta_{\tau',\tau}}{\Delta t}\left[
	\frac{1}{2}
	c_{\bm{0},\bm{e}_i+\bm{e}_j}(\tau_+)
	+\frac{1}{4}
	\sum_{k,l}c_{\bm{e}_k+\bm{e}_l,\bm{e}_i+\bm{e}_j}(\tau_+)B_{kl}(\tau_+,\tau_+)
	\right] \label{eq:hatC1_extended}
\end{eqnarray}
For $\theta_i^1$ one finds by the same method, \begin{align}
		\nonumber
		-\theta_i^1(\tau) & = \frac{1}{\Delta t}\frac{\partial \Gamma^1}{\partial \mu_i(\tau)} = \frac{\partial c_{\bar{0},\bar{0}}(\tau_+)}{\partial \mu_i(\tau)}
+ \sum_{k,l}  \left[  \frac{1}{2} \frac{\partial c_{\bm{e}_k+\bm{e}_l,\bm{0}}(\tau_+) }{\partial \mu_i(\tau)} B_{kl}(\tau_+,\tau_+)
		+ \frac{\partial c_{\bm{e}_k,\bm{e}_l}(\tau_+)}{\partial \mu_i(\tau) } R_{lk}(\tau,\tau_+) \right] \\ {\label{quadOP_theta_tilde_o1}}
&= c_{\bar{0},\bar{e}_i}(\tau_+) + \sum_{k,l} \left[ \frac{1}{2} c_{\bm{e}_k+\bm{e}_l,\bm{e_i}}(\tau_+) B_{kl}(\tau_+,\tau_+)  + c_{\bm{e}_k,\bm{e}_l+\bm{e}_i}(\tau_+) R_{lk}(\tau,\tau_+) \right]
\end{align}
For the physical solution, which obeys $R_{ij}(\tau_-,\tau) = 0$ on account of causality and $B_{ij}(\tau,\tau)  = 0$ from~\cref{eq:B0} $\forall\ i,j,\tau$, one obtains
\begin{equation}
			-\theta_i^1(\tau)  = 0
\end{equation}
as expected.
One verifies similarly that the physical value of $\hat C$  vanishes as stated in the main text, by substituting~\cref{eq:B0,c_on_property} into~\cref{eq:hatC1_extended}: \begin{equation}
	\begin{split}
-\hat{C}^1_{ij}(\tau,\tau') = 0
	\end{split}
\end{equation}

	\subsection{Second order in $\alpha$}{\label{SM:sec:second_order_extended}}

The second order correction to the extended Plefka free energy using the expression for $\Gamma^2$ from~\cref{gamma_2_exponent_calculation} but now with the additional order parameters and conjugated fields, is given by,
\begin{equation}
		\begin{split}
			\Gamma^2 = \Bigg\langle \Big\lbrace  \Delta t  \sum_\tau H_{\text{int}} &- \left\langle  \Delta t  \sum_\tau H_{\text{int}} \right\rangle + \Delta t \sum_{i,\tau} \left[ \tilde{\theta}_{i}^1(\tau)\delta\tilde{\phi}_i(\tau) + \theta_{i}^1(\tau_-)\delta\phi_i(\tau_-)  \right] \\
			&+ \Delta t \sum_{i,j,\tau,\tau'}
			\hat{R}_{ij}^1(\tau,\tau')\left( \delta\phi_j(\tau')\delta\tilde{\phi}_i(\tau) - R_{ji}(\tau',\tau) \right) \\
			&+\frac{1}{2}\Delta t \sum_{i,j,\tau,\tau'}
			\hat{B}_{ij}^1(\tau,\tau')\left( \delta\tilde{\phi}_i(\tau)\delta\tilde{\phi}_j(\tau') - B_{ij}(\tau,\tau') \right) \\
			&+\frac{1}{2}\Delta t \sum_{i,j,\tau,\tau'}
			\hat{C}_{ij}^1(\tau,\tau')\left( \delta{\phi}_i(\tau)\delta{\phi}_j(\tau') - C_{ij}(\tau,\tau') \right)
			\Big\rbrace^2 	\Bigg\rangle
		\end{split}
	\end{equation}
Since $\hat{R}_{ij}^1(\tau,\tau')$, $\hat{B}_{ij}^1(\tau,\tau')$ and $\hat{C}_{ij}^1(\tau,\tau')$ from~\cref{eq:hatR1_extended,eq:hatB1_extended,eq:hatC1_extended} respectively are local in time,
one can set $\tau'=\tau_-$ or $\tau'=\tau$ respectively in the two-time sums.
The preceding expression can then be written as,
	\begin{equation}{\label{eq:Gamma2_def}}
		\begin{split}
			\Gamma^2 & = \left\langle \left[ \Delta t  \sum_\tau H_{\text{int}}^* - \Delta t  \sum_\tau \left\langle H_{\text{int}}^*\right\rangle \right]^2 \right\rangle  = \left\langle \left(\Delta t  \sum_\tau H_{\text{int}}^*\right)^{2} \right\rangle - \left\langle \Delta t  \sum_\tau H_{\text{int}}^* \right\rangle^2
		\end{split}
	\end{equation}
	in terms of the modified interaction Hamiltonian
	\begin{equation}
H_{\text{int}}^* = \sum_{\bar{m},\bar{n}} \frac{c_{\bar{m},\bar{n}}^*(\tau)}{\bm m! \bm n!} \prod_i \delta\tilde{\phi}_i^{m_i}(\tau) \delta\phi_i^{n_i}(\tau_-)
	\end{equation}
Here the $c_{\bar{m},\bar{n}}^*$ are related to the $c_{\bar{m},\bar{n}}$ as follows: \begin{subequations}
	\begin{align}
	\label{c_mn_star_def0}
		c^*_{\bar{0},\bar{0}}(\tau) & = 0\\
	\label{c_mn_star_def1}
		c^*_{\bar{e}_i,\bar{0}}(\tau) & = c_{\bar{e}_i,\bar{0}}(\tau) + \tilde{\theta}_i^1(\tau) = - \frac{1}{2}\sum_{k,l} c_{\bar{e}_i,\bar{e}_k + \bar{e}_l}(\tau) C_{kl}(\tau_-,\tau_-) -\sum_{k,l} c_{\bm{e}_k+\bm{e}_i,\bm{e}_l} R_{lk}(\tau_-,\tau) \\
		\label{c_mn_star_def2}
		c^*_{\bar{0},\bar{e}_i}(\tau) & = c_{\bar{0},\bar{e}_i}(\tau) + \theta_i^1(\tau_-) = -\sum_{k,l} \left[ \frac{1}{2} c_{\bm{e}_k+\bm{e}_l,\bm{e_i}}(\tau) B_{kl}(\tau,\tau)  + c_{\bm{e}_k,\bm{e}_l+\bm{e}_i}(\tau) R_{lk}(\tau_-,\tau) \right] \\
		\label{c_mn_star_def3}
		c^*_{\bar{e}_i,\bar{e}_j}(\tau) & = c_{\bar{e}_i,\bar{e}_j}(\tau) + \Delta t \,\hat{R}_{ij}^1(\tau,\tau_-) = -\sum_{k,l} c_{\bm{e}_k+\bm{e}_i,\bm{e}_l+\bm{e}_j}(\tau)R_{lk}(\tau_-,\tau)\\
		\label{c_mn_star_def4}
		c^*_{\bar{e}_i+\bar{e}_j ,\bar{0}}(\tau) & = c_{\bar{e}_i+\bar{e}_j,\bar{0}}(\tau) + \Delta t \,\hat{B}_{ij}^1(\tau,\tau) =  -\frac{1}{2} 		\sum_{k,l}c_{\bm{e}_i+\bm{e}_j,\bm{e}_k+\bm{e}_l}(\tau)C_{kl}(\tau_-,\tau_-) \\
		\label{c_mn_star_def5}
		c^*_{\bar{0},\bar{e}_i+\bar{e}_j}(\tau) & = c_{\bar{0},\bar{e}_i+\bar{e}_j}(\tau) + \Delta t \,\hat{C}_{ij}^1(\tau_-,\tau_-) =
		-\frac{1}{2} \sum_{k,l}c_{\bm{e}_k+\bm{e}_l,\bm{e}_i+\bm{e}_j}(\tau)B_{kl}(\tau,\tau) \\
\label{c_mn_star_def6}
		c^*_{\bar{m},\bar{n}}(\tau) & = c_{\bar{m},\bar{n}}(\tau) \hspace{10pt} \text{otherwise}
\end{align}
\end{subequations}

We also note for later that if one inserts the physical solutions for $R$ and $B$ then the above coefficients simplify considerably, to
\begin{subequations}
	\begin{align}
	\label{c_mn_star_def_physical1}
		c^*_{\bar{e}_i,\bar{0}}(\tau) & = - \frac{1}{2}\sum_{k,l} c_{\bar{e}_i,\bar{e}_k + \bar{e}_l}(\tau) C_{kl}(\tau_-,\tau_-) \\
		\label{c_mn_star_def_physical2}
		c^*_{\bar{0},\bar{e}_i}(\tau) & = 0 \\
		\label{c_mn_star_def_physical3}
		c^*_{\bar{e}_i,\bar{e}_j}(\tau) & = 0\\
		\label{c_mn_star_def_physical4}
		c^*_{\bar{e}_i+\bar{e}_j ,\bar{0}}(\tau) & =  -\frac{1}{2} 		\sum_{k,l}c_{\bm{e}_i+\bm{e}_j,\bm{e}_k+\bm{e}_l}(\tau)C_{kl}(\tau_-,\tau_-) \\
		\label{c_mn_star_def_physical5}
		c^*_{\bar{0},\bar{e}_i+\bar{e}_j}(\tau) & = 0
\end{align}
\end{subequations}
and these simplifications apply independently of the values of $\mu$ and $\tilde{\mu}$. In particular, together with~\cref{c_mn_star_def0} we have that,
\begin{equation}
c^*_{\bm{0},\bm{n}}(\tau)=0 \qquad \forall\ \bm{n}\in \mathcal{S}
\label{cstar0n}
\end{equation}
at physical values of $R$ and $B$. We recall here that for a binary interaction network, the set $\mathcal{S}$ defined in the main text can only contain index vectors $\bm{n}$ of the form $\bm{n}=\bm{0}$, $\bm{e}_i$ and $\bm{e}_i+\bm{e}_j$.

Returning to the general case, we can write out~\cref{eq:Gamma2_def} as
	\begin{equation}
		\begin{split}
\Gamma^2 = (\Delta t)^2 \sum_{\tau,\tau'} \sum_{\bar{m},\bar{n}} \sum_{\bar{m}',\bar{n}'} & \frac{c^*_{\bar{m},\bar{n}}(\tau)}{\bm m! \bm n!} \frac{c^*_{\bar{m}',\bar{n}'}(\tau')}{\bm m'! \bm n'!}
			\Bigg[   \left\langle \prod_i \delta\tilde{\phi}_i^{m_i}(\tau) \delta\tilde{\phi}_i^{m_i'}(\tau') \delta \phi_i^{n_i}(\tau_-) \delta\phi_i^{n_i'}(\tau_-')\right\rangle\\
			&- \left\langle  \prod_i \delta\tilde{\phi}_i^{m_i}(\tau) \delta \phi_i^{n_i}(\tau_-) \right\rangle \left\langle  \prod_i \delta\tilde{\phi}_i^{m_i'}(\tau') \delta\phi_i^{n_i'}(\tau_-')\right\rangle
			\Bigg]
		\end{split}
		\label{eq:Gamma2_gen}
	\end{equation}
The averages can again be expressed in terms of Wick pairings; the subtraction in the second line then means that only pairings containing at least one factor that couples $\tau$ and $\tau'$ should be retained. We will use this constraint below without writing the subtraction explicitly. Note also that in the following, we will not explicitly keep any terms in $\Gamma^2$ which,
after taking the appropriate derivatives, will contain factors of physically zero-valued order parameters: such terms vanish in the final physical solution. In $\Gamma^2$ we therefore only need to work up to first order in any factors that vanish for the physical solution. This excludes e.g.\ terms with $\bm{m}=\bm{m}'=\bm{0}$ in~\cref{eq:Gamma2_gen}, for which from~\cref{c_mn_star_def2,c_mn_star_def5} the prefactor is quadratic in such ``unphysical'' factors. In the resulting effective fields, i.e.\ after derivatives have been taken, we only need zeroth order contributions in the unphysical factors, i.e.\ we can set those factors to zero.

	We now proceed to calculate $\theta^2$ and $\tilde{\theta}^2$ by taking the appropriate derivatives of~\cref{eq:Gamma2_gen}. $\tilde{\theta}_{i}^2(\tau)$ is given by
	\begin{equation}
		\begin{split}
			-\tilde{\theta}_{i}^2(\tau) = \frac{1}{\Delta t}\frac{\partial \Gamma^2}{\partial \tilde{\mu}_i(\tau)}
\end{split}
	\end{equation}
or more explicitly
\begin{equation}
		\begin{split}
			-\tilde{\theta}_{i}^2(\tau) &= 2\Delta t \sum_{\tau'}  \sum_{\bar{m},\bar{n}} \sum_{\bar{m}',\bar{n}'} \frac{1}{\bm m! \bm n!} \frac{\partial c^*_{\bar{m},\bar{n}}(\tau)}{\partial \tilde{\mu}_i(\tau)} \frac{c^*_{\bar{m}',\bar{n}'}(\tau')}{\bm m'! \bm n'!} \left\langle \prod_i  \delta\tilde{\phi}_i^{m_i}(\tau) \delta\tilde{\phi}_i^{m_i'}(\tau') \delta \phi_i^{n_i}(\tau_-) \delta\phi_i^{n_i'}(\tau_-')\right\rangle
\end{split}
	\end{equation}
To simplify this, we begin by noticing from~\cref{cstar0n} that $\bm{m}'$ must be nonzero.
This restriction
then implies $\tau'<\tau$, in order to make contractions with response fields at $\tau'$ causal. This further forces $\bar{m}=\bm{0}$, giving
	\begin{equation}
		\begin{split}
			-\tilde{\theta}_{i}^2(\tau) &= 2\Delta t \sum_{\tau'(<\tau)}\sum_{\bar{n},\bar{n}'} \sum_{\bar{m}'\neq \bar{0}} \frac{1}{\bm n!}\frac{\partial c^*_{\bar{0},\bar{n}}(\tau)}{\partial \tilde{\mu}_i(\tau)} \frac{c^*_{\bar{m}',\bar{n}'}(\tau')}{\bm m'! \bm n'!}  \left\langle \prod_i \delta\tilde{\phi}_i^{m_i'}(\tau') \delta \phi_i^{n_i}(\tau_-) \delta\phi_i^{n_i'}(\tau_-')\right\rangle \\
		\end{split}
	\end{equation}
But from~\cref{cstar0n}
the first derivative then vanishes for all $\bm{n}\in\mathcal{S}$, so that overall 
$\tilde{\theta}^2_i(\tau) = 0\ \forall\,i,\tau$.

One can similarly calculate $\theta_i^2$. We expect this to be zero on general grounds but let us check explicitly. Starting from
	\begin{equation}
-\theta_{i}^2(\tau) = \frac{1}{\Delta t}\frac{\partial \Gamma^2}{\partial \mu_i(\tau)}
\end{equation}
one finds
	\begin{equation}
		\begin{split}
			-\theta_{i}^2(\tau) &= 2\Delta t \sum_{\tau'}  \sum_{\bar{m},\bar{n}} \sum_{\bar{m}',\bar{n}'} \frac{1}{\bm m! \bm n!}\frac{\partial c^*_{\bar{m},\bar{n}}(\tau)}{\partial \mu_i(\tau)} \frac{c^*_{\bar{m}',\bar{n}'}(\tau')}{\bm m'! \bm n'! } \left\langle  \prod_i \delta\tilde{\phi}_i^{m_i}(\tau) \delta\tilde{\phi}_i^{m_i'}(\tau') \delta \phi_i^{n_i}(\tau_-) \delta\phi_i^{n_i'}(\tau_-')\right\rangle \\
\end{split}
	\end{equation}
For this average to be non-zero it must not contain acausal response function factors, so either $\bar{m}$ or $\bar{m}'$ must be zero. But since $c^*_{0,\bar{n}} \ \forall \ \bar{n}$ for the physical solution from~\cref{cstar0n}, the case $\bar{m}' = \bar{0}$ is again excluded and so $\bar{m}=\bar{0}$. We thus have
	\begin{equation}
		\begin{split}
			-\theta_{i}^2(\tau) &= 2\Delta t \sum_{\tau'}
\sum_{\bar{n},\bar{n}'} \sum_{\bar{m}'\neq \bar{0}}
\frac{\partial c^*_{\bar{0},\bar{n}}(\tau)}{\partial \mu_i(\tau)} \frac{c^*_{\bar{m}',\bar{n}'}(\tau')}{\bm m'! \bm n'! } \left\langle  \prod_i \delta\tilde{\phi}_i^{m_i'}(\tau') \delta \phi_i^{n_i}(\tau_-) \delta\phi_i^{n_i'}(\tau_-')\right\rangle \\
		\end{split}
	\end{equation}
By the same argument as above the derivative of $c^*_{\bar{0},\bar{n}}(\tau)$ vanishes so that ${\theta}^2_i(\tau) = 0\ \forall\,i,\tau$.

To calculate $\hat{R}_{ij}^2$ we have to take derivatives w.r.t.\ acausal response functions with $\tau'<\tau$ as follows:
	\begin{align}{\label{eq:hatR2_intermediate}}
			&-\hat{R}^2_{ij}(\tau,\tau') =	\frac{1}{(\Delta t)^2}\frac{\partial \Gamma^2}{\partial R_{ji}(\tau',\tau)} \\
&= \sum_{\tau'',\tau'''} \sum_{\bar{m},\bar{n}} \sum_{\bar{m}',\bar{n}'} \frac{c^*_{\bar{m},\bar{n}}(\tau'')}{\bm m! \bm n!} \frac{c^*_{\bar{m}',\bar{n}'}(\tau''')}{\bm m'! \bm n'!}
			\frac{\partial}{\partial R_{ji}(\tau',\tau)}
			\left\langle \prod_i  \delta\tilde{\phi}_i^{m_i}(\tau'') \delta\tilde{\phi}_i^{m_i'}(\tau''') \delta \phi_i^{n_i}(\tau_-'') \delta\phi_i^{n_i'}(\tau_-''')\right\rangle
	\end{align}
Here we have already discarded the $R$-dependence of the $c^*$ implied by~\cref{c_mn_star_def1,c_mn_star_def2,c_mn_star_def3}. This can be justified as follows: derivatives w.r.t.\ these occurrences of $R$ would be multiplied by the Gaussian average $\langle\ldots\rangle$ above. But this average can be nonzero for the physical solution only for $\bm{m}=\bm{m}'=\bm{0}$ as otherwise there would be factors $\delta\tilde{\phi}$ at the latest time, giving acausal response functions. In the prefactor one factor of either $c^*_{\bm{0},\bm{n}}(\tau'')$ or $c^*_{\bm{0},\bm{n}'}(\tau''')$ would then remain, which vanishes for the physical solution from~\cref{cstar0n}.

In~\cref{eq:hatR2_intermediate}, we first notice that terms with either $\bm{m}=\bm{0}$ or $\bm{m}'=\bm{0}$ vanish because of the prefactors. Thus both these index vectors have to be of the form $\bm{e}_k$ or $\bm{e}_k+\bm{e}_l$. But the second option is not allowed simultaneously for $\bm{m}$ and $\bm{m}'$, because then the average $\langle\ldots\rangle$ would contain two factors of $\delta\tilde{\phi}$ at the latest time and therefore be quadratic in acausal response functions. These considerations give
	\begin{equation}{\label{eq:hatR2_intermediate2}}
		\begin{split}
			&-\hat{R}^2_{ij}(\tau,\tau') =\sum_{\tau'',\tau'''} \sum_{\bar{n},\bar{n'}}\\
&\Biggl[
\sum_{kl}
\frac{c^*_{\bar{e}_k,\bar{n}}(\tau'')}{\bm n!} \frac{c^*_{\bar{e}_l,\bar{n}'}(\tau''')}{\bm n'!}
			\frac{\partial}{\partial R_{ji}(\tau',\tau)}
			\left\langle \delta\tilde{\phi}_k(\tau'') \delta\tilde{\phi}_l(\tau''')
						\prod_i  \delta \phi_i^{n_i}(\tau_-'') \delta\phi_i^{n_i'}(\tau_-''')\right\rangle
\\
&{}+
2\sum_{k\leq l}\sum_{p} \frac{c^*_{\bar{e}_k+\bar{e}_l,\bar{n}}(\tau'')}{(1+\delta_{kl})\bm n!} \frac{c^*_{\bar{e}_p,\bar{n}'}(\tau''')}{\bm n'!}
			\frac{\partial}{\partial R_{ji}(\tau',\tau)}
\left\langle
\delta\tilde{\phi}_k(\tau'') \delta\tilde{\phi}_l(\tau'') \delta\tilde{\phi}_p(\tau''')
\prod_i  \delta \phi_i^{n_i}(\tau_-'') \delta\phi_i^{n_i'}(\tau_-''')\right\rangle
\Biggr]
		\end{split}
	\end{equation}
As explained in the main text one can symmetrize the sum $2\sum_{k\leq l}(1+\delta_{kl})^{-1}\ldots$ to $\sum_{kl}\ldots$ in the second line. We will apply this symmetrization in the following without further comment.

\newcommand{\dph}{\delta\phi}
\newcommand{\dpt}{\delta\tilde{\phi}}
\newcommand{\tam}{\tau_-}
\newcommand{\tap}{\tau_+}

One next needs to evaluate the remaining averages. In the first line one needs at least two factors of $\dph$ to obtain Wick pairings involving the response function $R$ ($\sum_i (n_i+n'_i)=2$); the other option is to have four factors of $\dph$ ($\sum_i (n_i+n_i')=4$). Starting with the first case, this can be realized by e.g.\ $\bm{n}=\bm{e}_p+\bm{e}_q$ and $\bm{n}'=\bm{0}$, giving as the first contribution from the first line
\begin{equation}
L_{11}=
\sum_{\tau'',\tau'''}
\frac{1}{2}\sum_{klpq}
c^*_{\bar{e}_k,\bm{e}_p+\bm{e}_q}(\tau'') c^*_{\bar{e}_l,\bar{0}}(\tau''')
\frac{\partial}{\partial R_{ji}(\tau',\tau)}
\left\langle
\delta\tilde{\phi}_k(\tau'') \delta\tilde{\phi}_l(\tau''')
\delta\phi_p(\tau_-'') \dph_q(\tau_-'')
\right\rangle
\label{eq:L11}
\end{equation}
In the remaining average only two Wick pairings contain factors of $R$:
\begin{equation}
\left\langle
\delta\tilde{\phi}_k(\tau'') \delta\tilde{\phi}_l(\tau''')
\delta\phi_p(\tau_-'') \dph_q(\tau_-'')
\right\rangle =
R_{pk}(\tam'',\tau'')R_{ql}(\tam'',\tau''')+
R_{pl}(\tam'',\tau''')R_{qk}(\tam'',\tau'')+\ldots
\end{equation}
The derivative of this w.r.t.\ $R_{ji}(\tau',\tau)$ is
\begin{eqnarray}
&&\delta_{pj}\delta_{ki}\delta_{\tam''\tau'}\delta_{\tau''\tau}
R_{ql}(\tam'',\tau''')+
R_{pk}(\tam'',\tau'')
\delta_{qj}\delta_{li}\delta_{\tam''\tau'}\delta_{\tau'''\tam''}\\
&&{}+\delta_{pj}\delta_{li}\delta_{\tam''\tau'}\delta_{\tau'''\tau}
R_{qk}(\tam'',\tau'')+
R_{pl}(\tam'',\tau''')\delta_{qj}\delta_{ki}\delta_{\tam''\tau'}\delta_{\tau''\tau}
\nonumber
\end{eqnarray}
The second and third term still contain an acausal response as a factor so can be discarded. Inserting into~\cref{eq:L11} gives
\begin{eqnarray}
\!\!\!\!\!\!\!L_{11}\!\!\!\!\!\!&=&\!\!\!\!\!\!
\frac{1}{2}
\sum_{\tau''\tau'''klpq}
c^*_{\bar{e}_k,\bm{e}_p+\bm{e}_q}(\tau'') c^*_{\bar{e}_l,\bar{0}}(\tau''')
[\delta_{pj}\delta_{ki}\delta_{\tam''\tau'}\delta_{\tau''\tau}
R_{ql}(\tam'',\tau''')+
R_{pl}(\tam'',\tau''')\delta_{qj}\delta_{ki}\delta_{\tam''\tau'}\delta_{\tau''\tau}]\\&=&\!\!\!\!\!\!
\frac{1}{2}\frac{\delta_{\tau'\tam}}{\Delta t}\Delta t\sum_{\tau'''}
\left[
\sum_{lq}
c^*_{\bar{e}_i,\bm{e}_j+\bm{e}_q}(\tau) c^*_{\bar{e}_l,\bar{0}}(\tau''')
R_{ql}(\tam,\tau''')+
\sum_{lp}
c^*_{\bar{e}_i,\bm{e}_p+\bm{e}_j}(\tau) c^*_{\bar{e}_l,\bar{0}}(\tau''')
R_{pl}(\tam,\tau''')\right]
\end{eqnarray}
The two sums in square brackets are equal and the sum over $\tau'''$ is restricted by causality, so that overall
\begin{eqnarray}
L_{11}&=&\frac{\delta_{\tau'\tam}}{\Delta t}\,\Delta t\sum_{\tau''(<\tau)}
\sum_{lp}
c^*_{\bar{e}_i,\bm{e}_p+\bm{e}_j}(\tau) c^*_{\bar{e}_l,\bar{0}}(\tau'')
R_{pl}(\tam,\tau'')\\
&=&
- \frac{1}{2}\frac{\delta_{\tau'\tam}}{\Delta t}\,\Delta t\sum_{\tau''(<\tau)}
\sum_{klpq}
c_{\bar{e}_i,\bm{e}_p+\bm{e}_j}(\tau)
R_{pl}(\tam,\tau'')c_{\bar{e}_l,\bar{e}_k + \bar{e}_q}(\tau'') C_{kq}(\tau''_-,\tau''_-)
\label{eq:L11_final}
\end{eqnarray}
where in the final step we have inserted~\cref{c_mn_star_def_physical1}. We observe that $L_{11}$ is a time-local contribution to $-\hat{R}^2_{ij}(\tau,\tau')$, but with a magnitude that is controlled by an integral over the past.

Returning to~\cref{eq:hatR2_intermediate2}, a second way of obtaining two factors of $\dph$ in the first line is to have $\bm{n}=\bm{0}$, $\bm{n}'=\bm{e}_p+\bm{e}_q$, giving a second contribution
\begin{eqnarray}
L_{12}&=&
\sum_{\tau'',\tau'''}
\frac{1}{2}\sum_{klpq}
c^*_{\bar{e}_k,\bm{0}}(\tau'') c^*_{\bar{e}_l,\bm{e}_p+\bm{e}_q}(\tau''')
\frac{\partial}{\partial R_{ji}(\tau',\tau)}
\left\langle
\delta\tilde{\phi}_k(\tau'') \delta\tilde{\phi}_l(\tau''')
\delta\phi_p(\tau_-''') \dph_q(\tau_-''')
\right\rangle
\label{eq:L12}
\end{eqnarray}
This differs from $L_{11}$ only in the prefactors and in the time arguments ($\tam'''$ rather than $\tam''$) of the last two $\dph$ factors. If we therefore relabel $k\leftrightarrow l$ and $\tau''\leftrightarrow \tau'''$, we have
\begin{eqnarray}
L_{12}&=&
\sum_{\tau'',\tau'''}
\frac{1}{2}\sum_{klpq}
c^*_{\bar{e}_l,\bm{0}}(\tau''') c^*_{\bar{e}_k,\bm{e}_p+\bm{e}_q}(\tau'')
\frac{\partial}{\partial R_{ji}(\tau',\tau)}
\left\langle
\delta\tilde{\phi}_k(\tau'') \delta\tilde{\phi}_l(\tau''')
\delta\phi_p(\tau_-'') \dph_q(\tau_-'')
\right\rangle
\label{eq:L12b}
\end{eqnarray}
showing that $L_{12}=L_{11}$.

The third contribution to the first line of~\cref{eq:hatR2_intermediate2} with two factors of $\dph$ in the average is the one from $\bm{n}=\bm{e}_p$, $\bm{n}'=\bm{e}_q$:
\begin{eqnarray}
L_{13}&=&
\sum_{\tau'',\tau'''}
\sum_{klpq}
c^*_{\bar{e}_k,\bm{e}_p}(\tau'') c^*_{\bar{e}_l,\bm{e}_q}(\tau''')
\frac{\partial}{\partial R_{ji}(\tau',\tau)}
\left\langle
\delta\tilde{\phi}_k(\tau'') \delta\tilde{\phi}_l(\tau''')
\delta\phi_p(\tau_-'') \dph_q(\tau_-''')
\right\rangle
\label{eq:L13}
\end{eqnarray}
Here the prefactors both vanish for the physical solution from~\cref{c_mn_star_def_physical3}, giving $L_{13}=0$.

Finally we have in the first line of~\cref{eq:hatR2_intermediate2} the term that is quartic in $\dph$, corresponding to $\bm{n}=\bm{e}_p+\bm{e}_q$ and $\bm{n}'=\bm{e}_r+\bm{e}_s$:
\begin{equation}
L_{14}=
\frac{1}{4}\sum_{\tau''\tau'''klpqrs}
c_{\bar{e}_k,\bm{e}_p+\bm{e}_q}(\tau'') c_{\bar{e}_l,\bm{e}_r+\bm{e}_s}(\tau''')
\frac{\partial}{\partial R_{ji}(\tau',\tau)}
\left\langle
\delta\tilde{\phi}_k(\tau'') \delta\tilde{\phi}_l(\tau''')
\delta\phi_p(\tau_-'') \dph_q(\tau_-'')
\delta\phi_r(\tau_-''') \dph_s(\tau_-''')
\right\rangle
\label{eq:L14}
\end{equation}
Here we have already been able to drop the asterisks on the prefactors, because of~\cref{c_mn_star_def6}. Of the 15 Wick pairings in the sixth-order average, the three that pair $\dpt$ with $\dpt$ can be dropped because they do not give response functions $R$. The others can be grouped according to which time arguments are paired, with terms differing by a relabelling $p\leftrightarrow q$ or $r\leftrightarrow s$ giving the same contribution so that
\begin{eqnarray}
&&\!\!\!\!\!\!\!\!\!\!\!\!\!\!\!\!\!\!\!\!\!
\left\langle
\delta\tilde{\phi}_k(\tau'') \delta\tilde{\phi}_l(\tau''')
\delta\phi_p(\tau_-'') \dph_q(\tau_-'')
\delta\phi_r(\tau_-''') \dph_s(\tau_-''')
\right\rangle
= \\
&& 4 R_{pk}(\tam'',\tau'') R_{rl}(\tam''',\tau''')C_{qs}(\tam'',\tam''')
\nonumber
\\&&{}+2 R_{pk}(\tam'',\tau'') R_{ql}(\tam'',\tau''')C_{rs}(\tam''',\tam''')
+2 R_{rk}(\tam''',\tau'') R_{sl}(\tam''',\tau''')C_{pq}(\tam'',\tam'')
\nonumber\\
&&{}+4 R_{rk}(\tam''',\tau'') R_{pl}(\tam'',\tau''')C_{qs}(\tam'',\tam''') + \ldots
\nonumber
\end{eqnarray}
The first line is quadratic in unphysical factors (acausal responses) so can be discarded. The two terms in the second line can be combined because they become equal under relabelling $(k,q,p,\tau'')\leftrightarrow (l,r,s,\tau''')$, which also leaves the $c$-prefactors unchanged. The derivative w.r.t.\ $R_{ji}(\tau',\tau)$ is then
\begin{eqnarray}
&& 4
\delta_{pj}\delta_{ki}\delta_{\tam''\tau'}\delta_{\tau''\tau}R_{ql}(\tam'',\tau''')C_{rs}(\tam''',\tam''')
+
4
\delta_{qj}\delta_{li}\delta_{\tam''\tau'}\delta_{\tau'''\tau}
R_{pk}(\tam'',\tau'')
C_{rs}(\tam''',\tam''')\\
&&{}+
4
\delta_{rj}\delta_{ki}\delta_{\tam'''\tau'}\delta_{\tau''\tau}R_{pl}(\tam'',\tau''')C_{qs}(\tam'',\tam''')+
4
\delta_{pj}\delta_{li}\delta_{\tam''\tau'}\delta_{\tau'''\tau}
R_{rk}(\tam''',\tau'') C_{qs}(\tam'',\tam''')\nonumber
\end{eqnarray}
The second terms contains an acausal response as a factor and can be dropped. Inserting the rest into~\cref{eq:L14} and carrying out the sums gives
\begin{eqnarray}
L_{14}&=&
\sum_{\tau''\tau'''klpqrs}
c_{\bar{e}_k,\bm{e}_p+\bm{e}_q}(\tau'') c_{\bar{e}_l,\bm{e}_r+\bm{e}_s}(\tau''')
[
\delta_{pj}\delta_{ki}\delta_{\tam''\tau'}\delta_{\tau''\tau}R_{ql}(\tam'',\tau''')C_{rs}(\tam''',\tam''')\\
&&{}+
\delta_{rj}\delta_{ki}\delta_{\tam'''\tau'}\delta_{\tau''\tau}R_{pl}(\tam'',\tau''')C_{qs}(\tam'',\tam''')
\nonumber\\
&&{}+
\delta_{pj}\delta_{li}\delta_{\tam''\tau'}\delta_{\tau'''\tau}
R_{rk}(\tam''',\tau'') C_{qs}(\tam'',\tam''')]
\nonumber\\
&=&
\frac{\delta_{\tau'\tam}}{\Delta t}\,\Delta t \sum_{\tau'''lqrs}
c_{\bar{e}_i,\bm{e}_j+\bm{e}_q}(\tau) R_{ql}(\tam,\tau''')c_{\bar{e}_l,\bm{e}_r+\bm{e}_s}(\tau''')C_{rs}(\tam''',\tam''')\\
&&{}+\sum_{lpqs}
c_{\bar{e}_i,\bm{e}_p+\bm{e}_q}(\tau) c_{\bar{e}_l,\bm{e}_j+\bm{e}_s}(\tap')
R_{pl}(\tam,\tap')C_{qs}(\tam,\tau')
\nonumber\\
&&{}+
\sum_{kqrs}
c_{\bar{e}_k,\bm{e}_j+\bm{e}_q}(\tap') c_{\bar{e}_i,\bm{e}_r+\bm{e}_s}(\tau)
R_{rk}(\tam,\tap') C_{qs}(\tau',\tam)
\nonumber
\end{eqnarray}
Comparing with eq.~(\ref{eq:L11_final}) shows that the first line in this expression precisely cancels $L_{11}+L_{12}=2L_{11}$. The second and third line are seen to be equal after relabelling $(l,p,q)\leftrightarrow (k,r,s)$ so that overall the first line of~\cref{eq:hatR2_intermediate2}
equals
\begin{equation}
L_{11}+L_{12}+L_{13}+L_{14}=
2\sum_{lpqs}
c_{\bar{e}_i,\bm{e}_p+\bm{e}_q}(\tau)
R_{pl}(\tam,\tap')C_{qs}(\tam,\tau')
c_{\bar{e}_l,\bm{e}_j+\bm{e}_s}(\tap')
\label{eq:L1_final}
\end{equation}

Moving on to the second line of~\cref{eq:hatR2_intermediate2}, one sees that one or three factors of $\dph$ are required in the average to make this nonzero; but in the first case all Wick pairings contain a factor of $B$, so vanish. The choice $\bm{n}=\bm{e}_q$, $\bm{n}'=\bm{e}_r+\bm{e}_s$ gives then as the first contribution to the second line
\begin{equation}
L_{21} = \frac{1}{2}\sum_{\tau''\tau'''klpqrs} c_{\bar{e}_k+\bar{e}_l,\bm{e}_q}(\tau'') c_{\bar{e}_p,\bm{e}_r+\bm{e}_s}(\tau''')
\frac{\partial}{\partial R_{ji}(\tau',\tau)}
\left\langle
\delta\tilde{\phi}_k(\tau'') \delta\tilde{\phi}_l(\tau'') \delta\tilde{\phi}_p(\tau''')
\delta \phi_q(\tau_-'') \delta\phi_r(\tau_-''')\delta\phi_s(\tau_-''')\right\rangle
\end{equation}
To avoid Wick pairings with unphysical factors of $B$, each $\dpt$ needs to be paired with a $\dph$, leaving only six pairings. Four out of these contain two factors of acausal response $R$ and so can also be dropped, leaving only
\begin{eqnarray}
&&\left\langle
\delta\tilde{\phi}_k(\tau'') \delta\tilde{\phi}_l(\tau'') \delta\tilde{\phi}_p(\tau''')
\delta \phi_q(\tau_-'') \delta\phi_r(\tau_-''')\delta\phi_s(\tau_-''')\right\rangle
= \nonumber \\
&&\ =\ R_{rk}(\tam''',\tau'')R_{sl}(\tam''',\tau'')R_{qp}(\tam'',\tau''') +
R_{sk}(\tam''',\tau'')R_{rl}(\tam''',\tau'')R_{qp}(\tam'',\tau''') + \ldots
\end{eqnarray}
If we now differentiate w.r.t.\ $R_{ji}(\tau',\tau)$, we must not have two factors of $R$ with opposite ordering of $\tau''$ and $\tau'''$ left as at least one of them would be acausal. The only derivative that contributes is therefore the one of the $R_{qp}$ factor, giving
\begin{eqnarray}
L_{21} &=& \frac{1}{2}\sum_{klrs} c_{\bar{e}_k+\bar{e}_l,\bm{e}_j}(\tap') c_{\bar{e}_i,\bm{e}_r+\bm{e}_s}(\tau)
[R_{rk}(\tam,\tap')R_{sl}(\tam,\tap') +
R_{sk}(\tam,\tap')R_{rl}(\tam,\tap')]
\\
&=& \sum_{klrs} c_{\bar{e}_i,\bm{e}_r+\bm{e}_s}(\tau)
R_{rk}(\tam,\tap')R_{sl}(\tam,\tap')c_{\bar{e}_k+\bar{e}_l,\bm{e}_j}(\tap')
\end{eqnarray}
because the two terms in the first line are equal.

The second contribution to the second line comes from the index vector assignment
$\bm{n}=\bm{e}_q+\bm{e}_r$, $\bm{n}'=\bm{e}_s$, which yields
\begin{equation}
L_{22} = \frac{1}{2}\sum_{\tau''\tau'''klpqrs} c_{\bar{e}_k+\bar{e}_l,\bm{e}_q}(\tau'') c_{\bar{e}_p,\bm{e}_r+\bm{e}_s}(\tau''')
\frac{\partial}{\partial R_{ji}(\tau',\tau)}
\left\langle
\delta\tilde{\phi}_k(\tau'') \delta\tilde{\phi}_l(\tau'') \delta\tilde{\phi}_p(\tau''')
\delta \phi_q(\tau_-'') \delta\phi_r(\tau_-'')\delta\phi_s(\tau_-''')\right\rangle
\end{equation}
and differs from $L_{21}$ only in the time argument of $\dph_r$. Because of this there are, out of the six Wick pairings not involving a factor of $B$, now four that contain at most one factor of an acausal response function:
\begin{eqnarray}
&&\left\langle
\delta\tilde{\phi}_k(\tau'') \delta\tilde{\phi}_l(\tau'') \delta\tilde{\phi}_p(\tau''')
\delta \phi_q(\tau_-'') \delta\phi_r(\tau_-'')\delta\phi_s(\tau_-''')\right\rangle
= \nonumber \\
&&\ =\ R_{rk}(\tam'',\tau'')R_{sl}(\tam''',\tau'')R_{qp}(\tam'',\tau''') +
R_{sk}(\tam''',\tau'')R_{rl}(\tam'',\tau'')R_{qp}(\tam'',\tau''')\nonumber\\
&&{}+R_{qk}(\tam'',\tau'')R_{sl}(\tam''',\tau'')R_{rp}(\tam'',\tau''') +
R_{sk}(\tam''',\tau'')R_{ql}(\tam'',\tau'')R_{rp}(\tam'',\tau''') + \ldots
\end{eqnarray}
To get a nonzero result after taking the derivative w.r.t.\ $R_{ji}(\tau',\tau)$, this derivative needs to act on the acausal response function in each term, i.e.\ the one with time arguments $(\tam'',\tau'')$. But this always leaves a product of two response functions with opposite ordering of their time arguments, so that overall $L_{22}=0$.

Adding $L_{21}$ to~\cref{eq:L1_final}, we finally have for the effective response field at $O(\alpha^2)$:
\begin{eqnarray}{\label{eq:hatR2_eqn}}
-\hat{R}^2_{ij}(\tau,\tau') &=&
2\sum_{klpq}c_{\bar{e}_i,\bm{e}_k+\bm{e}_l}(\tau) R_{kp}(\tam,\tap')C_{lq}(\tam,\tau')c_{\bar{e}_p,\bm{e}_j+\bm{e}_q}(\tap')
\nonumber
\\
&&{}+
\sum_{klpq} c_{\bar{e}_i,\bm{e}_k+\bm{e}_l}(\tau)R_{kp}(\tam,\tap')R_{lq}(\tam,\tap')c_{\bar{e}_p+\bar{e}_q,\bm{e}_j}(\tap')
\end{eqnarray}
where we have relabelled the summation indices so that $k,l$ correspond to the later time $\tau$ and $p,q$ to the earlier time $\tau'$, to make the correspondence between the two terms easier to see. Diagrammatically, one would write,

	\begin{equation}{\label{hatR2_T2}}
	\begin{split}
		-\hat{R}^2_{ij}(\tau,\tau')
&=  \sum_{klpq}  \left[ \; 2\quad \parbox{40mm}{ \begin{fmfgraph*}(100,40)
				\fmfleft{e1}
				\fmfright{e2}
\fmfv{label=\leglabel{i},label.dist=0,label.angle=180}{e1}
\fmfv{label=\leglabel{j},label.dist=2}{e2}
				\fmf{end_arrow_dash_two,tension=2}{v1,e1}
				\fmf{dashes,tension=2}{e2,v2}
				\fmf{fermion,left,tension=1.1,label=$R_{kp}$,label.dist=2}{v2,v1}
				\fmf{plain,right,tension=1.1,label=$C_{lq}$,label.dist=2}{v2,v1}
				\fmfv{label=$\tau$,label.dist=7,label.angle=105,decor.shape=circle, decor.filled=full, decor.size= 4thick}{v1}
				\fmfv{label=$\tau'$,label.dist=7, label.angle=70,decor.shape=circle, decor.filled=full, decor.size= 4thick}{v2}
		\end{fmfgraph*}} +
	 \quad \parbox{40mm}{ 	\begin{fmfgraph*}(100,40)
			\fmfleft{e1}
\fmfv{label=\leglabel{i},label.dist=0,label.angle=180}{e1}
			\fmfright{e2,e3,e4}
\fmfv{label=\leglabel{j},label.dist=2}{e3}
			\fmf{end_arrow_dash_two,tension=3.}{v1,e1}
\fmf{phantom}{e2,v2}
			\fmf{dashes}{e3,v2}
			\fmf{phantom}{e4,v2}
			\fmf{fermion,left,tension=1.5,label=$R_{kp}$,label.dist=2}{v2,v1}
			\fmf{fermion,right,tension=1.5,label=$R_{lq}$,label.dist=2}{v2,v1}
			\fmfv{label=$\tau$,label.dist=7,label.angle=105,decor.shape=circle, decor.filled=full, decor.size= 4thick}{v1}
			\fmfv{label=$\tau'$,label.dist=7, label.angle=70,decor.shape=circle, decor.filled=full, decor.size= 4thick}{v2}
\end{fmfgraph*}}
		\right] \\[10pt]
	\end{split}
\end{equation}
We can neglect the first term for the final $\hat R^2$ expression used in the work, see~\cref{main-hatR2} in the main text and the preceding discussion.

The $O(\alpha^2)$ contribution to the field $\hat{B}$ conjugate to the order parameter $B$ capturing $\dpt\,\dpt$ correlations can be calculated in an analogous matter, starting from \begin{align}
			&-\hat{B}^2_{ij}(\tau,\tau') =	\frac{1}{(\Delta t)^2}\frac{\partial \Gamma^2}{\partial B_{ij}(\tau',\tau)}\\
&= \sum_{\tau'',\tau'''} \sum_{\bar{m},\bar{n}} \sum_{\bar{m}',\bar{n}'} \frac{c^*_{\bar{m},\bar{n}}(\tau'')}{\bm m! \bm n!} \frac{c^*_{\bar{m}',\bar{n}'}(\tau''')}{\bm m'! \bm n'!}
			\frac{\partial}{\partial B_{ij}(\tau',\tau)}
\left\langle \prod_i  \delta\tilde{\phi}_i^{m_i}(\tau'') \delta\tilde{\phi}_i^{m_i'}(\tau''') \delta \phi_i^{n_i}(\tau_-'') \delta\phi_i^{n_i'}(\tau_-''')\right\rangle
\end{align}
where derivatives of the prefactors again do not contribute. Following then steps closely analogous to those above, one finds contributions to $\hat{B}^2_{ij}(\tau,\tau')$ of the form
\begin{equation}
\sum_{klpq}
c_{\bar{e}_i,\bar{e}_k + \bar{e}_l}(\tau) c_{\bar{e}_p+\bar{e}_j,\bar{e}_q}(\tau') C_{lq}(\tau_-,\tau'_-) R_{kp}(\tau_-,\tau')
\end{equation}
and
\begin{equation}
\sum_{klpq} c_{\bar{e}_i,\bar{e}_k + \bar{e}_l}(\tau) c_{\bar{e}_j,\bar{e}_p+\bar{e}_q}(\tau') C_{lq}(\tau_-,\tau'_-)  C_{kp}(\tau_-,\tau'_-)
\end{equation}
which diagrammatically correspond to
\begin{equation}
\parbox{37mm}{ 	\begin{fmfgraph*}(100,40)
		\fmfleft{e1}
		\fmfright{e2}
\fmfv{label=\leglabel{i},label.dist=0,label.angle=180}{e1}
		\fmfv{label=\leglabel{j},label.dist=0}{e2}
		\fmf{end_arrow_dash_two,tension=3.}{v1,e1}
		\fmf{end_arrow_dash_two,tension=3.}{v2,e2}
		\fmf{fermion,left,tension=1.5,label=$R_{kp}$,label.dist=2}{v2,v1}
		\fmf{plain,right,tension=1.5,label=$C_{lq}$,label.dist=2}{v2,v1}
		\fmfv{label=$\tau$,label.dist=7,label.angle=105,decor.shape=circle, decor.filled=full, decor.size= 4thick}{v1}
		\fmfv{label=$\tau'$,label.dist=7, label.angle=70,decor.shape=circle, decor.filled=full, decor.size= 4thick}{v2}
\end{fmfgraph*}}
\end{equation}
and
\vspace*{10pt}
\begin{equation}
	\parbox{40mm}{ 	\begin{fmfgraph*}(100,40)
			\fmfleft{e1}
			\fmfright{e2}
\fmfv{label=\leglabel{i},label.dist=0,label.angle=180}{e1}
			\fmfv{label=\leglabel{j},label.dist=0}{e2}
			\fmf{end_arrow_dash_two,tension=3}{v1,e1}
			\fmf{end_arrow_dash_two,tension=3}{v2,e2}
			\fmf{plain,left,tension=1.5,label=$C_{kp}$,label.dist=2}{v2,v1}
			\fmf{plain,right,tension=1.5,label=$C_{lq}$,label.dist=2}{v2,v1}
			\fmfv{label=$\tau$,label.dist=7,label.angle=105,decor.shape=circle, decor.filled=full, decor.size= 4thick}{v1}
			\fmfv{label=$\tau'$,label.dist=7, label.angle=70,decor.shape=circle, decor.filled=full, decor.size= 4thick}{v2}
	\end{fmfgraph*}}
\end{equation}
\end{fmffile}

As explained in the main text, we ignore both of these contributions as they contain factors of $C$, which is higher order in $\alpha$ (at least $O(\alpha$)) and so has no $O(1)$ contribution.

Finally we note that the $\hat C^2$ field can also be calculated explicitly. This vanishes as expected on general grounds, so we omit the details.

\newcommand{\be}{\begin{equation}}
\newcommand{\ee}{\end{equation}}
\newcommand{\bea}{\begin{eqnarray}}
\newcommand{\eea}{\end{eqnarray}}
\newcommand{\Hi}{H_{\rm int}}
\newcommand{\hg}{\hat{G}}
\newcommand{\eff}{_{\rm eff}}
\newcommand{\T}{^{\rm T}}
\newcommand{\half}{\frac{1}{2}}

\section{Relation of extended Plefka calculation to field theory and diagrammatics}
\label{app:structure}

We provide a compact version of the calculations for the extended Plefka free energy in this section, to make the combinatorics and structure somewhat more transparent as well as the connection to diagrammatic (2PI) field theory approaches.

\subsection{Setup}

Starting from the Doi-Peliti path integral for reaction networks, we collect all $\phi$ and $\tilde{\phi}$ into a vector $x$, via an additional index $\sigma=1,2$ to distinguish the field types so that $x_{\sigma,i,\tau}=\phi_i(\tau)$ for $\sigma=1$ and $x_{\sigma,i,\tau}=\tilde{\phi}_i(\tau)$ for $\sigma=2$. We use Roman letters $a=(\sigma,i,\tau)$ as a shorthand for the overall index on $x$. Then the general structure of the problem is given by a baseline Hamiltonian
\be
H_0 = -\half x\T G_0^{-1}x + k\T x
\ee
with the linear term coming from the creation terms in~\cref{main-eq:H0} baseline Hamiltonian in the main text.
Assuming at most binary reactions, one then has in addition interaction terms up to fourth order:
\be
\Hi = c_0 + c_1\T y + \half y\T c_2y + \frac{1}{6} c_3yyy + \frac{1}{24}c_4yyyy
\label{Hint}
\ee
Here the tensors $c_3$ and $c_4$ are assumed to be symmetric and contracted with the appropriate powers of $y=x-\mu$, where $\mu=\langle x\rangle$ is the average of $x$ in the full distribution $P(x)\propto e^{H_0+\Hi}$. The $c_n$ for $n\leq 3$ are $\mu$-dependent with $\partial_\mu c_n = c_{n+1}$ (after symmetrization as necessary).

\subsection{Extended Plefka expansion}

In an extended Plefka calculation we construct the action with linear and quadratic constraint terms
\be
\Xi = H_0 + \alpha \Hi + \theta\T (x-\mu) + \half \tr \hg(yy\T-G)
\label{Xi}
\ee
where $\hg$ can be chosen to be symmetric because antisymmetric terms would drop out anyway. The quadratic constraint ensures that $G=\langle yy\T\rangle$ is the connected correlation function propagator. Writing out the component indices in the trace term explicitly to get $\hg_{ab}(y_b y_a - G_{ba})/2$, with summation over repeated indices implied, shows that $\hg_{ab}$ is the conjugate variable to $G_{ba}$.

The Plefka free energy (field theoretically, the effective or Schwinger action)
\be
\Gamma = \extr_{\theta,\hg} \ln \int dx\,e^{\Xi}
\ee
is calculated perturbatively as
\be
\Gamma = \Gamma^0 + \alpha \Gamma^1 + \frac{\alpha^2}{2}\Gamma^2 + \ldots
\ee
The conjugate fields are
\be
\theta = -\partial_\mu \Gamma, \qquad \half\hg = -\partial_G \Gamma
\ee
(in principle there is a transpose in the expression for $\hg$ but the matrix is symmetric anyway by assumption) and are expanded in $\alpha$ in the analogous way, giving the effective fields
\be
\theta\eff = -\alpha \theta^1-\frac{\alpha^2}{2}\theta^2, \qquad
\hg\eff = -\alpha \hg^1-\frac{\alpha^2}{2}\hg^2
\ee

The resulting effective Hamiltonian for the $\alpha=0$ theory is, up to irrelevant constants,
\be
H\eff = H_0 + (k+\theta\eff)\T x + \half \tr \hg\eff yy\T
\ee
giving a Gaussian distribution $e^{H\eff}$ of $x$ with mean (found by maximization of $H\eff$ w.r.t.\ $x$) and covariance
\be
\mu=G_0(k+\theta\eff), \qquad G = (G_0^{-1}-\hg\eff)^{-1}
\label{mean_cov_alpha0}
\ee
This means that $\hg\eff$ plays the role of a self-energy in a Schwinger-Dyson equation. For the concrete case of dynamical field theories, $G_0^{-1}$ is normally a differential operator so it can be useful to write alternatively
\be
G_0^{-1}\mu = k+\theta\eff, \qquad
G_0^{-1}G = 1 + \hg\eff G
\label{Schwinger-Dyson}
\ee

Note that, as an alternative to~(\ref{Xi}), one could constrain the disconnected correlators, i.e.\ write
\be
\Xi = H_0 + \alpha \Hi + \theta'{}\T (x-\mu) + \half \tr \hg(xx\T-G-\mu\mu\T)
\label{Xi2}
\ee
The quadratic terms in $x$ are then as before, while the linear terms become equal if
\be
\theta' = \theta - \hg \mu
\ee
so that correspondingly
\be
\theta'\eff = \theta\eff - \hg\eff\mu
\label{thetaprime}
\ee
In this way of setting up the theory, the effective Hamiltonian at $\alpha=0$ becomes
\be
H\eff = H_0 + \theta'{}\T\eff x + \half \tr \hg\eff xx\T
\ee
and the mean of the corresponding Gaussian distribution obeys
\be
k+\theta\eff' + (\hg\eff-G_0^{-1})\mu =0
\label{mean_alpha0_again}
\ee
Because of~(\ref{thetaprime}) this retrieves the previous relation~(\ref{mean_cov_alpha0}), as it should: in both cases first and second order statistics of $x$ are being constrained, with the only difference being in the parametrization. The advantage of~(\ref{mean_alpha0_again}) from a field-theoretic perspective is that it can be written as
\be
\mu = (G_0^{-1}-\hg\eff)^{-1}(k+\theta\eff') = G(k+\theta\eff')
\ee
so that $\theta\eff'$ is the 1-point vertex function, from which the mean of $x$ is calculated by attaching a propagator line.

\subsection{First order in $\alpha$}

From the general Plefka identity from~\cref{main-eq:Gamma1} in the main text,
with the average taken over the $\alpha=0$ theory,
\be
\Gamma^1 = \langle \Hi\rangle = c_0 + \half \tr c_2 G + \frac{1}{8}c_4 GG
\ee
giving
\be
-\theta^1 = \partial_\mu \Gamma^1 = c_1 + \frac{1}{2}c_3 G
\ee
because $c_4$ does not depend on $\mu$, while
\be
-\half \hg^1 = \half c_2 + \frac{1}{4}c_4 G, \qquad -\hg^1 = c_2 + \frac{1}{2}c_4 G
\ee
The second term here is the typical one-loop diagram from $\phi^4$-theory, with the expected symmetry factor of $1/2$. Similarly the second term in $\theta^1$ is the simplest tadpole diagram.

To linear order in $\alpha$ one therefore has
\be
\theta\eff = \alpha c_1 + \frac{\alpha}{2} c_3 G, \qquad
\hg\eff = \alpha c_2 + \frac{\alpha}{2}c_4 G
\ee
and one has to solve first for $G$ using
\be
G_0^{-1}G
 = 1 + (\alpha c_2 + \frac{\alpha}{2}c_4 G)G
\ee
then insert $G$ into $\theta\eff$ to eventually find $\mu$.

\subsection{Second order in $\alpha$}

Again from the general Plefka identities, see~\cref{gamma_2_exponent_calculation}
one has,
\be
\Gamma^2 = \left\langle
\left(\Delta \frac{d\Xi}{d\alpha}\right)^2
\right\rangle = \left\langle
\left(\frac{d\Xi}{d\alpha}\right)^2
\right\rangle -
\left\langle
\frac{d\Xi}{d\alpha}
\right\rangle^2
\label{Gamma2}
\ee
where the derivative is taken at $\alpha=0$, giving
\bea
\frac{d\Xi}{d\alpha}&=&\Hi + (\theta^1)\T(x-\mu)+\half \tr \hg^1(yy\T-G)
\eea
Inserting the explicit forms of $\Hi$, $\theta^1$ and $\hg^1$ yields
\bea
\frac{d\Xi}{d\alpha}&=&
c_0^* + c_1^*{}\T y
+\half y\T c_2^* y
+\frac{1}{6}c_3 yyy+\frac{1}{24}c_4 yyyy
\eea
where
\be
c_0^* = c_0-\half \tr \hg^1 G, \qquad
c_1^* = - \half c_3 G, \qquad
c_2^* = - \frac{1}{2}c_4 G
\ee
The constant term $c_0^*$ will drop out when the variance in~(\ref{Gamma2}) is calculated. The two lowest square terms in this variance are
\bea
\langle (c_1^*{}\T y)^2\rangle &=& c_1^*{}\T Gc_1^* \ = \
c_{1,a}^*c_{1,b}^*G_{ab}
\\
\left \langle \left(\Delta \half y\T c_2^* y\right)^2 \right \rangle &=& \frac{1}{4}\,2c_{2,ab}^*c_{2,cd}^* G_{ac}G_{bd}
= \frac{1}{2}c_{2,ab}^*c_{2,cd}^* G_{ac}G_{bd}
\eea
where the factor 2 in the second line comes from the two equal Wick pairings; the third is removed by the subtraction of the mean. Summation over repeated indices is implied.

Similarly one has for the higher order square terms
\bea
\left \langle \left(\frac{1}{6} c_3yyy\right)^2 \right \rangle &=& \frac{1}{36}c_{3,abc}c_{3,def}[6G_{ad}G_{be}G_{cf}+9
G_{ab}G_{cd}G_{ef}]
\\
\left \langle \left(\Delta \frac{1}{24} c_4yyyy\right)^2 \right \rangle &=& \frac{1}{576}c_{4,abcd}c_{4,efgh}[24G_{ae}G_{bf}G_{cg}G_{dh}+72 G_{ab}G_{ef}G_{cg}G_{dh}]
\eea
where the terms in the first line corresponds to three respectively one pairing between the first and second factor in the square on the l.h.s., and to four and two pairings in the second line.

Finally there are the two cross terms that give even powers of $y$:
\bea
2\left\langle \left(c_1^*{}\T y\right)\left(\frac{1}{6}c_3yyy\right)
\right\rangle  &=&
2\,\frac{1}{6}\,3c_{1,a}^* c_{3,bcd}G_{ab}G_{cd} \\
2\left\langle \left(\Delta \half y c_2^*{}\T y\right)\left(\Delta \frac{1}{24}c_4yyyy\right)
\right\rangle  &=&
2\,\frac{1}{48}\,12 c_{2,ab}^*c_{4,cdef}G_{ac}G_{bd}G_{ef}
\eea
Overall one has therefore for the $O(\alpha^2)$ coefficient in the expansion of the Plefka free energy:
\bea
\Gamma^2 &=&
c_{1,a}^*c_{1,b}^*G_{ab}
+ \frac{1}{2}c_{2,ab}^*c_{2,cd}^* G_{ac}G_{bd}+
c_{1,a}^* c_{3,bcd}G_{ab}G_{cd}
\nonumber\\
&&{}+ \frac{1}{2} c_{2,ab}^*c_{4,cdef}G_{ac}G_{bd}G_{ef}
+
\frac{1}{6}c_{3,abc}c_{3,def}G_{ad}G_{be}G_{cf}+
\frac{1}{4} c_{3,abc}c_{3,def}
G_{ab}G_{cd}G_{ef}
\nonumber \\
&&{}+ \frac{1}{24}c_{4,abcd}c_{4,efgh}G_{ae}G_{bf}G_{cg}G_{dh}
+ \frac{1}{8}c_{4,abcd}c_{4,efgh} G_{ab}G_{ef}G_{cg}G_{dh}
\eea
To simplify one can now insert the explicit expressions for $c_1^*$ and $c_2^*$:
\bea
\Gamma^2 &=&
\frac{1}{4}c_{3,acd}c_{3,bef}G_{ab}G_{cd}G_{ef}
+ \frac{1}{8}c_{4,abef}c_{4,cdgh}G_{ac}G_{bd}G_{ef}G_{gh}-\frac{1}{2}
c_{3,aef}c_{3,bcd}G_{ab}G_{cd}G_{ef}
\nonumber\\
&&{}- \frac{1}{4} c_{4,abgh}c_{4,cdef}G_{ac}G_{bd}G_{ef}G_{gh}
+
\frac{1}{6}c_{3,abc}c_{3,def}G_{ad}G_{be}G_{cf}+
\frac{1}{4} c_{3,abc}c_{3,def}
G_{ab}G_{cd}G_{ef}
\nonumber \\
&&{}+ \frac{1}{24}c_{4,abcd}c_{4,efgh}G_{ae}G_{bf}G_{cg}G_{dh}
+ \frac{1}{8}c_{4,abcd}c_{4,efgh} G_{ab}G_{ef}G_{cg}G_{dh}
\eea
One observes that the three terms where exactly one factor of $G$ connects two $c_3$-vertices cancel (with prefactors $1/4-1/2+1/4=0$). Similarly the three $c_4 c_4$ terms where exactly two factors of $G$ connect the two $c_4$-vertices cancel (prefactors $1/8-1/4+1/8$). This leaves for $\Gamma^2$ only diagrams with at least three propagator connections between two vertices (2PI diagrams):
\bea
\Gamma^2 &=&
\frac{1}{6}c_{3,abc}c_{3,def}G_{ad}G_{be}G_{cf}
+ \frac{1}{24}c_{4,abcd}c_{4,efgh}G_{ae}G_{bf}G_{cg}G_{dh}
\eea

From $\Gamma^2$ one can now obtain the effective fields. Because $c_4$ is independent of $\mu$, only the first term contributes to $\theta^2$:
\be
-\theta^2_g = \partial_{\mu_g}\Gamma^2 = \frac{1}{3}c_{4,gabc}c_{3,def}G_{ad}G_{be}G_{cf}
\ee
For the self-energy contribution one gets
\be
-\half \hg^2_{ij} = \partial_{G_{ij}}\Gamma^2
= \frac{1}{2}c_{3,ibc}c_{3,jef}G_{be}G_{cf}
+ \frac{1}{6}c_{4,ibcd}c_{4,jfgh}G_{bf}G_{cg}G_{dh}
\ee
or after relabelling
\be
-\hg^2_{ab} = c_{3,acd}c_{3,bef}G_{ce}G_{df}
+ \frac{1}{3}c_{4,acde}c_{4,bfgh}G_{cf}G_{fg}G_{eh}
\ee
Combining with the $O(\alpha)$ result gives the effective field to $O(\alpha^2)$:
\be
\theta\eff{}_{,a}
=-\alpha\theta^1_a - \frac{\alpha^2}{2}\theta^2_a =
\alpha c_{1,a} + \frac{\alpha}{2}c_{3,abc}G_{bc} + \frac{\alpha^2}{6}c_{4,abcd}c_{3,efg}G_{be}G_{cf}G_{dg}
\ee
and for the self-energy to $O(\alpha^2)$:
\be
\hg\eff{}_{,ab} = - \alpha \hg^1_{ab} - \frac{\alpha^2}{2} \hg^2_{ab} =
\alpha c_{2,ab} + \frac{\alpha}{2}c_{4,abcd}G_{cd}+
\frac{\alpha^2}{2} c_{3,acd}c_{3,bef}G_{ce}G_{df}
+ \frac{\alpha^2}{6}c_{4,acde}c_{4,bfgh}G_{cf}G_{fg}G_{eh}
\label{Geff_final}
\ee
The prefactors are consistent with the usual diagrammatic rules, e.g.\ the last term in the self-energy has three equivalent internal propagator lines so a symmetry factor of $1/3!$; the second term similarly has two such lines.

\subsection{Translation back to Doi-Peliti notation}

To illustrate the connection of the above expressions to the original notation in the main text, consider the third order term in~(\ref{Hint}). This is originally
\be
\Delta t\sum_\tau \sum_{k\leq l} \sum_m \frac{1}{1+\delta_{kl}}[
c_{\bm{e}_m,\bm{e}_k+\bm{e}_l}(\tau)\delta\tilde{\phi}_m(\tau)\delta\phi_k(\tau_-)\delta\phi_l(\tau_-)
+
c_{\bm{e}_k+\bm{e}_l,\bm{e}_m}(\tau)\delta\tilde{\phi}_k(\tau)\delta\tilde{\phi}_l(\tau)\delta\phi_m(\tau_-)]
\ee
or after symmetrization in $k,l$
\be
\Delta t \half \sum_{\tau klm}[
c_{\bm{e}_m,\bm{e}_k+\bm{e}_l}(\tau)\delta\tilde{\phi}_m(\tau)\delta\phi_k(\tau_-)\delta\phi_l(\tau_-)
+
c_{\bm{e}_k+\bm{e}_l,\bm{e}_m}(\tau)\delta\tilde{\phi}_k(\tau)\delta\tilde{\phi}_l(\tau)\delta\phi_m(\tau_-)]
\ee
In the general form above this has to equal
\be
\frac{1}{6}\sum_{abc}c_{3,abc}y_a y_b y_c
\ee
where we recall that each index corresponds to a triple of original indices, e.g.\ $a=(\alpha,k,\tau)$. The above equality can be achieved by setting
\be
c_{3,abc} =
6\Delta t \half \sum_{\tau klm}[
c_{\bm{e}_m,\bm{e}_k+\bm{e}_l}(\tau)
\delta_{a,(2,m,\tau)}\delta_{b,(1,k,\tau_-)}\delta_{c,(1,l,\tau_-)}
+
c_{\bm{e}_k+\bm{e}_l,\bm{e}_m}(\tau)
\delta_{a,(2,k,\tau)}\delta_{b,(2,l,\tau)}\delta_{c,(1,m,\tau_-)}
]
\ee
To satisfy the assumption of symmetry in $(a,b,c)$ this still has to be symmetrized, giving (note that only three rather than six permutations are needed because of the symmetry between $k$ and $l$)
\bea
c_{3,abc} &=&
\Delta t \sum_\tau \sum_{klm}[
c_{\bm{e}_m,\bm{e}_k+\bm{e}_l}(\tau)\times
\\
&&\times(\delta_{a,(2,m,\tau)}\delta_{b,(1,k,\tau_-)}\delta_{c,(1,l,\tau_-)}
+\delta_{b,(2,m,\tau)}\delta_{a,(1,k,\tau_-)}\delta_{c,(1,l,\tau_-)}
+\delta_{c,(2,m,\tau)}\delta_{a,(1,k,\tau_-)}\delta_{b,(1,l,\tau_-)})
\nonumber\\&&{}+
c_{\bm{e}_k+\bm{e}_l,\bm{e}_m}(\tau)\times
\nonumber\\
&&\times(\delta_{a,(2,k,\tau)}\delta_{b,(2,l,\tau)}\delta_{c,(1,m,\tau_-)}+
\delta_{b,(2,k,\tau)}\delta_{c,(2,l,\tau)}\delta_{a,(1,m,\tau_-)}+
\delta_{a,(2,k,\tau)}\delta_{c,(2,l,\tau)}\delta_{b,(1,m,\tau_-)})\nonumber
]
\eea
One sees, then, that the nonzero entries of the symmetrized $c_3$ are just the original coefficients $c$, up to a factor of $\Delta t$. By a similar argument one finds for the fourth order coefficients
\bea
c_{4,abcd}&=&\Delta t\sum_{\tau k l p q} c_{\bm{e}_k+\bm{e}_l,\bm{e}_p+\bm{e}_q}(\tau) \times \nonumber\\
&&\times(
\delta_{a,(2,k,\tau)}\delta_{b,(2,l,\tau)}
\delta_{c,(1,p,\tau_-)}
\delta_{d,(1,q,\tau_-)}
+
\delta_{a,(2,k,\tau)}\delta_{c,(2,l,\tau)}
\delta_{b,(1,p,\tau_-)}
\delta_{d,(1,q,\tau_-)}\nonumber\\
&&{}+
\delta_{a,(2,k,\tau)}\delta_{d,(2,l,\tau)}
\delta_{b,(1,p,\tau_-)}
\delta_{c,(1,q,\tau_-)}
+
\delta_{b,(2,k,\tau)}\delta_{c,(2,l,\tau)}
\delta_{a,(1,p,\tau_-)}
\delta_{d,(1,q,\tau_-)}
\nonumber\\
&&{}+
\delta_{b,(2,k,\tau)}\delta_{d,(2,l,\tau)}
\delta_{a,(1,p,\tau_-)}
\delta_{c,(1,q,\tau_-)}
+
\delta_{c,(2,k,\tau)}\delta_{d,(2,l,\tau)}
\delta_{a,(1,p,\tau_-)}
\delta_{b,(1,q,\tau_-)}
)
\eea
so that these again match, for the nonzero elements, the coefficients defined in the main text.

For the propagator $G$ one has e.g.\ $G_{(1,i,\tau),(2,j,\tau')}=\langle \delta\tilde{\phi}_i(\tau)\delta\phi_j(\tau') \rangle = R_{ij}(\tau,\tau')$ so that one can write $G$ in block form as
\be
G = \left( \begin{array}{cc}
\bm{C} & \bm{R}\\
\bm{R}\T & \bm{B}
\end{array}\right)
\ee
where the diagonal blocks contain the correlations $C_{ij}(\tau,\tau') =
\langle \delta\phi_i(\tau)\delta\phi_j(\tau') \rangle$ and  $B_{ij}(\tau,\tau')=\langle \delta\tilde{\phi}_i(\tau)\delta\tilde{\phi}_j(\tau') \rangle$, with $\bm{B}=0$ for the physical solution of the dynamics. For the self-energy $\hg\eff$ one can similarly identify the block structure as
\be
\hg\eff = (\Delta t)^2 \left( \begin{array}{cc}
\hat{\bm{C}} & \hat{\bm{R}}\T\\
\hat{\bm{R}} & \hat{\bm{B}}
\end{array}\right)
\ee
In the Schwinger-Dyson equation~(\ref{Schwinger-Dyson}) the nontrivial term on the r.h.s.\ is then
\be
\hg\eff G = (\Delta t)^2 \left( \begin{array}{cc}
\hat{\bm{C}} & \hat{\bm{R}}\T\\
\hat{\bm{R}} & \hat{\bm{B}}
\end{array}\right) \left( \begin{array}{cc}
\bm{C} & \bm{R}\\
\bm{R}\T & \bm{B}
\end{array}\right) = (\Delta t)^2
\left( \begin{array}{cc}
\hat{\bm{C}}\bm{C} + \hat{\bm{R}}\T\bm{R}\T &
\hat{\bm{C}}\bm{R} + \hat{\bm{R}}\T\bm{B}
\\
\hat{\bm{R}}\bm{C} + \hat{\bm{B}}\bm{R}\T
&
\hat{\bm{R}}\bm{R} + \hat{\bm{B}}\bm{B}\T
\end{array}\right)
\ee
so that the $(2,2)$ block (with $\bm{B}=0$) matches the equation~\cref{main-FD_eqn_mixed_plefka}
for $R$ in the main text and similarly the $(2,1)$ block matches the equation~\cref{main-C_update_multispecies_explicit_plefka}
for $C$.

With the above identifications one can translate the general formulas from this section into the notation from the main text. Taking the $O(G^2)$ term from~(\ref{Geff_final}) as an example, the $(2,1)$-component of $\hg\eff{}_{,ab}$ is the response part of the self-energy as shown above, denoted $\hat{R}^{\rm eff}_{ij}(\tau,\tau')$ in the original notation.
Picking out the corresponding terms, i.e.\ the ones with $a=(2,i,\tau)$ and $b=(1,j,\tau')$ gives for $\alpha=1$ the contribution
\bea
\hg\eff{}_{,ab} &=& \half \sum_{cdef}
\Bigl\{\Delta t \sum_{\tau''} \sum_{klm}[
c_{\bm{e}_m,\bm{e}_k+\bm{e}_l}(\tau'')\delta_{a,(2,m,\tau'')}\delta_{c,(1,k,\tau''_-)}\delta_{d,(1,l,\tau''_-)}
\\&&{}+
c_{\bm{e}_k+\bm{e}_l,\bm{e}_m}(\tau'')
(\delta_{a,(2,k,\tau'')}\delta_{c,(2,l,\tau'')}\delta_{d,(1,m,\tau''_-)}+
\delta_{a,(2,k,\tau'')}\delta_{d,(2,l,\tau'')}\delta_{c,(1,m,\tau''_-)}
)
]
\Bigr\}\times \nonumber\\
&&\times \Bigl\{\Delta t \sum_{\tau'''} \sum_{pqr}[
c_{\bm{e}_r,\bm{e}_p+\bm{e}_q}(\tau''')(
\delta_{e,(2,r,\tau''')}\delta_{b,(1,p,\tau'''_-)}\delta_{f,(1,q,\tau'''_-)}+
\delta_{f,(2,r,\tau''')}\delta_{b,(1,p,\tau'''_-)}\delta_{e,(1,q,\tau'''_-)}
)
\nonumber\\&&{}+
c_{\bm{e}_p+\bm{e}_q,\bm{e}_r}(\tau''')\delta_{e,(2,p,\tau''')}\delta_{f,(2,q,\tau''')}\delta_{b,(1,r,\tau'''_-)}]\Bigr\}G_{ce}G_{df}
\nonumber
\eea
The Kronecker deltas force $\tau''=\tau$ and $\tau'''=\tau'_+$. For the field type indices of $(c,d)$ the expression above contains the options $(1,1)$, $(1,2)$, $(2,1)$; similarly for $(e,f)$ the three terms are $(1,2)$, $(2,1)$, $(2,2)$. Only the terms with $(c,d)=(1,1)$ lead to a nonzero product $G_{ce}G_{df}$, which otherwise contains at least one factor of $B=G_{22}$ or a product of response functions with opposite time ordering, of which one must therefore be acausal. All terms in the second line can therefore be omitted, giving
\bea
\hg\eff{}_{,ab} &=& \frac{(\Delta t)^2}{2} \sum_{cdef}
\sum_{kl}
c_{\bm{e}_i,\bm{e}_k+\bm{e}_l}(\tau)\delta_{c,(1,k,\tau_-)}\delta_{d,(1,l,\tau_-)}
\times \nonumber\\
&&\times \Bigl\{\sum_{pqr}[
c_{\bm{e}_r,\bm{e}_p+\bm{e}_q}(\tau'_+)(
\delta_{e,(2,r,\tau'_+)}\delta_{b,(1,p,\tau')}\delta_{f,(1,q,\tau')}+
\delta_{f,(2,r,\tau'_+)}\delta_{b,(1,p,\tau')}\delta_{e,(1,q,\tau')}
)
\nonumber\\&&{}+
c_{\bm{e}_p+\bm{e}_q,\bm{e}_r}(\tau'_+)\delta_{e,(2,p,\tau'_+)}\delta_{f,(2,q,\tau'_+)}\delta_{b,(1,r,\tau')}]\Bigr\}G_{ce}G_{df}
\\
&=& \frac{(\Delta t)^2}{2}
\sum_{kl}
c_{\bm{e}_i,\bm{e}_k+\bm{e}_l}(\tau) \Bigl\{\sum_{qr}
c_{\bm{e}_r,\bm{e}_j+\bm{e}_q}(\tau'_+)
G_{(1,k,\tau_-),(2,r,\tau'_+)}G_{(1,l,\tau_-),(1,q,\tau')}
\nonumber\\
&&{}+\sum_{qr}c_{\bm{e}_r,\bm{e}_j+\bm{e}_q}(\tau'_+)G_{(1,k,\tau_-),(1,q,\tau')}G_{(1,l,\tau_-),(2,r,\tau'_+)}
\nonumber\\
&&{}+
\sum_{pq}
c_{\bm{e}_p+\bm{e}_q,\bm{e}_j}(\tau'_+)
G_{(1,k,\tau_-),(2,p,\tau'_+)}G_{(1,l,\tau_-),(2,q,\tau'_+)}
\Bigr\}
\eea
Dividing by $(\Delta t)^2$ and writing the propagator entries in terms of $R$ and $C$ gives the corresponding contribution to $\hat{R}^{\rm eff}_{ij}(\tau,\tau')$ as
\bea
\hat{R}^{\rm eff}_{ij}(\tau,\tau') &=& \frac{1}{2}\sum_{kl}
c_{\bm{e}_i,\bm{e}_k+\bm{e}_l}(\tau) \Bigl\{\sum_{qr}
c_{\bm{e}_r,\bm{e}_j+\bm{e}_q}(\tau'_+)
R_{kr}(\tau_-,\tau'_+)C_{lq}(\tau_-,\tau')
\nonumber\\
&&{}+\sum_{qr}
c_{\bm{e}_r,\bm{e}_j+\bm{e}_q}(\tau'_+)
C_{kq}(\tau_-,\tau') R_{lr}(\tau_-,\tau'_+)
\nonumber\\
&&{}+
\sum_{pq}
c_{\bm{e}_p+\bm{e}_q,\bm{e}_j}(\tau'_+)
R_{kp}(\tau_-,\tau'_+)R_{lq}(\tau_-,\tau'_+)\Bigr\}\\
&=&
\frac{1}{2}\sum_{klpq}
c_{\bm{e}_i,\bm{e}_k+\bm{e}_l}(\tau) \Bigl\{2
R_{kp}(\tau_-,\tau'_+)C_{lq}(\tau_-,\tau') c_{\bm{e}_p,\bm{e}_j+\bm{e}_q}(\tau'_+)
\nonumber\\&&{}
+
R_{kp}(\tau_-,\tau'_+)R_{lq}(\tau_-,\tau'_+)c_{\bm{e}_p+\bm{e}_q,\bm{e}_j}(\tau'_+)\Bigr\}
\eea
because the two first lines with the terms of the form $RC$ are identical up to relabelling of $k$ and $l$. This result is consistent with~\cref{eq:hatR2_eqn}.
Again the symmetry factors are consistent with the usual rules, i.e.\ the $RR$-term has a factor of $1/2$ because of the symmetry between the two response function propagators, while the $RC$-term does not.

One can proceed similarly to find $\hat{B}^{\rm eff}$, by picking out the terms from $\hg\eff{}_{,ab}$ with $a=(2,i,\tau)$ and $b=(2,j,\tau')$; we omit the details here for brevity and because, as explained in the text, we drop the $O(\alpha^2)$ contributions to this effective field within in our approximation methods.

\section{Computational complexity}

\subsection{Computational complexity of gSBR}{\label{sec:computational_complexity_gSBR}}

The computational complexity of gSBR is dominated by evaluating the operator inverse resulting from the geometric bubbles series, cf~\cref{main-eq_hatR2_contnuous} of the main paper.
In discrete time with constant $\Delta t$ there are $\zeta=t/\Delta t$ time steps, so $\chi$ is a matrix of dimension $P^2(\zeta+1)$ where $P$ is the total number of species. In a naive implementation, computing its inverse then has a computational complexity of $O(P^4 \zeta^3)$. This comes from an $O((P^2 \zeta)^2)$ effort for the inversion of the (temporally) triangular matrix $\chi$, which needs to be performed for every one of the $\zeta$ time steps. Numerically, it is therefore helpful to reduce the factor $P^2$ in the size of the matrix $\chi$ by going back to ordered species index pairs $k\leq l$, compensating with the appropriate combinatorial factors. Depending on the reaction network structure, there can also be many species pairs $kl$ that give zero entries in $\chi$, reducing its effective size further.

\begin{figure}
	\includegraphics[width=0.45\linewidth]{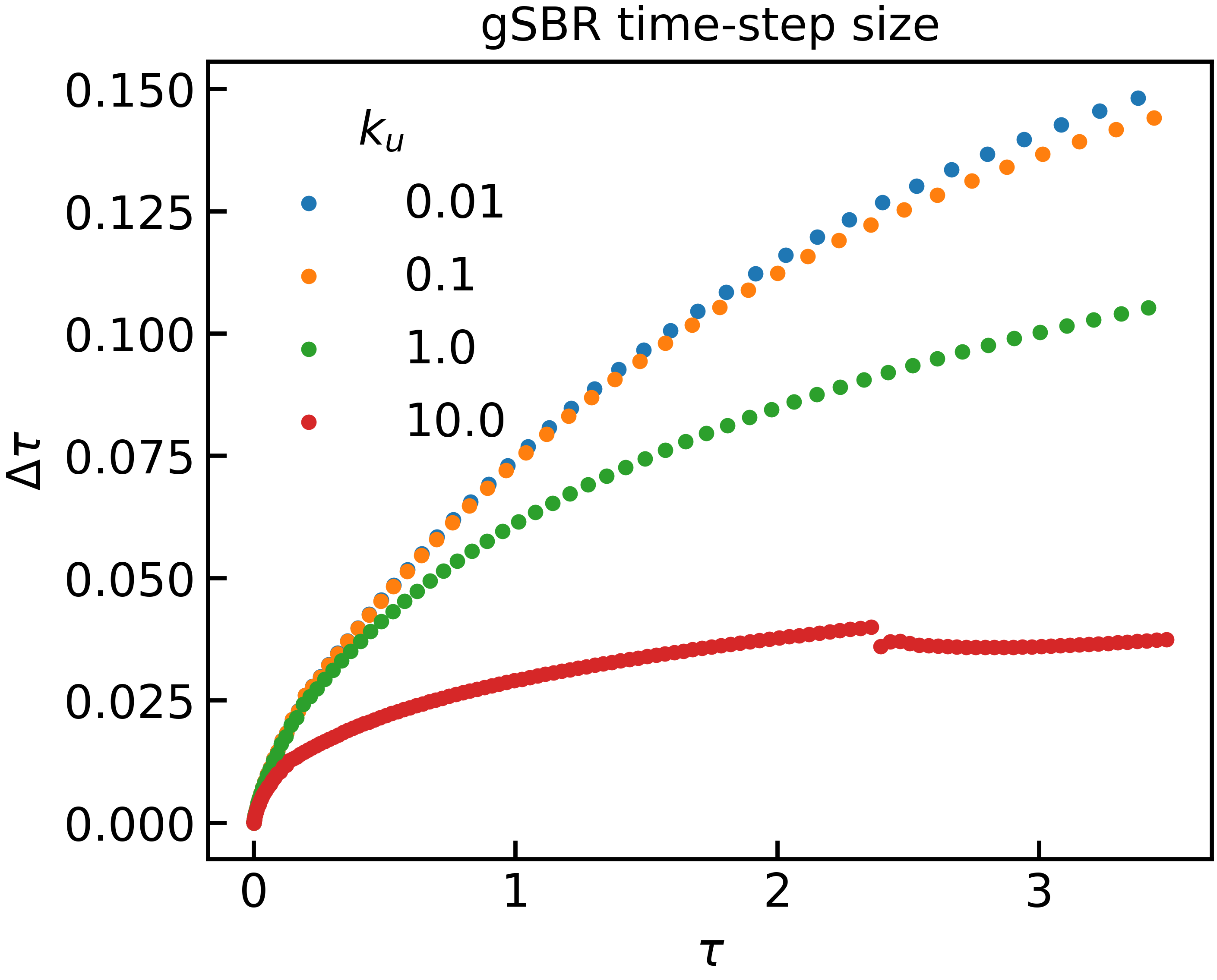}
	\includegraphics[width=0.45\linewidth]{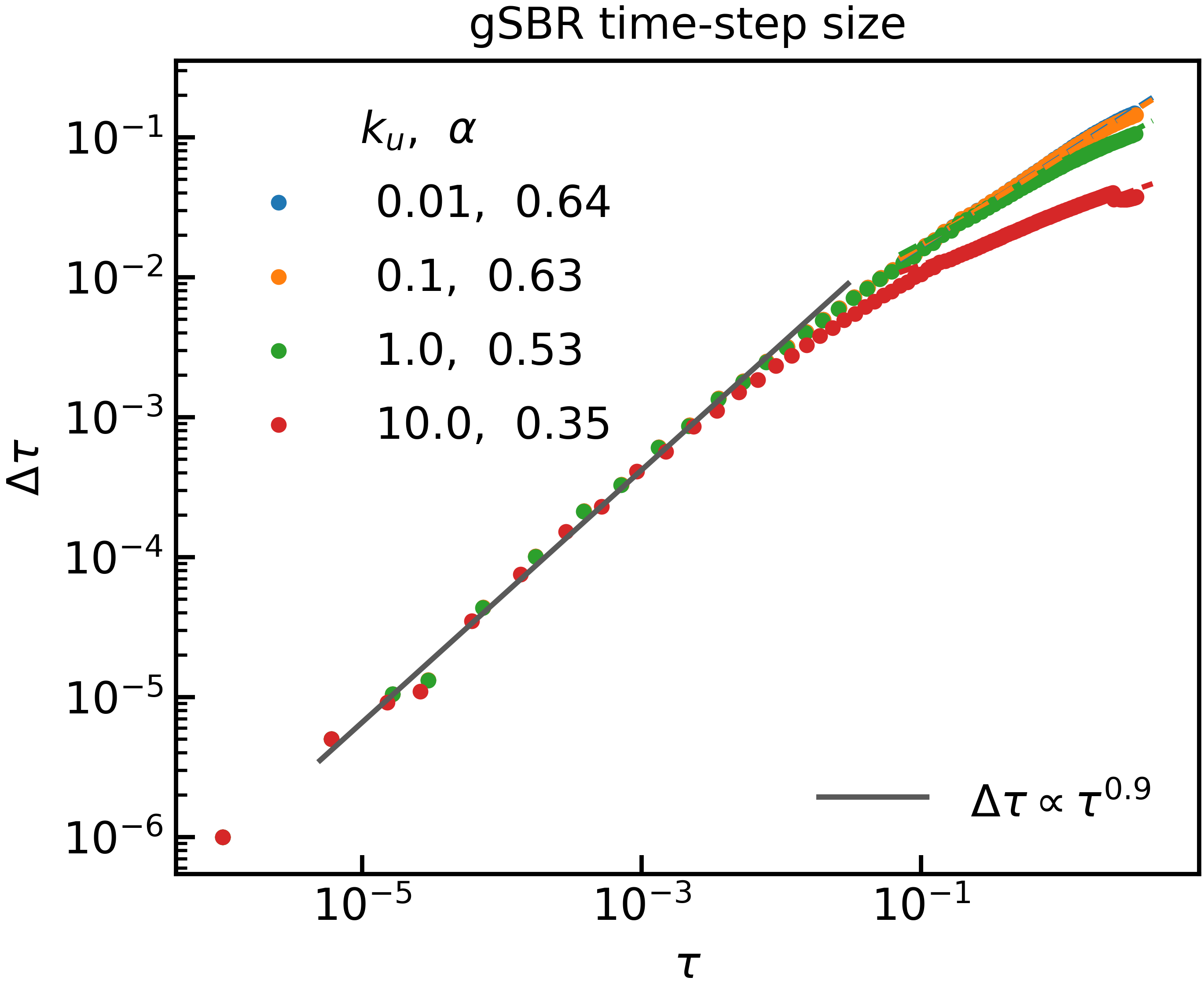}
	\caption[time-step size gene]{Time-step size as a function of time of the dynamics according to the adaptive time KB solver with the gSBR method for the gene regulation model with the standard parameter set at different $k_u$ values on (left) linear scale and (right) log-log scale. For larger $k_u$ values the solver takes smaller time steps. On the log-log scale the step size follows a power law, $\Delta \tau \propto \tau^{\alpha}$: during the initial ramp all $k_u$ collapse onto a common
		$\alpha \approx 0.9$ (grey line), while at late times the fitted exponent
		(dashed lines, values in the legend) decreases from $\alpha = 0.64$ at
		$k_u = 10^{-2}$ to $\alpha = 0.35$ at $k_u = 10$. Since $\alpha < 1$, the
		number of steps needed to reach a time $T$ grows only as $T^{1-\alpha}$,
		which is the origin of the speed-up over a uniform grid quoted in the~\cref{sec:computational_complexity_gSBR}.}
	\label{fig_gene_time}
\end{figure}

In practice the solver does not use a constant $\Delta t$ but the adaptive
step-size control of the Kadanoff--Baym integrator, and this changes how
$\zeta$ grows with the integration time. Empirically the step size follows a
power law in the elapsed time, $\Delta t(t) \simeq A\,t^{\alpha}$, with
$\alpha \approx 0.9$ during the initial transient --- i.e.\ an essentially
logarithmic grid --- crossing over to $\alpha \approx 0.35$--$0.65$ once the
dynamics has relaxed, the smaller value corresponding to the faster networks
(cf.\ \cref{fig_gene_time}). The number of steps required to reach a time $T$
is therefore only sublinear in $T$,
\begin{equation}
	\zeta(T) = \int_{t_0}^{T} \frac{d t}{\Delta t(t)}
	\simeq \frac{T^{1-\alpha} - t_0^{1-\alpha}}{A\,(1-\alpha)} ,
\end{equation}
rather than $\zeta = T/\Delta t$, and the saving over a uniform grid of
spacing $\Delta t(t_0)$ grows as a power of the dynamic range,
$\zeta_{\mathrm{unif}}/\zeta \simeq (1-\alpha)\,(T/t_0)^{\alpha}$.
For the gene-regulation network (cf.\ \cref{sec:example_SM_gene,fig_gene_time}) ($P = 3$, $T = 3.5$, absolute and relative tolerances $10^{-3}$ and $10^{-2}$) the adaptive solver uses $\zeta = 67$
steps at $k_u = 10^{-2}$ and $\zeta = 144$ at $k_u = 10$, whereas a uniform
grid resolving the same early-time transient,
$\Delta t = \Delta t(t = 10^{-2}) \approx 3 \times 10^{-3}$, would need
$\zeta \approx 1.1 \times 10^{3}$, between $10$ and $17$ times more steps.
Given the $O(P^4 \zeta^3)$ cost and $O(P^4 \zeta^2)$ memory of the inversion,
this amounts to \textit{three orders of magnitude fewer operations and two orders of
magnitude less memory at equal accuracy}.

\subsection{SBR series summation}{\label{sec:simpleSBR_sum}}
In the simplified version of gSBR that we call SBR, there is no mixing of the responses from different reaction channels.
This is computationally less expensive, since one ends up with geometric series involving matrices of reduced space.
We refer the reader to~\cite{harsh_accurate_2023,harsh_dynamics_2024} for the details of this SBR geometric series; while this is very similar to the gSBR geometric series described in the main text, there are some subtle differences regarding the set of index vectors to be summer over for the $\hat R$ and $\tilde \theta$ fields.

	\section{Numerical solution of master equation}{\label{sec:app_ME}}

	In this section, we summarize our method for the numerical solutions of the master equation. For the particular case of single species and single step master equations -- where only one molecule can be created or destroyed in one time step -- a more efficient method~\cite{smith_general_2015} is available, which relies on calculating only the eigenvalues (analytically or numerically) of the master operator. We instead use the significantly more general finite space projection method~\cite{munsky_finite_2006} here. The errors of this method are well controlled and its convergence to the true solution can be shown for many cases~\cite{munsky_finite_2006}.

One starts from the master equation for the time evolution of the probabilities $P(\bm n,\tau)$ of having $\bm n$ number of molecules in the reaction volume, see~\cref{main-Master_equation}
of the main text.
To make the state space finite and thus allow a numerical approach to the problem, one imposes an upper bound on the maximum copy number, calling this $\bm n^{\text{max}}=(n_1^{\text{max}}, \ldots, n_P^{\text{max}})$. The state-space of the system is thus of size
$\prod_i (n^{\text{max}}_i +1)$. Clearly the $n^{\text{max}}_i$ have to be chosen large enough to cover the bulk of the probability throughout the dynamics. For the numerics for the gene regulation system (\cref{sec:example_SM_gene}), for example, we use $(n_G^{\text{max}},n_{G^*}^{\text{max}},n_P^{\text{max}}) = (10,10,20)$.

	 Calling the vector of state probabilities
	$| P(\tau) \rangle = | P(\bm 0,\tau), P(\bm e_i,\tau), \ldots, P(\bm n^{\text{max}},\tau) \rangle$, the master equation can then be written in matrix-vector form as
	\begin{equation}
		\partial_\tau |P(\tau) \rangle = M |P(\tau)\rangle
		\label{master_eqn_matrix_form}
	\end{equation}
	with an appropriately defined master operator $M$, a matrix with elements $M_{\bm n \bm n'}$; the off-diagonal entries give the rates for transitions from $\bm n'$ to $\bm n$. To ensure probability conservation, any reactions that would violate the upper bound $\bm n \leq \bm n^{\text{max}}$ are removed. In the concrete example of the gene regulation system of~\cref{main-sec:gene_regulation} in the main paper and~\cref{sec:example_SM_gene}, this, for example, applies to the $G \xrightarrow{k_t} G+P$ reaction from $n_P=n^\text{max}_P$ to $n_P=n^\text{max}_P+1$.

	The initial distribution over the state space is a product Poisson distribution with mean $\nzero{i}$ such that, \begin{equation}
	P(\bm n,0) = \mathcal{N}^{-1} \prod_i e^{- \nzero{i}} \frac{ \nzero{i}^{n_{i}} }{n_i!}
\end{equation}

	Due to the truncation of the state space an additional normalization factor $\mathcal{N} = \sum_{\bm n=0}^{\bm n^{\text{max}}} P(n,0)$ is in principle necessary here, though for sensibly chosen $n^\text{max}$ this will be very close to unity.

	Given the initial condition, $|P(\tau)\rangle$ can be found by integrating the system \cref{master_eqn_matrix_form} of linear differential equations using any standard algorithm, in the simplest case Euler integration with fixed time step $\Delta t$. Alternatively, and this is the approach we use throughout, one can write the solution as  $ | P(\tau) \rangle =  e^{M \tau} | P(0) \rangle$ and evaluate the matrix exponential by diagonalizing $M$. To do this one decomposes
	\begin{equation}
		M = R \Lambda L^{\rm T}
	\end{equation}
	where the columns of the matrices $R$ and $L$ are the right and left eigenvectors, respectively, normalized as $RL^{\rm T}=\mathbb{I}$, and $\Lambda$ is a diagonal matrix collecting the corresponding eigenvalues. We perform this diagonalization using standard numpy linear algebra packages for eigendecomposition. The matrix exponential is then 	$e^{M\tau} = R e^{\Lambda \tau} L^{\rm T} $ and accordingly
	\begin{equation}
		|P(\tau)\rangle = R e^{\Lambda \tau} L^{\rm T}  | P(0)\rangle
	\end{equation}

	Once $|P(\tau)\rangle$ is known, averages of one-time observables such as the mean copy number or its second moment can be evaluated straightforwardly as
	\begin{equation}
		\langle n_i(\tau) \rangle = \sum_{\bm n} n_i P(\bm n,\tau),  \qquad  \langle n_i(\tau) n_j(\tau) \rangle = \sum_{\bm n} n_i n_j P(\bm n,\tau)
	\end{equation}
	To measure two-time quantities such as the connected number correlator, $\langle \delta n_i(\tau) \delta n_j(\tau')\rangle$ we use
	\begin{equation}
		\langle \delta n_i(\tau) \delta n_j(\tau')\rangle = \langle \bm{1} | \hat{n}_i\, e^{M(\tau-\tau')} \hat{n}_j\, e^{M \tau'} | P(0) \rangle - \langle n_i(\tau) \rangle \langle n_j(\tau') \rangle
	\end{equation}
	where $\hat{n}_i$ is the number operator for species $i$, a diagonal matrix with the number of particles of species $i$ ($n_i=0,1,\ldots n^\text{max}_i$) for each state on the diagonal.

	To find the response function $R_{ij}(\tau,\tau')$, finally, we use the following procedure. Going back to the definition of the response as caused by temporary perturbations in the creation rate $k_{1j}$, we perturb $k_{1j}$ to $k_{1j}'=k_{1j}+\Delta k_{1j}$ at time $\tau'$ for a time $\Delta t$ and call $M'=M+\Delta M$ the perturbed master operator. The response function for $\tau>\tau'$ is then
	\begin{equation}
		\Delta t \,\Delta k_{1j}\, R_{ij}(\tau,\tau') = \langle \bm{1} | \hat{n}_i \, e^{M(\tau-(\tau'+\Delta t))} e^{M' \Delta t} e^{M \tau'} | P(0) \rangle - \langle \bm{1} | \hat{n}_i\, e^{M(\tau-(\tau'+\Delta \tau))} e^{M \Delta \tau} e^{M \tau'} | P(0) \rangle
	\end{equation}
	In the limit $\Delta t \rightarrow 0$ this simplifies to
	\begin{equation}
		R_{ij}(\tau,\tau') = \langle \bm{1} | \hat{n}_i\,e ^{M(\tau-\tau')} \frac{\Delta M}{\Delta k_{1j}} e^{M \tau'} | P(0) \rangle
	\end{equation}
	The result is independent of $\Delta k_{1j}$ as it should be because $\Delta M$ is directly proportional to $\Delta k_{1j}$.

	\section{Example systems}{\label{sec:examples_SM}}

	\subsection{Gene regulation}{\label{sec:example_SM_gene}}

		\begin{figure}
		\includegraphics[width=\linewidth]{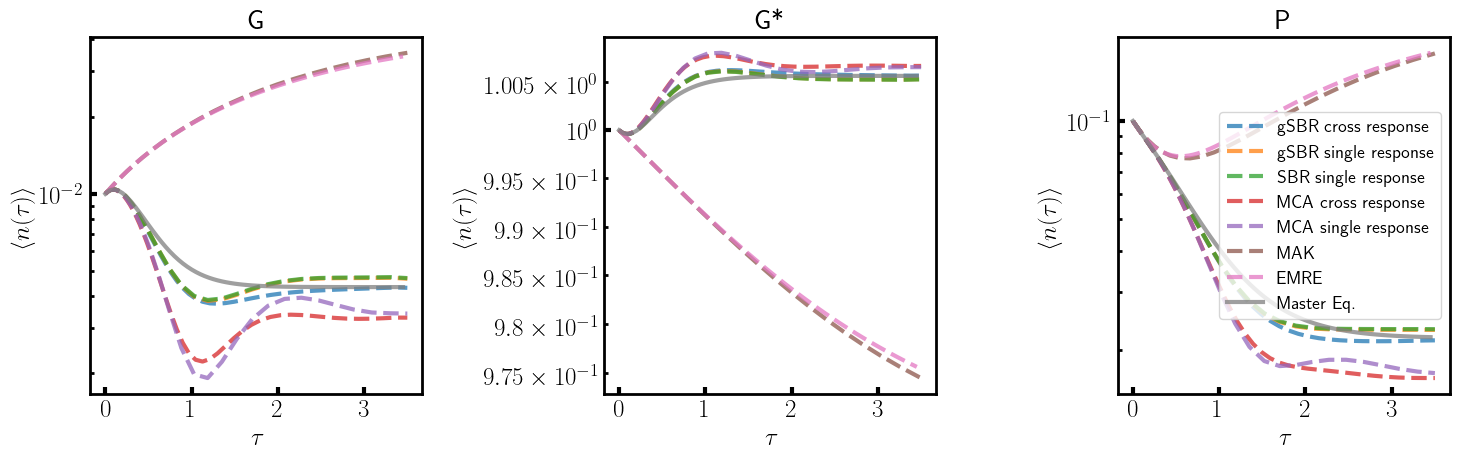}\\
		\includegraphics[width=\linewidth]{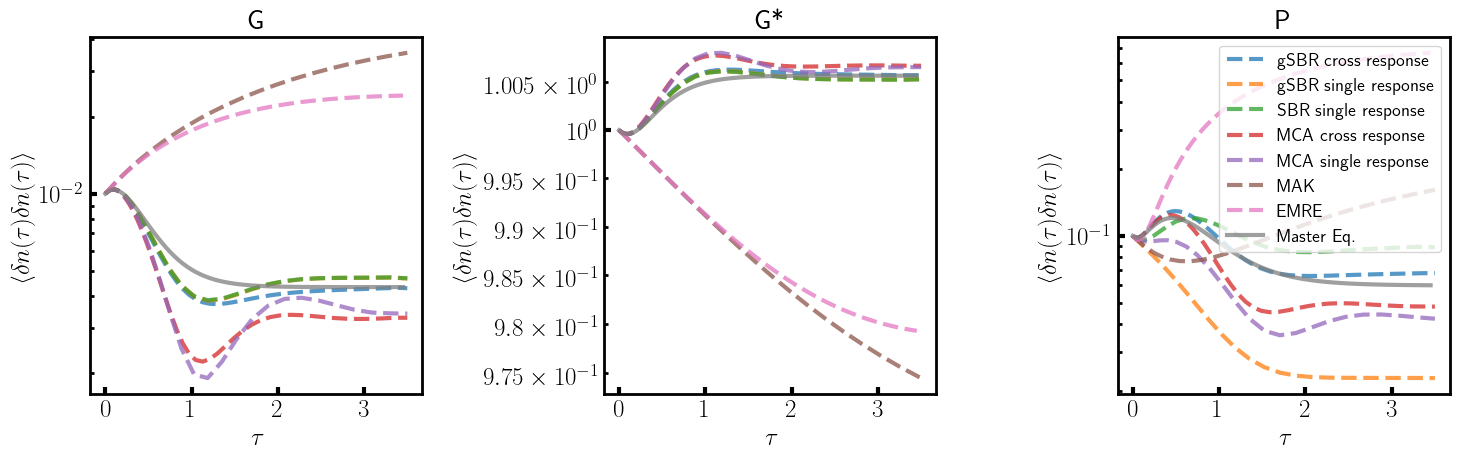}
		\caption[gene comparisons all methods]{Mean copy numbers (top row) and variances (bottom row) for the gene regulation system with the standard parameter set, as predicted by different methods. The moment closure method diverges rapidly as $\tau$ increases and is not plotted. The gSBR methods with and without cross species response functions perform best, followed by SBR, MCA and EMRE, the latter lying very close to the MAK results.}
		\label{fig_gene_5}
	\end{figure}

The gene regulation system treated in the main text in~\cref{main-sec:gene_regulation} is defined in the Julia Catalyst.jl framework as follows:
\begin{minted}[fontsize=\small]{julia}
	gene_system = @reaction_network
	begin
	@species G(t) = 0.01 G_star(t) = 1. P(t) = 0.1
	@parameters k_t = 10. k_b = 1. k_u = 0.1 k_2P = 2.
	(k_t), G --> G + P
	(k_b,k_u), G + P <--> G_star
	(k_2P), P --> 0
	end
\end{minted}

\begin{minted}[fontsize=\small]{julia}
	tstart= 0.
	tmax = 3.5
	atol = 1e-3
	rtol = 1e-2

	structure = Achedmy.ReactionStructure(gene_system)
	variables = Achedmy.ReactionVariables(structure,"cross")
	sol = Achedmy.solve_dynamics!(structure, variables, "gSBR", tmax, tstart, atol, rtol)
\end{minted}

A systematic comparison of results from a broad range of prediction methods for the means and variances of copy numbers in the gene regulation system is shown in~\cref{fig_gene_5}.

\subsection{Enzyme kinetics}{\label{sec:example_SM_enzyme}

The enzyme kinetics of the main text in~\cref{main-sec:enzyme} is defined in the Julia Catalyst.jl framework as follows:

\begin{minted}[fontsize=\small]{julia}
	enzyme_system = @reaction_network begin
	@species S(t) = 1. E(t) = 0.9 C(t) = 0.1 X(t) = 0.1
	@parameters k_f = 1. k_b = 0.1 k_d = 1. k_2X = 1. k_2S = 1. k_1S = 1.
	(k_f,k_b), S + E <--> C
	(k_d), C --> E + X
	(k_2X), X --> 0
	(k_2S,k_1S), S <--> 0
	end
\end{minted}

\begin{minted}[fontsize=\small]{julia}
	tstart= 0.
	tmax = 10.
	atol = 1e-3
	rtol = 1e-2

	structure = Achedmy.ReactionStructure(enzyme_system)
	variables = Achedmy.ReactionVariables(structure,"cross")
	sol = Achedmy.solve_dynamics!(structure, variables, "gSBR", tmax, tstart, atol, rtol)
\end{minted}

	\begin{figure}
	\includegraphics[width=\linewidth]{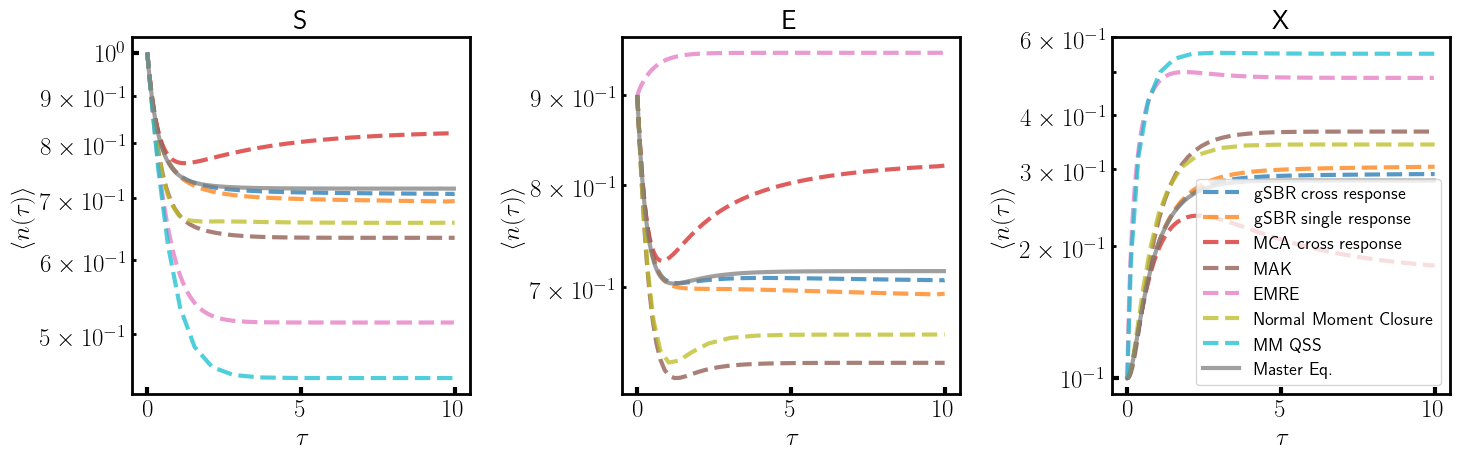}\\
	\includegraphics[width=\linewidth]{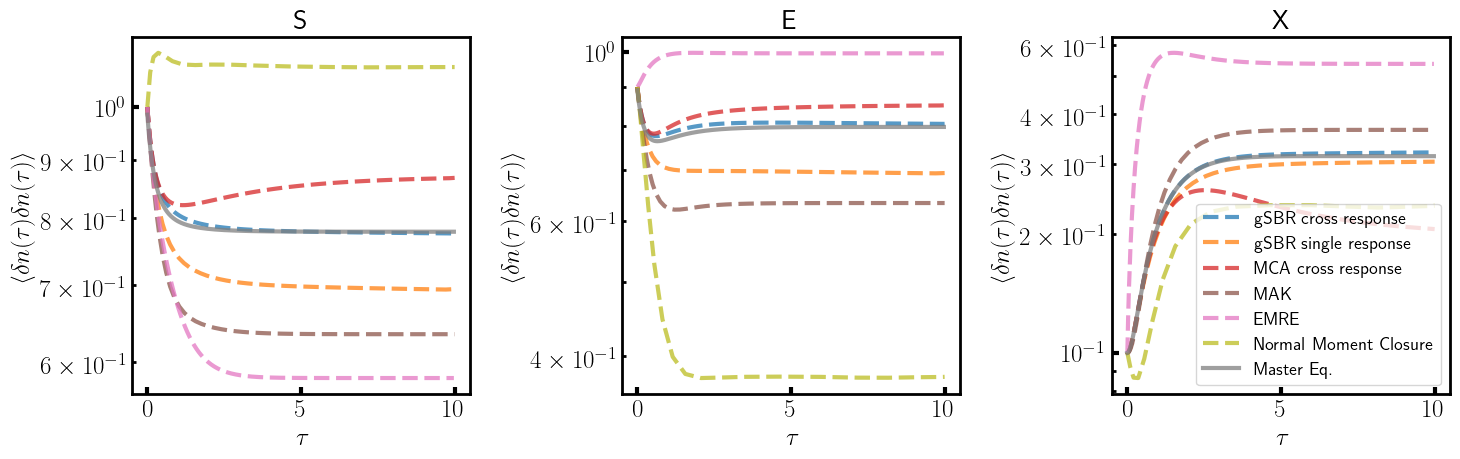}
	\caption[enzyme comparisons all methods]{Mean copy numbers (top row) and variances (bottom row) for the enzyme kinetics system with the standard parameter set, from different dynamical approximation methods. The gSBR methods with and without cross-species response functions perform the best, followed by MCA, MAK, EMRE and MM QSS. The SBR methods are unstable and not plotted here. Since the MM QSS is an approximate coarse grained approach that treats \textit{E} and \textit{C }in quasi steady state, it performs worst.}
	\label{fig_enzyme_3}
\end{figure}

	The well known Michaelis-Menten Quasi Steady State~\cite{MichaelisMenten2007} approximation for the dynamics of this system is given by the following dynamical equations for the mean copy numbers:
\begin{eqnarray}{\label{qss1}}
			\partial_\tau{\mu}_S &=  k_{1S} -k_{2S}\mu_S -k_{\text{max}} \frac{\mu_S}{\mu_S + K_m} \\
			\partial_\tau{\mu}_X &= k_{\text{max}} \frac{\mu_S}{\mu_S + K_m} - k_{2X} \mu_X
	\end{eqnarray}
	where $K_m$ is the Michaelis-Menten constant and $k_{\text{max}}$ is the maximum reaction velocity given by
	\begin{equation}{\label{qss2}}
		\begin{split}
			K_m &= \frac{k_d + k_b}{k_f} \\
			k_{\text{max}} &= k_d \mu_{E}^{\text{total}}
		\end{split}
	\end{equation}
Here $\mu_{E}^{\text{total}} = \mu_E + \mu_C$ is the mean total enzyme copy number, which stays conserved throughout the dynamics. The enzyme and the complex follow the substrate in QSS, with their concentrations given explicitly by
	\begin{equation}{\label{qss3}}
		\begin{split}
			\mu_E = \frac{K_m \mu_E^{\text{total}}}{K_m + \mu_S} \quad \quad
			\mu_C = \frac{\mu_S \mu_E^{\text{total}}}{K_m + \mu_S}
		\end{split}
	\end{equation}

	A systematic comparison of results from a broad range of prediction methods including the Michaelis-Menten Quasi Steady State (MM-QSS) approximation for the means and variances of copy numbers in the enzyme system is shown in~\cref{fig_gene_5}. Equal-time cross correlators between $SX$ and $EX$ for enzyme kinetics system from gSBR, LNA and the ground truth solution for three forward rates are shown in~\cref{fig_enz_2}.

		\begin{figure}
		\includegraphics[width=\linewidth]{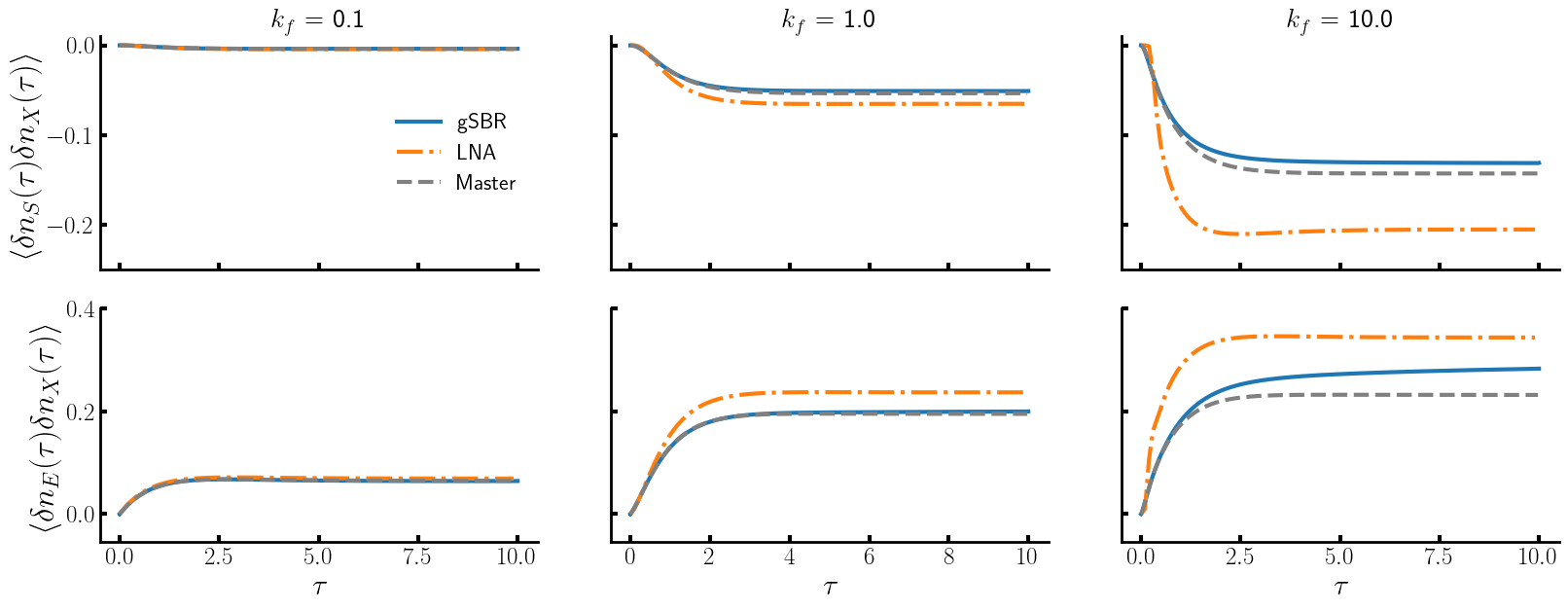}
		\caption[Cross correlators for enzyme kinetics system for three forward rates]{Equal-time cross-species correlators $\langle \delta n_S(\tau) \delta n_X(\tau) \rangle$ (top row),  $\langle \delta n_E(\tau) \delta n_X(\tau) \rangle$ (bottom row) as a function of rescaled time $\tau$ for $k_f = 0.1$ (left), $k_f = 1$ (middle) and $k_f = 10$ (right). The correlators are plotted as predicted by the gSBR, LNA and the ground truth master equation solution. In all the cases, gSBR is more faithful at reproducing the cross correlations than the LNA, with deviations at intermediate to long times.
		}
		\label{fig_enz_2}
	\end{figure}

	\subsection{SIR infection dynamics}{\label{sec:example_SM_SIR}

	The SIR infection dynamics of the main text in~\cref{main-sec:SIR} is defined in the Julia Catalyst.jl framework as follows:

	\begin{figure}
		\includegraphics[width=\linewidth]{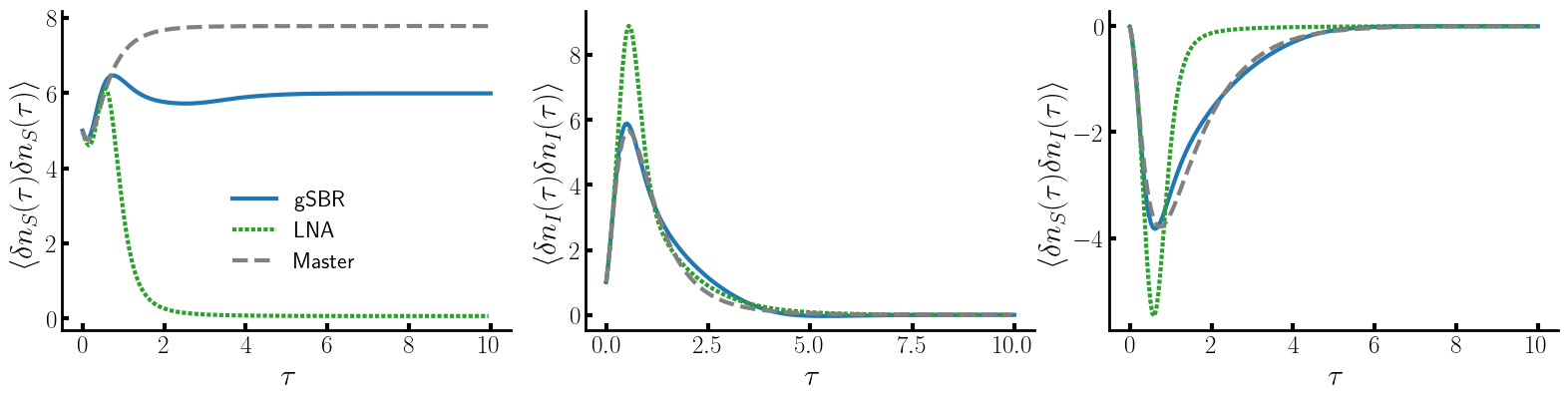}\\
		\caption[Variances and correlators in the SIR system]{Equal-time correlators $\langle \delta n_S(\tau) \delta n_S(\tau) \rangle$ (left), $\langle \delta n_I(\tau) \delta n_I(\tau) \rangle$ (middle) and $\langle \delta n_S(\tau) \delta n_I(\tau) \rangle$ (right) as functions of time $\tau$ as obtained using the gSBR, the LNA and the Master equation methods. The LNA wrongly predicts that the variance of S falls sharply to zero while the gSBR estimates the true variance qualitatively much better: it first increases and then stays large because of the bimodal steady state distribution. The variance of \textit{I} peaks at intermediate times and then decays to zero. The \textit{SI} cross correlator is negatively peaked at intermediate times because of the anti-correlation of \textit{S} and \textit{I}, but decays to zero at long times. These latter trends are reproduced by both approximation methods; however, the LNA overestimates the peak value in each case.}
		\label{SIR_3}
	\end{figure}

\begin{minted}[fontsize=\small]{julia}
	SIR_system = @reaction_network begin
	@species S(t) = 5. I(t) = 1.
	@parameters k_i = 0.75 k_2I = 1.
	(k_i), S + I --> I+I
	(k_2I), I --> 0
	end
\end{minted}

\begin{minted}[fontsize=\small]{julia}
	tstart= 0.
	tmax = 10.
	atol = 1e-3
	rtol = 1e-2
	dtmax = 0.05 #maximum time step allowed

	structure = Achedmy.ReactionStructure(SIR_system)
	variables = Achedmy.ReactionVariables(structure,"cross")
	sol = Achedmy.solve_dynamics!(structure, variables, "gSBR", tmax, tstart, atol, rtol, dtmax)
\end{minted}

We consider the variances of \textit{S }and \textit{I }and the equal-time cross-correlator between \textit{S }and \textit{I} in~\cref{SIR_3}. The LNA predicts that the steady state variance of \textit{S }falls sharply to zero after a small increase while the gSBR yields a qualitatively much more accurate prediction. The LNA overestimates the peaks in the variance of \textit{I}and the \textit{SI} cross-correlators as a function of time, while the gSBR closely reproduces the ground truth dynamics. The negative \textit{SI} cross correlation at short times is consistent from the negligible diagonal probability values in~\cref{main-SIR_1} (middle) of the main text.

\clearpage

\bibliographystyle{unsrtnat}
    \bibliography{paper}